\documentclass[journal]{IEEEtran}
\usepackage[utf8]{inputenc}
\usepackage[T1]{fontenc}

\ifCLASSINFOpdf
\else
   \usepackage[dvips]{graphicx}
\fi

\usepackage{cite}
\usepackage{amsmath,amssymb,amsfonts}
\usepackage{graphicx,color}
\usepackage{nameref}
\usepackage{ragged2e}
\usepackage{array} 
\usepackage{booktabs}
\usepackage{graphics}
\usepackage{mathrsfs}
\usepackage{subeqnarray}
\usepackage{xcolor}
\usepackage[ruled,vlined]{algorithm2e}
\usepackage{algorithmic}
\usepackage{mathtools}

\usepackage[ruled,vlined]{algorithm2e}
\usepackage{algorithmic}
\usepackage{mathtools}

\DeclareMathOperator*{\argmin}{arg\,min}
\newcommand{\pinv}{\dag}
\newcommand{\ma}[1]{\boldsymbol{#1}}
\newcommand{\compl}{\mathbb{C}}
\newcommand{\real}{\mathbb{R}}
\newcommand{\integer}{\mathbb{Z}}
\newcommand{\diagof}[1]{{\rm diag\!}\left\{ #1 \right\}}
\newcommand{\opvec}[1]{{\rm vec}\left\{ #1 \right\}}
\newcommand{\vecd}[1]{{\rm vecd}\left\{ #1 \right\}}
\newcommand{\ten}[1]{\mathcal #1}
\newcommand{\fronorm}[1]{\left|\left|#1\right|\right|_{\rm F}}
\newcommand{\fronormbig}[1]{\bigl|\bigl|#1\bigr|\bigr|_{\rm F}}
\newcommand{\krp}{\diamond}
\newcommand{\kron}{\otimes}
\newcommand{\trans}{{{\rm T}}}

\newcommand{\unf}[2]{\left[ \ten{#1} \right]_{(#2)}}
\newcommand{\mat}[1]{\ensuremath{{\mathbf{#1}}}}
\newcommand{\vet}[1]{\mathit{\boldsymbol{#1}}}
\newcommand{\lowint}[1]{\left\lceil{#1}\right\rceil}

\newtheorem{remark}{Remark}
\newtheorem{proposition}{Proposition}

\usepackage{tikz}
\newcommand*\circled[1]{%
  \begin{tikzpicture}[baseline=-0.6ex]
    \node[draw,circle,inner sep=0.5pt]{\scalebox{0.5}{#1}};
  \end{tikzpicture}}
\newcommand{\mwcontract}[3]{\hspace{2pt}\circled{#1}_{#2}^{#3}\hspace{1pt}}

\begin{document}
\bstctlcite{IEEEexample:BSTcontrol}
\title{Semi-Blind Receivers for\\~Hybrid Reflecting and Sensing RIS
\thanks{This work is partially supported by the National Institute of Science and Technology (INCT-Signals), sponsored by Brazil's National Council for Scientific and Technological Development (CNPq) under grant 406517/2022-3, and CAPES/Brazil. A. L. F. de Almeida is partially supported by CNPq under grant 312491/2020-4.}
}

\author{Amarilton L. Magalhães\IEEEauthorrefmark{1}\IEEEauthorrefmark{4}, André L. F. de Almeida\IEEEauthorrefmark{1}\\Federal University of Ceará\IEEEauthorrefmark{1}, Federal Institute of Education, Science and Technology of Ceará\IEEEauthorrefmark{4}, Brazil}

\maketitle

\begin{abstract}
Recent research has delved into advanced designs for reconfigurable intelligent surfaces (RIS) with integrated sensing functions. One promising concept is the hybrid RIS (HRIS), which blends sensing and reflecting meta-atoms. This enables HRIS to process signals, aiding in channel estimation (CE) and symbol detection tasks. This paper formulates novel semi-blind receivers for HRIS-aided wireless communications that enable joint symbol and CE at the HRIS and BS. The proposed receivers exploit a tensor coding at the transmit side, while capitalizing on the multilinear structures of the received signals. We develop iterative and closed-form receiver algorithms for joint estimation of the uplink channels and symbols at both the HRIS and the BS, enabling joint channel and symbol estimation functionalities. The proposed receivers offer symbol decoding capabilities to the HRIS and ensure ambiguity-free separate CE without requiring an \emph{a priori} training stage. We also study identifiability conditions that provide a unique joint channel and symbol recovery, and discuss the computational complexities and tradeoffs involved in the proposed semi-blind receivers. Our findings demonstrate the competitive performances of the proposed solutions at the HRIS and the BS and unveil distinct performance trends based on the possible combinations of HRIS-BS receiver pairs. Finally, extensive numerical results elucidate the interplay between power splitting, symbol recovery, and CE accuracy in HRIS-assisted communications. Such insights are pivotal for optimizing receiver design and enhancing system performance in future HRIS deployments.
\end{abstract}
\begin{IEEEkeywords}
Hybrid RIS, joint channel and symbol estimation, reconfigurable surfaces, semi-blind receivers, tensor modeling.
\end{IEEEkeywords}
\IEEEpeerreviewmaketitle

\section{Introduction}
\IEEEPARstart{R}{ecently}, reconfigurable intelligent surface (RIS) has been envisioned as a key enabling technology for deploying future wireless networks, for example, the sixth generation (6G) \cite{tishchenko2025theemergence, katwe2024overview, alexandropoulos2024hybrid, basar2024reconfigurable,chen2023reconfigurable, pan2021reconfigurable}. RIS is a large array of passive reflecting elements mounted on a planar surface that can independently interact with the impinging electromagnetic waves by means of software-controllable phase shifts \cite{you2025next, huang2020holographic, wu2020towards,basar2024reconfigurable, alexandropoulos2024hybrid}. Several applications for RIS can be found in the literature, such as coverage for users located in dead zones and co-channel interference suppression for users located at the edges of cells \cite{wu2020towards,wu2020beamforming}, improvement of the physical layer security \cite{dong2020secure}, integration with unmanned aerial vehicles (UAVs) and other aerial platforms \cite{alfattani2021aerial}, simultaneous wireless information and power transfer (SWIPT) \cite{wu2020towards}, and integrated sensing and communications (ISAC) \cite{chepuri2023integrated}, to mention a few.

In this context, accurate channel state information (CSI) is crucial in optimizing RIS-assisted systems \cite{han2023csi, renzo2020smart}. Its acquisition is necessary and challenging for designing the RIS reflection coefficients as well as the precoder/beamformer at both the base station (BS) and user terminal (UT) \cite{swindlehurst2022channel, schroeder2022two, jin2021channel, wu2020beamforming}. In general, channel estimation (CE) in multiple-input multiple-output (MIMO) RIS-assisted wireless communication systems faces two main challenges. The first is related to a notable increase in the required number of pilots compared to conventional systems, driven by a large number of RIS elements, leading to a significantly large number of channel coefficients \cite{zhu2023sensing, yang2023separate, luo2023reconfigurable, zhang2023channel, hu2022semipassive, lin2021tensor}. The second one is the unavailability of estimating the separate channels from the RIS-assisted one, namely UT-RIS and RIS-BS links, since the passive RIS (PRIS) acts only as a signal reflector \cite{yang2023separate, zhang2023channel, hu2022semipassive, liu2020deep, lin2021tensor} and only the cascaded channel is estimated\footnote{\linespread{0.7}\selectfont The cascaded (RIS-assisted, concatenated, composite, combined, or compound) channel comprises the joint effect of both UT-RIS and RIS-BS channels. Sometimes, the cascaded CE is achieved through its decoupled version, whose estimated matrices are affected by scaling ambiguities, as shown in \cite{wei2021channel} and \cite{dearaujo2021channel}. This necessitates complementary techniques to acquire the knowledge required for scaling removal.} so that all the receiver processing is done only at the BS or the UT. Solutions in the literature that have addressed these challenges are usually sorted into two families of methods: the first one incorporates novel algorithms to leverage the channel structure while preserving the original hardware properties of the PRIS. In contrast, the second approach involves modifying its hardware architecture to allow for additional signal processing capabilities at the RIS \cite{zhu2023sensing}. This paper relies on the second approach.

Given the passive nature of the RIS, most CE-related work commonly falls into cascaded CE, which is sufficient for applications like rate maximization and beamforming design. In contrast, a scaling ambiguity-free separate CE is preferred for applications like channel sounding, user localization, and mobility tracking, as highlighted in \cite{choi2023ajoint}. For instance, in mobility scenarios, \cite{zegrar2020ageneral} argues that separate CE facilitates channel tracking by identifying the behavior of the individual links under temporal variations. As pointed out therein, cascaded CE complicates tracking as changes occur in either the UT-RIS, RIS-BS, or both links. Moreover, in scenarios where the UT-RIS channel changes more rapidly than the RIS-BS one, the former must be estimated more often, while the latter not, highlighting the importance of recovering such channels individually instead of the combined one \cite{swindlehurst2022channel, nadeem2020asymptotic}. On the other hand, some designs depend on the availability of the individual channels, such as in \cite{li2022joint}, \cite{ye2020joint}, and \cite{sun2022energy}. The importance of estimating the involved channels separately is reinforced in \cite{zhang2023channel, boiadjieva2023joint, yang2023active, hu2021twotimescale, chen2021low}.

From a hardware perspective, a notable study was carried out in \cite{taha2021enabling}. Specifically, the authors in \cite{taha2021enabling} proposed a RIS structure by sparsely replacing some passive reflecting elements with active sensors connected to a single receive RF-chain each, thereby enabling baseband processing at the RIS controller. This receive RF-chain is comprised of a low noise amplifier, a downconverter mixer (from RF to baseband), and an analog-to-digital converter \cite{alexandropoulos2020ahardware}. These active elements merely sense the impinging signal without reflection. Adopting this hybrid architecture\footnote{\linespread{0.7}\selectfont This architecture has been referred to in the literature using different terminologies, such as \emph{hybrid semi-passive} RIS, \emph{sensing} RIS, \emph{receiving} RIS, or simply \emph{hybrid} RIS. Such an architecture should not be confused with the one considered in this paper, as will be clear later.}, the authors introduced a CE scheme based on compressed sensing and deep learning, achieving minimal pilot overhead and facilitating the CE process at the sacrifice of increased hardware complexity and power consumption. Inspired by the pioneering research in \cite{taha2021enabling}, such a hybrid architecture has been comprehensively investigated in subsequent works, such as in \cite{liu2020deep, chen2021low, lin2021tensor, hu2022semipassive} and references therein.

In contrast to RIS, another metasurface-based technology called dynamic metasurface antenna (DMA) has emerged to enable low-cost, extremely large antenna arrays \cite{shlezinger2021dynamic}. Despite the differences between the RIS and DMA operation, it is attainable to envision a hybrid meta-atom capable of reflecting and sensing since their meta-atoms share similarities \cite{alexandropoulos2024hybrid}. Motivated by DMAs, a novel metasurface was introduced in \cite{alamzadeh2021reconfigurable}, where the physical structure of each meta-atom is modified to couple small portions of the incoming wave. Relying on this paradigm, \cite{alexandropoulos2024hybrid} introduced the so-called hybrid RIS (HRIS) architecture, outlining its prospects/obstacles. These meta-atoms are integrated into sampling waveguides similarly to \cite{alamzadeh2021reconfigurable}, and the sensed signals are forwarded \emph{via} analog combining to RF-chains, whose outputs enable signal processing in the digital domain while retaining their reconfigurable reflection functionalities. The relationship of both reflected and sensed portions is dictated by the coupling level, controlled by changing either the substrate-integrated waveguide or the annular slot sizes \cite{alexandropoulos2024hybrid}. Based on this architecture and leveraging transmitted pilots, the work \cite{zhang2023channel} exploits the signal processing capabilities at the HRIS to estimate the UTs-HRIS channels from the sensed signal part. In contrast, the HRIS-BS channel is estimated at the BS from the reflected one. This is accomplished by exploiting a feedback control link (CL) between the HRIS and the BS, through which the BS acquires the UT-HRIS channel matrix estimated at the HRIS. This CL is used to reconfigure the reflection patterns of the RIS \cite{basar2024reconfigurable,Sokal_2023,wu2020towards}. In \cite{zhang2023channel}, error-free transmission over a high-throughput CL was considered. Detailed studies of different advanced/hybrid architectures are provided in \cite{tishchenko2025theemergence} and \cite{basar2024reconfigurable}.

Tensor decompositions have been successfully applied to model wireless communication systems \cite{Almeida_Elsevier_2007,teseandre}, including blind/semi-blind receivers \cite{sidiropoulos2000blind}, space-time (ST)/space-time-frequency (STF) coding schemes \cite{favier2012tensor, favier2014tensor}. These works have highlighted the effectiveness of tensor decompositions and their powerful uniqueness properties to harness the multidimensional nature of received signals and channels for deriving receiver algorithms capable of operating semi-blindly under less restrictive requirements than competing (matrix-based) methods, while offering good performance/complexity tradeoffs. See \cite{Chen2021, Miron2020, Sidiropoulos2017, Almeida2016} and references therein for an overview. Recent works have introduced tensor modeling to the context of passive RIS (PRIS)-aided communications to solve problems such as CE, semi-blind joint CE and symbol detection, and channel tracking \cite{dearaujo2021channel, ardah2021trice, wei2021channel, dearaujo2022semiblind, dearaujo2023semiblind, gomes2023channel}. Among these works, parallel factor (PARAFAC) decomposition, also known as canonical polyadic decomposition (CPD) \cite{comon2009tensor}, was applied to solve the CE estimation problem in a PRIS approach \cite{dearaujo2021channel, wei2021channel, sepideh2024anefficient}, and also \cite{lin2021tensor,magalhaes2025closed} in a hybrid semi-passive one. More recently, \cite{gomes2023channel} proposed PARAFAC-based algorithms for CE accounting for RIS operating under imperfections from real-world effects. In particular, without requiring prior CE via training sequences, \cite{dearaujo2022semiblind} and \cite{dearaujo2023semiblind} introduced data-aided semi-blind CE methods for PRIS-aided communications using Khatri-Rao ST coding (KRSTC), integrating symbol detection and CE through closed-form and iterative receivers, respectively. In \cite{dearaujo2023semiblind}, the authors exploited the PARATUCK tensor decomposition \cite{harshman1996uniqueness, dealmeida2009space}, while a generalized version was presented in \cite{dearaujo2022semiblind}. In the work \cite{magalhaes2025reducing}, an algebraic framework was derived to reduce the computational complexity of the iterative receiver proposed in \cite{dearaujo2023semiblind}. Nonetheless, KRSTC thresholds the number of streams to the number of transmitting antennas.

Different from the aforementioned works, which are mostly restricted to PRIS, where the cascaded CE problem is concentrated at the BS, this paper shows that the estimation of the individual channels and the transmitted symbols can be achieved jointly at both the HRIS and BS in a semi-blind fashion by resorting to tensor modeling. Moreover, in contrast to \cite{zhang2023channel}, which relies on pilot-assisted CE at the HRIS, our approach expands upon this by incorporating joint symbol and CE directly at the HRIS. This is achieved iteratively or in closed form using simple algorithms without the need for pilot training. As will be discussed later, empowering HRIS with symbol-decoding capabilities is useful in several scenarios. Part of this work has been presented in a conference paper \cite{magalhaes2025closed}, which was limited to a pair of closed-form receivers. This work goes beyond our previous study by i) developing a series of new semi-blind ``HRIS-BS'' receiver pairs; ii) presenting detailed derivations of the proposed algorithms; iii) delving into identifiability, uniqueness, and computational complexity of the proposed solutions, while discussing their tradeoffs, and iv) providing an extensive numerical performance evaluation.

The contributions of this paper are summarized as follows:

\emph{First}, leveraging the HRIS architecture \cite{alexandropoulos2024hybrid} and assuming a one-way structured time domain protocol, we formulate the received signals at both the HRIS and BS using a new tensor formalism that disentangles the received signal into effective channel tensors and coded signal tensors following PARAFAC and/or Tucker decompositions. Exploiting these tensor models allows the HRIS to jointly estimate the associated channel and decode the transmitted symbols in a semi-blind fashion. By transmitting data symbols in advance during the CE stage, our approach can improve data rate and reduce symbol decoding delay compared to pilot-only methods.

\emph{Second}, capitalizing on the proposed tensor models, we derive semi-blind joint symbol and CE methods for HRIS-aided MIMO wireless communication systems. More specifically, we formulate iterative and closed-form receiver pairs split between HRIS and BS to solve the semi-blind CE problem effectively. The proposed receivers eliminate the need for training sequences and additional steps for scaling ambiguity removal on the estimated channels and symbols, while partially circumventing the path-loss effects induced by the cumulated UT-HRIS and HRIS-BS links.

\emph{Third}, we study identifiability at both the HRIS and the BS, derive a set of conditions that ensure a unique channel and symbol recovery, and discuss the computational complexities and trade-offs involved by the proposed semi-blind receivers. 

\emph{Finally}, extensive numerical results showcase the interplay between power splitting, symbol recovery, and CE accuracy in HRIS-assisted communications. Our findings demonstrate competitive performances among receivers at the HRIS and BS and uncover distinct performance trends based on the combinations of HRIS-BS receiver pairs. We also dive into a brief exploration of scenarios where a joint symbol and CE at the HRIS are useful.

The rest of the paper is organized as follows. Section \ref{sec:sys} describes the system and signal models at the HRIS and the BS, including the transmission protocol and the main assumptions. Section \ref{sec:tenmodel} derives the corresponding tensor signal models and develops the core equations associated with the receiver design. The proposed semi-blind receivers for the HRIS and the BS are detailed in Sections \ref{sec:receiverststc} and \ref{sec:receiverskrstc}. Sections \ref{sec:identifiability}, \ref{sec:uniqueness}, and \ref{sec:cost} discuss identifiability, uniqueness, and computational complexity, respectively. Section \ref{sec:pairs} presents the evaluated combinations of the ``HRIS-BS'' semi-blind receivers. Section \ref{sec:results} contains our numerical results, and Section \ref{sec:discussion} discusses the potential use cases benefiting from the proposed joint channel/symbol estimation at the HRIS. Finally, conclusions are drawn in Section \ref{sec:conclusion}.

\subsection{Notation and Properties}
We utilize lowercase $a$, bold lowercase $\vet{a}$, bold uppercase $\mat{A}$, and calligraphic $\ten{A}$ to denote scalars, column vectors, matrices, and tensors, respectively. The $(i,j)$-th element of $\mat{A}$ is denoted as $[\mat{A}]_{i,j}$. Transpose, conjugate, and Moore-Penrose pseudo-inverse of $\mat{A}$ are denoted as $\mat{A}^\trans$, $\mat{A}^\ast$, and $\mat{A}^\pinv$, respectively. The operator $\diagof{\vet{a}}$ constructs a diagonal matrix from $\vet{a}$. $\lowint{a}$ is the smallest integer greater than or equal to $a$, and the Frobenius norm is indicated by $\fronorm{\cdot}$. The symbols $\krp$, and $\kron$ represent the Khatri-Rao and Kronecker matrix products, respectively. Stated $\mat{A}\! \in\!\compl^{I \times J}$, the vectorization operator, denoted as $\opvec{\mat{A}}$, yields the vector $\vet{a} \in \compl^{JI \times 1}$. Conversely, the reverse operation, $\mathrm{unvec}_{I \times J}(\vet{a})$, restores the matrix $\mat{A}$. A tensor $\ten{A}\! \in \!\compl^{I_1 \times I_2 \times \cdots \times I_P}$ is a multidimensional array with order $P$. Unfolding is the procedure that reshapes a tensor into a matrix. For instance, a 3rd-order tensor can be \textit{matricized} such that one mode varies along the rows and the other two along the columns. This is referred to as $n$-mode unfolding, $n=\{1,2,3 \}$. The 1-mode, 2-mode, and 3-mode unfoldings of $\ten{A} \in \compl^{I_1 \times I_2 \times I_3}$ are respectively given by
\begin{align}
    \label{nmode1}\unf{A}{1} &= [\ten{A}_{\cdot\cdot 1},\cdots,\ten{A}_{\cdot\cdot K}] \in \compl^{I_1 \times I_3I_2},\\
    \label{nmode2}\unf{A}{2} &= [\ten{A}_{\cdot\cdot 1}^\trans,\cdots,\ten{A}_{\cdot\cdot K}^\trans] \in \compl^{I_2 \times I_3I_1},\\
    \label{nmode3}\unf{A}{3} &= [\opvec{\ten{A}_{\cdot\cdot 1}},\cdots,\opvec{\ten{A}_{\cdot\cdot K}}]^\trans \in \compl^{I_3 \times I_2I_1}.
\end{align}
In addition, $\ten{I}_{3,P} \!\in \!\real^{P \times P \times P}$ is the 3rd-order identity tensor. Consider two $P$-th order tensors $\ten{A} \!\in \!\compl^{I_1 \times \cdots \times I_p \times \cdots \times I_P}$ and $\ten{B} \!\in\! \compl^{J_1 \times \cdots \times J_q \times \cdots \times J_P}$, such that $I_P\! = \!J_P$ and $I_p \!= \!J_q$. We define the \textit{mode-wise contraction} operation as a contraction between slices of $\ten{A}$ and $\ten{B}$. For simplicity, we assume this operation affects the $P$-mode of such tensors, which gives
\begin{equation}\label{def:mw1}
    \hspace{-2ex} \ten{A}\mwcontract{P}{p}{q}\ten{B}\! \doteq \!\ten{C}\! \,\,\in\! \compl^{\scriptscriptstyle I_1\! \times\! \cdots\! \times I_{p\!-\!1}\! \times \!I_{p\!+\!1}\! \times\! \cdots\! \times\! I_{P\!-\!1}\! \times\! J_1\! \times \!\cdots\! \times\! J_{q\!-\!1}\! \times \!J_{q\!+\!1} \!\times\! \cdots\! \times\! J_P},
\end{equation}
where the $P$-mode slice of the $(2P\!-\!3)$-th order tensor $\ten{C}$ results from the tensor contraction between the $P$-mode slices of $\ten{A}$ and $\ten{B}$, involving mode $p$ of $\ten{A}$ and mode $q$ of $\ten{B}$. For instance, the mode-wise contraction of two 3rd-order tensors $\ten{A}$ and $\ten{B}$ is accomplished by
\begin{equation}\label{def:mw2}
    \ten{C}_{\cdot\cdot j} = \left(\ten{A}\mwcontract{3}{p}{q}\ten{B}\right)_{\cdot\cdot j} = \ten{A}_{\cdot\cdot j} \bullet_p^q \ten{B}_{\cdot\cdot j}, \quad j=1,\cdots,J_P \; .
\end{equation}
Throughout this paper, we make use of the following identities:
\begin{equation}\label{prop:vec}
	\opvec{\mat{ABC}} = (\mat{C}^\trans \kron \mat{A})\opvec{\mat{B}};
\end{equation}
\begin{equation}\label{prop:vecd}
	\opvec{\mat{ABC}} = \left(\mat{C}^\trans \krp \mat{A}\right)\vecd{\mat{B}}, \hspace{0.1cm} \mathrm{\mbox{for }} \mat{B} \mathrm{\mbox{ diagonal}};
\end{equation}
\begin{equation}\label{prop:kronkron}
	\mat{AB} \kron \mat{CD} = (\mat{A} \kron \mat{C})(\mat{B} \kron \mat{D});
\end{equation}
\begin{equation}\label{prop:diagab}
	\diagof{\vet{a}}\vet{b} = \diagof{\vet{b}}\vet{a}, \quad \mathrm{\mbox{for }} \vet{a},\vet{b} \in \compl^{P \times 1};
\end{equation}
\begin{equation}\label{prop:kronabvec}
	\vet{a} \kron \vet{b} = \opvec{ba^\trans} \in \compl^{PQ \times 1},
\end{equation}
where $\vet{a} \in \compl^{P \times 1}$ and $\vet{b} \in \compl^{Q \times 1}$.

Additionally, Tables \ref{tab:acro} and \ref{tab:notations} summarize the most essential acronyms and variable notations used in this paper, along with their meanings.

\begin{table}[!t]
    \caption{List of main acronyms and their meanings.}
	\label{tab:acro}
	\centering
    \begin{tabular}{>{\centering\arraybackslash}m{2cm} >{\centering\arraybackslash}m{5cm}}
        \midrule
		Acronym&Definition\\
		\midrule
        RIS&Reconfigurable intelligent surface\\
        HRIS&Hybrid RIS\\
        PRIS&Passive RIS\\
        BS&Base station\\
        UT&User terminal\\
        CE&Channel estimation\\
        CL&Control link\\
        KRSTC&Khatri-Rao space-time coding\\
        TSTC&Tensor space-time coding\\
        KronF&Kronecker factorization\\
        KRF&Khatri-Rao factorization\\
        BALS&Bilinear alternating least-squares\\
        TALS&Trilinear alternating least-squares\\
        \midrule
	\end{tabular}
\end{table}

\begin{table}[!t]
    \caption{List of main notations and their meanings.}
	\label{tab:notations}
	\centering
	\begin{tabular}{>{\centering\arraybackslash}m{2cm} >{\centering\arraybackslash}m{5cm}}
        \midrule
		Notation&Definition\\
		\midrule 
        $L$& number of UT antennas\\
        $M$& number of BS antennas\\
        $N$& number of HRIS elementos\\
        $N_c$& number of HRIS RF-chains\\
        $R$& number of user data streams\\
        $T$& number of symbol periods\\
        $K$& number of sub-frames\\
        $\ma{\Psi}$&reflecting phase-shift matrix\\
        $\ma{\Lambda}$&coding matrix (KRSTC)\\
        $\ten{W}$&coding tensor (TSTC)\\
        $\ten{T}_{\ma{\Phi}}$&sensing phase-shift tensor\\
        $\ma{\Phi}$&sensing phase-shift matrix\\
        $\mat{G}$&UT-HRIS channel matrix\\
        $\mat{H}$&HRIS-BS channel matrix\\
        $\ma{\Theta}$&combined channel matrix\\
        $\mat{X}$&symbol matrix\\
        $\mat{Q}$&composite matrix comprising symbols and UT-HRIS channel\\
        $\mat{Z}$&composite matrix comprising symbols and HRIS-BS channel\\
        $\ten{Y}^\mathrm{RC}$&detected signal tensor at the HRIS\\
        $\ten{Y}^\mathrm{BS}$&received signal tensor at the BS\\
        \midrule
	\end{tabular}
\end{table}

\section{System and Signal Models}\label{sec:sys}
\begin{figure}[!t]
	\centering
	\includegraphics[width=0.46\textwidth]{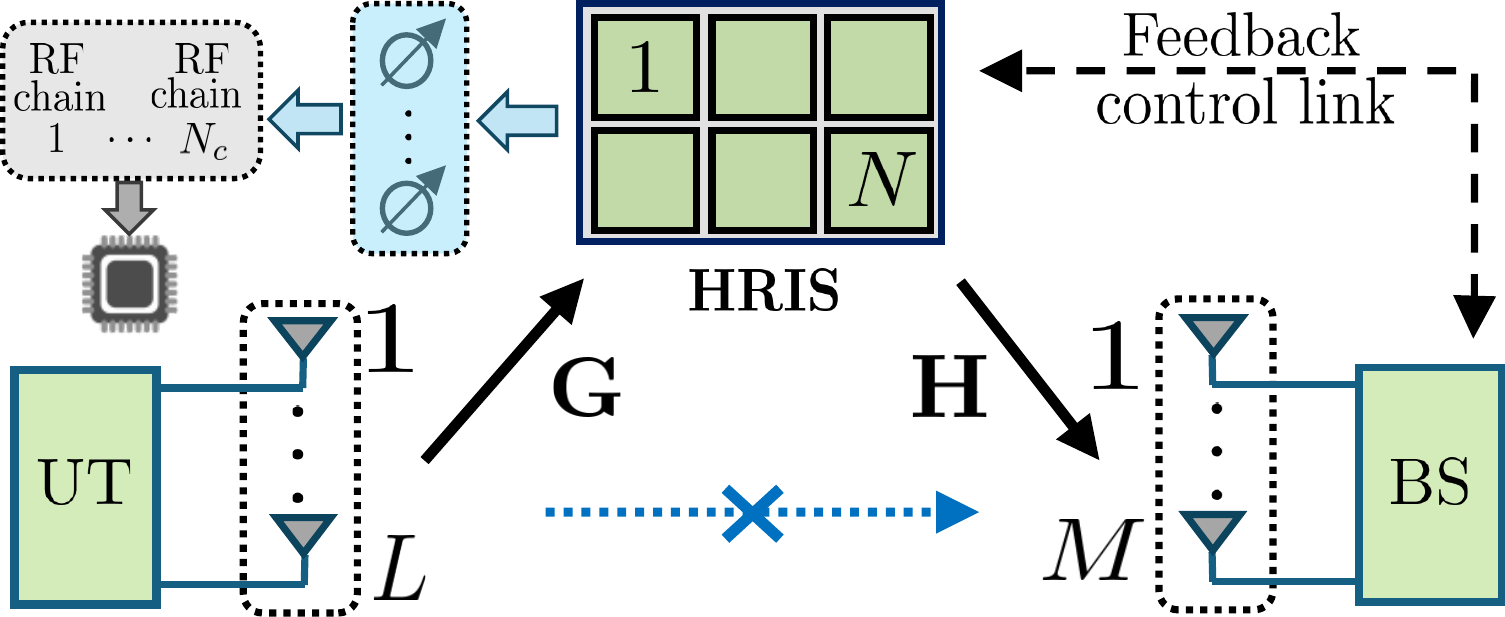}\\
	\caption{System model.}
	\label{fig:systemmodel}
\end{figure}

We consider a single-user HRIS-assisted MIMO communication system where the multi-antenna UT and BS are equipped with $L$ and $M$ antennas, respectively\footnote{\linespread{0.7}\selectfont Although we adopt a single user with $L$ antennas, the expressions derived in this work can be adapted for a scenario with $L$ single-antenna users.}. This work considers uplink communication\footnote{\linespread{0.7}\selectfont Although the primary focus of this paper is on uplink communication from the UT to BS, the results obtained can also be applied to downlink communication in the opposite direction by leveraging uplink-downlink channel reciprocity and simply reversing the roles of the transmitter and receiver.}. We suppose there is no direct link between the BS and UT due to blockages, and it is left out of the signal model. Hence, only non-LoS (NLoS) transmission is considered. In addition, the HRIS controller is linked to the BS \emph{via} a control feedback channel, which is assumed to be error-free. The UT communicates with the BS through the assistance of an HRIS comprising a metasurface of $N$ meta-atom elements connected \emph{via} analog combining to a digital controller through $N_c$ RF-chains \cite{alexandropoulos2024hybrid, zhang2023channel}, as depicted in Fig. \ref{fig:systemmodel}. We use the power split parameter $\rho_n(t)$ to represent the fraction of the signal reflected from the $n$-th HRIS meta-atom at the $t$-th time instant. Hence, $1-\rho_n(t)$ denotes the sensed portion forwarded to the RF-chains. $e^{j\psi_n(t)}$ is the controllable reflecting phase-shift of the $n$-th meta-atom at the $t$-th time instant, and $e^{j\phi_{n_c,n}(t)}$ is the phase-shift that models the joint effect on the wave captured by the $n$-th meta-atom element at the $t$-th time instant caused by the adjustable frequency response of the meta-material element by phase-shifting and the propagation inside the waveguide, which forwards to the $n_c$-th RF-chain. We consider $\psi_n(t),~\phi_{n_c,n}(t) \in [0,2\pi)$. As pointed out by \cite{zhang2023channel}, when the sensing elements are connected to multiple RF-chains, conventional phase-shifter networks are required to apply distinct phase-shifts to each chain. This reconfigurability reflects the external control over HRIS parameters ($\rho$, $\psi$, and $\phi$).

The design of the sensing phase-shifts depends on the degree of connectivity involving the sensing elements of the HRIS, and includes single-connected, partially-connected, or a more general fully-connected case. In this work, we consider a fully-connected HRIS architecture for generalization purposes. A single-connected one would be conceived by connecting each sensing element to an RF-chain, implying $\ma{\Phi}_k$ having a diagonal construction \cite{nguyen2024decision}. On the other hand, if we have a partially connected architecture, each group of $N_g$ elements would be connected to an RF-chain. Hence, the matrix $\ma{\Phi}_k$ would have a block diagonal structure \cite{gavras2025simultaneous}. These two architectures are particular cases of the full-matrix $\ma{\Phi}_k$ adopted in this paper.

A structured two-block time-domain transmission is adopted, during which the semi-blind CE  occurs in the first block of $T_s$ symbol periods, comprised of $K$ sub-frames of $T$ symbol periods each (i.e., $T_s\!=\!KT$). In contrast, the second block has $T_d$ symbol periods dedicated to pure data transmission. Note that during $T_s$ symbol periods, this structure spends the same time as that dedicated to addressing CE in \cite{zhang2023channel}, in which only pilots are transmitted. The key difference is that in our approach, data symbols are transmitted in advance during $T_s$ symbol periods, enhancing the data rate and reducing the overall symbol decoding delay. A quasi-static flat-fading channel with coherence time $T_c$ is assumed, where UT-HRIS and HRIS-BS channels remain constant during at least $T_s$ symbol periods, with $T_s \ll T_c$. Digital precoding/combiner design, RIS phase shift optimization, and signal processing in the second block fall outside the scope of this work.
\begin{figure}[!t]
	\centering
        \includegraphics[width=0.48\textwidth]{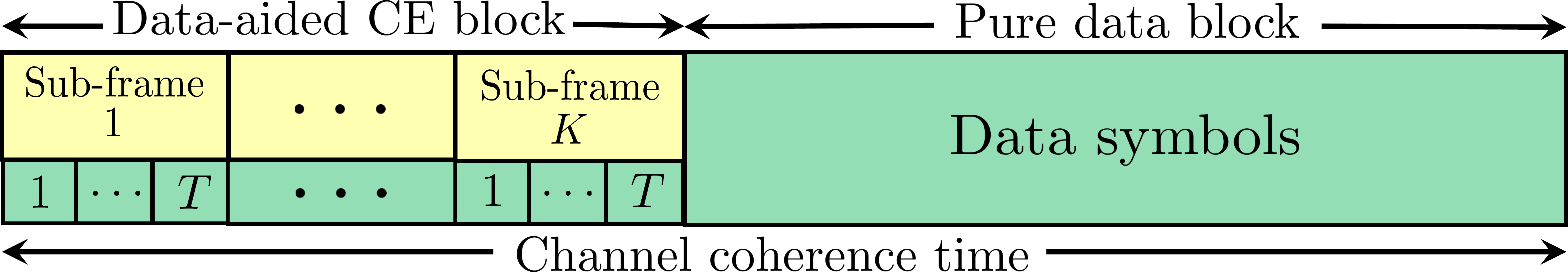}\\
 \caption{Transmission time structure.}
 	\label{fig:protocol}
\end{figure}

\begin{remark}
    We emphasize that the objective of this paper is not to compare passive and hybrid RIS architectures. Instead, it is assumed that, in certain scenarios, network engineers may opt to trade off the energy efficiency provided by passive RISs in favor of enabling additional functionalities. In such cases, the RIS is envisioned as an active network node, acting as an intermediate point capable of performing specific tasks and supporting more dynamic and autonomous network operations, which are not expected in a passive RIS architecture. For instance, the HRIS can act as a sensing receiver in an RIS-assisted ISAC scenario. Recall that in ISAC scenarios, BSs are allowed to act as dedicated sensing receivers, which may be integrated with or physically separated from the ISAC transmitters \cite{huang2022coordinated}. In addition, in RIS-assisted ISAC scenarios, due to its passive nature, the RIS typically requires a separate receiver unit to sense the environment and manage configuration control \cite{chepuri2023integrated, taghvaee2024fully}. In this context, \cite{strinati2024distributed} proposed a distributed ISAC architecture that enables cooperation among sensing nodes to understand their surrounding environment, allowing RISs with sensing capabilities (hybrid RISs) to replace sensing receiver nodes. Examples of such potential functionalities are discussed in Section \ref{sec:discussion}.
\end{remark}

\begin{remark}
    In the context of energy consumption, we assume that the adopted HRIS is capable of switching between two operational modes: \textit{hybrid mode} and \textit{passive mode}. In hybrid mode operation, the HRIS's active units related to the sensing process are turned on, i.e., adders, phase-shifters, and RF-chains (including low-noise amplifiers, downconverters, and analog-to-digital converters). When no sensing task is required, the HRIS can switch to the passive mode, and all the active units are turned off. This means that the active components are activated as needed to perform a channel estimation procedure, or even to the HRIS to extract some relevant information from the decoded/estimated symbols. In this paper, we consider the semi-blind joint symbol and channel estimation under the hybrid mode operation.
\end{remark}

\subsection{Tensor ST Coding (TSTC)}
Before transmission at the UT, the input symbols undergo a tensor ST coding scheme \cite{ximenes2014parafac}, which incorporates spatial multiplexing with spreading, enabling a linear combination of streams across both spatial and temporal dimensions. Consequently, our transmit signal model extends beyond that exploited in \cite{dearaujo2023semiblind}. This way, all the $R$ independent streams at the $t$-th symbol period of the $k$-th sub-frame ($k\!=\!1,\!\cdots\!,\!K$) are split across $L$ combiners. At the $l$-th combiner, each independent symbol $x_{r,t}$ ($r\!=\!1,\!\cdots\!,\!R$) is coded by $w_{l,r,k}$ ($l=1,\cdots,L$). After that, the $R$ coded signals are combined to yield $s_{l,t,k} = (1/\sqrt{L})\textstyle \sum_{r=1}^R{w_{l,r,k}x_{r,t}}$, to be forwarded to the $l$-th transmitting antenna. We assume that coefficients $w_{l,r,t,k}$ remain constant within the $k$-th sub-frame and may vary from one sub-frame to another, which means $w_{l,r,t,k} \!=\! w_{l,r,k}$, for $t=1,\cdots,T$. Collecting the coded symbols forwarded to all $L$ antennas, we have $\vet{s}_{t,k}\! =\! \mat{W}_k\vet{x}_t \!\in\! \compl^{L \times 1}$, in which $\vet{x}_t \!\doteq\![x_{1,t},\!\cdots\!,x_{R,t}]^\trans \!\in\! \compl^{R \times 1}$ is comprised by symbols coming from all the $R$ data streams at the $t$-th time instant, and $\mat{W}_k \in \compl^{L \times R}$ is the coding matrix of the $k$-th sub-frame gathering all $R$ inputs and $L$ outputs, whose entries are $[\mat{W}_k]_{l,r} = (1/\sqrt{L})w_{l,r,k}$. The HRIS parameters are assumed to be reconfigured just like the coefficients $w_{l,r,k}$, which leads to $\rho_{n,t,k} = \rho_{n,k}$, $\psi_{n,t,k} = \psi_{n,k}$ and $\phi_{n_c,n,t,k} = \phi_{n_c,n,k}$, following \cite{zhang2023channel}.

The portion of the signal transmitted by the $L$ UT antennas, sensed by the $N$ HRIS meta-atoms, and then forwarded to the $N_c$ RF-chains \emph{via} analog combining at the $t$-th time slot of the $k$-th sub-frame, is given by $\vet{y}^{\mathrm{RC}}_{t,k} = \ma{\Phi}_k\mat{G}\mat{W}_k\vet{x}_t + \vet{\nu}^{\mathrm{RC}}_{t,k} \in \compl^{N_c \times 1}$, where $\vet{\nu}^{\mathrm{RC}}_{t,k}$ is the associated additive noise at the HRIS, $\mat{G} \!\in\! \compl^{N \times L}$ is the UT-HRIS channel matrix and $\ma{\Phi}_k \in \compl^{N_c \times N}$ is the sensing phase shift matrix of the $k$-th sub-frame that yields the analog combining carried out by HRIS \cite{zhang2023channel}, in which $[\ma{\Phi}_k]_{n_c,n} \!= \!\sqrt{\!(1\!-\!\rho_{n,k})/N_c}e^{j\phi_{n_c,n,k}}$. Meanwhile, the received signal at the BS, corresponding to the portion of the signal reflected by the HRIS, is given by $\vet{y}^{\mathrm{BS}}_{t,k} = \mat{H}\diagof{\vet{\psi}_k}\mat{G}\mat{W}_k\vet{x}(t)  + \vet{\nu}^{\mathrm{BS}}_{t,k}\in \compl^{M \times 1}$, where $\vet{\nu}^{\mathrm{BS}}_{t,k}$ represents the noise at the BS, $\mat{H} \!\in\! \compl^{M \times N}$ is the HRIS-BS channel matrix, and $\vet{\psi}_k \doteq [\sqrt{\rho_{1,k}}e^{j\psi_{1,k}},\!\cdots\!,\!\sqrt{\rho_{N,k}}e^{j\psi_{N,k}}] \in \compl^{N \times 1}$ is the reflecting phase-shift beam. After $T$ time slots of the $k$-th sub-frame, we collect column-wise $\vet{y}^{\mathrm{RC}}_{t,k}$ and $\vet{y}^{\mathrm{BS}}_{t,k}$, into the matrices $\mat{Y}^{\mathrm{RC}}_k \doteq [\vet{y}^{\mathrm{RC}}_{1,k},\cdots,\vet{y}^{\mathrm{RC}}_{T,k}] \in \compl^{N_c \times T}$ and $\mat{Y}^{\mathrm{BS}}_k \doteq [\vet{y}^\mathrm{BS}_{1,k},\cdots,\vet{y}^\mathrm{BS}_{T,k}] \in \compl^{M \times T}$, respectively, to get
\begin{equation}\label{matyrckw}
	\mat{Y}^\mathrm{RC}_k = \ma{\Phi}_k\mat{G}\mat{W}_k\mat{X} + \mat{V}^\mathrm{RC}_k \in \compl^{N_c \times T}
\end{equation}
and
\begin{equation}\label{matybskw}
	\mat{Y}^\mathrm{BS}_k = \mat{H}\diagof{\vet{\psi}_k}\mat{G}\mat{W}_k\mat{X} + \mat{V}^\mathrm{BS}_k \in \compl^{M \times T},
\end{equation}
where $\mat{X} = [\vet{x}_1,\cdots,\vet{x}_T] \in \compl^{R \times T}$ is the symbol matrix, which collects $T$ symbol periods of the $R$ data streams. Here, $\mat{V}^\mathrm{RC}_k$ and $\mat{V}^\mathrm{BS}_k$ stand for the noise matrices associated with sub-frame $k$ at the HRIS and BS, respectively.

\subsection{Khatri-Rao ST Coding (KRSTC)}
We also consider KRSTC for comparison purposes. Following \cite{Sidropoulos_2002_KRST}, \cite{dearaujo2023semiblind} and the adopted time protocol, each independent symbol $x_{l,t}$ ($l=1,\cdots,L$) is coded by a coefficient $\lambda_{l,k}$ ($k=1,\cdots,K$) to yield $s_{l,t,k} = \lambda_{l,k}x_{l,t}$, which is then forwarded to the $l$-th transmitting antenna. Collecting the coded symbols for all $L$ antennas, we have $\vet{s}_{t,k} = \diagof{\vet{\lambda}_k}\vet{x}_t \in \compl^{L \times 1}$, in which $\vet{x}_t \doteq [x_{1,t},\cdots,x_{L,t}]^\trans \in \compl^{L \times 1}$ and $\vet{\lambda}_k \doteq [\lambda_{1,k},\cdots,\lambda_{L,k}]^\trans \in \compl^{L \times 1}$ is the coding vector of the $k$-th sub-frame. In this case, \eqref{matyrckw} and \eqref{matybskw} are, respectively, recast as
\begin{equation}\label{matyrckkr}
    \mat{Y}^\mathrm{RC}_k = \ma{\Phi}_k\mat{G}\diagof{\vet{\lambda}_k}\mat{X} + \mat{V}^\mathrm{RC}_k  \in \compl^{N_c \times T},
\end{equation}
\begin{equation}\label{matybskkr}
    \mat{Y}^{\mathrm{BS}}_k = \mat{H}\diagof{\vet{\psi}_k}\mat{G}\diagof{\vet{\lambda}_k}\mat{X} + \mat{V}^\mathrm{BS}_k  \in \compl^{M \times T},
\end{equation}
where the symbol matrix is recast as $\mat{X} \in \compl^{L \times T}$. Note that KRSTC is a special case of TSTC, where the coding matrix associated with the $k$-th sub-frame is diagonal. This implies $R=L$ and the absence of signal combining/multiplexing at the transmitter. The signal model of \eqref{matybskkr}, which follows the PARATUCK model, was considered in \cite{dearaujo2023semiblind} for the PRIS case.

\section{Tensor Signal Modeling}\label{sec:tenmodel}
In the following, we recast the received signals using a tensor approach. We can collect $\mat{Y}^\mathrm{RC}_k$, the signal sensed at the HRIS associated with the $k$-th sub-frame defined in \eqref{matyrckw}, to form the 3rd-order sensed signal tensor at the HRIS $\ten{Y^\mathrm{RC}} \doteq \mat{Y}^\mathrm{RC}_1 \sqcup_3 \mat{Y}^\mathrm{RC}_2 \sqcup_3 \cdots \sqcup_3 \mat{Y}^\mathrm{RC}_{K} \in \compl^{N_c \times T \times K}$, where $\sqcup_3$ indicates a concatenation along the third dimension, for $k=1,\cdots,K$. This way, $\mat{Y}^\mathrm{RC}_k$ can be viewed as a frontal slice of $\ten{Y^\mathrm{RC}}$, i.e., $\ten{Y^\mathrm{RC}}_{\cdot\cdot k} \in \compl^{N_c \times T}$. This matrix is constructed by fixing the third-mode index $k$ and varying the tensor along modes one and two. Likewise, we construct the 3rd-order received signal tensor at the BS $\ten{Y^\mathrm{BS}} \doteq \mat{Y}^\mathrm{BS}_1 \sqcup_3 \mat{Y}^\mathrm{BS}_2 \sqcup_3 \cdots \sqcup_3 \mat{Y}^\mathrm{BS}_{K} \in \compl^{M \times T \times K}$, for which $\mat{Y}^\mathrm{BS}_k$, defined in \eqref{matybskw}, matches to the $k$-th frontal slice $\ten{Y^\mathrm{BS}}_{\cdot\cdot k} \in \compl^{M \times T}$. The scalar representations of $\ten{Y^\mathrm{RC}}$ and $\ten{Y^\mathrm{BS}}$ are, respectively, given by
\begin{equation*}\label{scalaryrc}
	y_{n_c,t,k}^{\mathrm{RC}} \!=\!\!\sum\limits_{n=1}^N\!\sum\limits_{l=1}^L\!\sum\limits_{r=1}^R{\!\!\sqrt{\frac{1\!-\!\rho_{n,k}}{LN_c}}e^{j\phi_{n_c,n,k}}\!g_{n,l}w_{l,r,k}x_{r,t}} \!+\! \nu_{n_c,t,k}^{\mathrm{RC}}
\end{equation*}
and
\begin{equation*}\label{scalarybs}
	y_{m,t,k}^{\mathrm{BS}} \!=\!\! \sum\limits_{n=1}^N\!\sum\limits_{l=1}^L\!\sum\limits_{r=1}^R{\!\!\sqrt{\frac{\rho_{n,k}}{L}}h_{m,n}e^{j\psi_{n,k}}g_{n,l}w_{l,r,k}x_{r,t}} \!+\! \nu_{n_c,t,k}^{\mathrm{BS}}.
\end{equation*}
\begin{figure}[!t]
    \centering
    \includegraphics[width=0.4\textwidth]{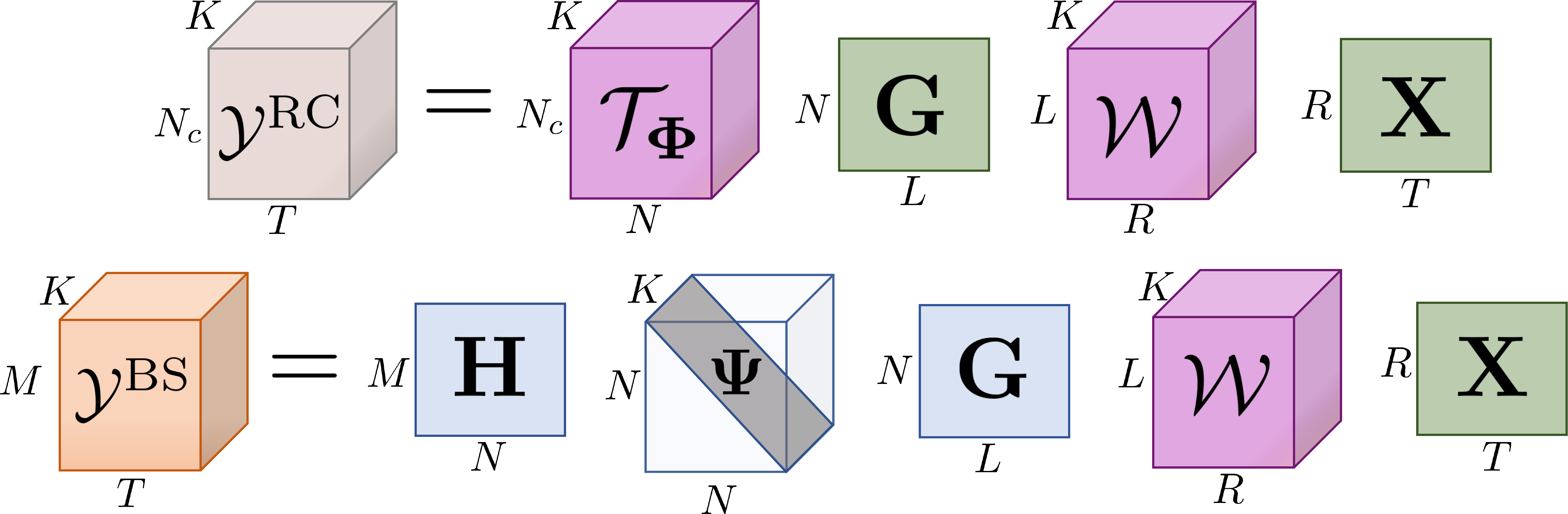}
    \caption{Tensor structures of the received signals at the HRIS and the BS.}
    \label{fig:ten}
\end{figure}
Indeed, these signals are three-way (3D) arrays having one spatial dimension ($N_c$ for the HRIS or $M$ for the BS) and two temporal dimensions ($T$ and $K$). These tensor models are exploited later to derive the proposed HRIS-BS semi-blind receiver pairs.

\subsection{Received Signal Tensors in a Decoupled Perspective}
In this section, we show how the received tensors at the HRIS and BS can be viewed as a combination of two tensors whose structures may follow PARAFAC or Tucker models in a decoupled perspective. 

Regarding the TSTC scheme, and upon closer examination, the noiseless part of $\ten{Y}^\mathrm{RC}$ can be viewed as a slice-wise product between the \textit{effective UT-HRIS channel} $\ten{C}$ and the \textit{coded symbol tensor} $\ten{S}$, and can be written using the $n$-mode product notation \cite{kolda2009tensor} as $\ten{C} \!=\! \ten{T}_{\ma{\Phi}} \times_1 \mat{I}_{N_c} \times_2 \mat{G}^\trans \times_3 \mat{I}_K \!\in\! \compl^{N_c \times L \times K}$ and $\ten{S} \!=\! \ten{W} \times_1 \mat{I}_L \times_2 \mat{X}^\trans \times_3 \mat{I}_K \!\in\! \compl^{L \times T \times K}$, respectively. At the same time, $\ten{T}_{\ma{\Phi}} \doteq \ma{\Phi}_1 \sqcup_3 \ma{\Phi}_2 \sqcup_3 \cdots \sqcup_3 \ma{\Phi}_K \in \compl^{N_c \times N \times K}$ and $\ten{W} \doteq \mat{W}_1 \sqcup_3 \mat{W}_2 \sqcup_3 \cdots \sqcup_3 \mat{W}_K \in \compl^{L \times R \times K}$ are, respectively, the \textit{sensing phase shift tensor} and the \textit{coding tensor}. Using the $K$-mode slice contraction operator \eqref{def:mw1}, we have
\begin{align}\label{tenyrc}
    \ten{Y}^\mathrm{RC} =& \: \ten{C}\mwcontract{3}{2}{1}\ten{S} + \ten{V}^\mathrm{RC}\nonumber,\\
    =& \left(\ten{T}_{\ma{\Phi}} \times_2 \mat{G}^\trans\right) \mwcontract{3}{2}{1} \left(\ten{W} \times_2 \mat{X}^\trans\right) + \ten{V}^\mathrm{RC},
\end{align}
where $\ten{V}^\mathrm{RC}$ represents the additive noise tensor at the HRIS. From this perspective, the sensed signal at the HRIS results from a mode-wise contraction of the tensors $\ten{C}$ and $\ten{S}$, which follow Tucker-(1,3) models \cite{FavierAlmeida2014}, respectively. We refer to the tensor model in \eqref{tenyrc} as a \textit{double Tucker} model. 

The noiseless part of the tensor $\ten{Y}^\mathrm{BS}$ results from a mode-wise contraction between the \textit{cascaded UT-HRIS-BS channel tensor} $\ten{T}_{\ma{\Omega}}$ and the \textit{coded symbol tensor}, where the first follows a PARAFAC model and can be written as $\ten{T}_{\ma{\Omega}} = \ten{I}_{3,N} \times_1 \mat{H} \times_2 \mat{G}^\trans \times_3 \ma{\Psi} \in \compl^{M \times L \times K}$, where $\ma{\Psi} \doteq [\vet{\psi}_1,\cdots,\vet{\psi}_K]^\trans \in \compl^{K \times N}$. Hence, $\bar{\ten{Y}}^\mathrm{BS}$ is given by
\begin{align}\label{tenybs}
   \hspace{-0.3cm}\ten{Y}^\mathrm{BS} =&\: \ten{T}_{\ma{\Omega}} \mwcontract{3}{2}{1} \ten{S} + \ten{V}^\mathrm{BS}\nonumber\\
   =&\! \left(\ten{I}_{3,N} \!\times_1 \!\mat{H} \!\times_2\! \mat{G}^\trans\!\! \times_3\! \ma{\Psi}\right) \!\mwcontract{3}{2}{1}\! \left(\ten{W} \!\times_2 \!\mat{X}^\trans\right) \!+\! \ten{V}^\mathrm{BS}\!\!,
\end{align}
where $\ten{V}^\mathrm{BS}$ is the corresponding additive noise tensor at the BS. Since the signal tensor received at the BS corresponds to the 3-mode contraction between a PARAFAC and a Tucker-(1,3) models, we refer to \eqref{tenybs} as a \textit{PARAFAC-Tucker} model. The received signal tensors at both the HRIS and BS are illustrated in Fig. \ref{fig:ten}. 

The mode-wise contraction formalism, applied in \eqref{tenyrc} and \eqref{tenybs}, makes it possible to decouple the tensor structures of the transmitted signals and their respective combined/effective channels, revealing their associated (PARAFAC/Tucker) tensor decompositions in a modularized fashion. 

If KRSTC is used instead of TSTC, the received signal tensors $\ten{Y}^\mathrm{RC}$ and $\ten{Y}^\mathrm{BS}$ can be built the same way, assuming $R=L$. The coded symbol tensor follows a PARAFAC model and is written as $\bar{\ten{S}}\! =\! \ten{I}_{3,L} \!\times_1\! \mat{I}_L \!\times_2 \!\mat{X}^\trans \!\times_3\! \ma{\Lambda} \!\in\! \compl^{L \times T \times K}$, where $\ma{\Lambda}\! \doteq\! [\vet{\lambda}_1,\!\cdots\!,\vet{\lambda}_K]^\trans \!\in\! \compl^{K \times L}$. In this case, the received signal tensor at the HRIS would be $\ten{Y}^\mathrm{RC}=\left(\ten{T}_{\ma{\Phi}} \times_2 \mat{G}^\trans\right) \! \mwcontract{3}{2}{1} \! \left(\ten{I}_{3, L} \times_2 \mat{X}^\trans \times_3 \ma{\Lambda}\right) + \ten{V}^\mathrm{RC}$, corresponding to the \textit{Tucker-PARAFAC} tensor structure. On the other hand, at the BS, the received signal tensor follows a PARATUCK-2 tensor structure, which is given by $\ten{Y}^\mathrm{BS} \! = \! \left(\ten{I}_{3, N} \! \times_1 \! \mat{H} \! \times_2 \! \mat{G}^\trans \! \times_3 \! \ma{\Psi}\right) \! \mwcontract{3}{2}{1} \!\left(\ten{I}_{3, L} \! \times_2 \! \mat{X}^\trans \! \times_3 \! \ma{\Lambda}\right) + \ten{V}^\mathrm{BS}$. It is clear that these tensors are special cases of \eqref{tenyrc} and \eqref{tenybs}.

\section{Semi-Blind Receivers Exploiting TSTC Scheme}\label{sec:receiverststc}
In this section, we develop the proposed semi-blind receivers for joint channel and symbol estimation at the HRIS and BS by exploiting the tensor signals derived in the previous section.

\textit{Optimization problem for the HRIS}: For the HRIS, consider the noisy sensed signal tensor $\ten{Y}^\mathrm{RC}$. Our goal is found estimates of the UT-HRIS channel $\mat{G}$ and the symbol matrix $\mat{X}$ by solving the following problem
\begin{equation}\label{optbsxgw}
    \min \limits_{\mat{G},\mat{X}} \Big\|\ten{Y}^\mathrm{RC} - \left(\ten{T}_{\ma{\Phi}} \times_2 \mat{G}^\trans\right) \mwcontract{3}{2}{1} \left(\ten{W} \times_2 \mat{X}^\trans\right)\Big\|^2_{\textrm{F}}.
\end{equation}
Starting from this cost function, we formulate two solutions to solve this problem by exploiting the different reshapings of the tensor signal structure. The first one resorts to an iterative alternating linear estimation scheme, while the second delivers closed-form estimates of the channel and symbols.

\subsection{HRIS-BALS Receiver (TSTC)}\label{hriswbalsxg}
Using the definition \eqref{def:mw2}, we apply the $\mathrm{vec}\{\cdot\}$ operator to the $k$-th frontal slice of $\ten{Y}^\mathrm{RC}$ defined in \eqref{tenyrc}, i.e., $\mathrm{vec}\{(\ten{C}\mwcontract{3}{1}{2}\ten{S})_{\cdot\cdot k} + \ten{V}^\mathrm{RC}_{\cdot\cdot k}\} = \mathrm{vec}\{\mat{Y}^\mathrm{RC}_k\}$ to define $\vet{y}^\mathrm{RC}_k$, given by
\begin{equation}
    \vet{y}^\mathrm{RC}_k = (\mat{X}^\trans \kron \mat{I}_{N_c})(\mat{W}_k^\trans \kron \ma{\Phi}_k)\vet{g} \in \compl^{TN_c \times 1} + \vet{\nu}_k^\mathrm{RC},
\end{equation}
where $\vet{g}\! \doteq \!\opvec{\mat{G}} \!\in\! \compl^{LN \times 1}$, and $\vet{\nu}^\mathrm{RC} \doteq \opvec{\mat{V}^\mathrm{RC}_k} \in \compl^{TN_c \times 1}$. We define  $\vet{y}^\mathrm{RC}\! \doteq\! \bigl[(\vet{y}^\mathrm{RC}_1)^\trans,\cdots,(\vet{y}^\mathrm{RC}_K)^\trans\bigr]^\trans \!= \!\mathrm{vec}\bigl\{\bigl[\ten{Y}^\mathrm{RC}\bigr]_{(3)}^\trans\bigr\} \!\in \!\compl^{KTN_c \times 1}$ by stacking $\vet{y}^\mathrm{RC}_k$ during the $K$ sub-frames, to get 
\begin{equation}\label{hriswvetyrc}
	\vet{y}^\mathrm{RC} = \bigl(\mat{I}_{K} \kron \mat{X}^\trans \kron \mat{I}_{N_c}\bigr)\mat{F}_\mathrm{g}\vet{g} + \vet{\nu}^\mathrm{RC},
\end{equation}
where $\mat{F}_\mathrm{g} \doteq \bigl[\mat{W}_1 \kron \ma{\Phi}_1^\trans,\cdots,\mat{W}_K \kron \ma{\Phi}_K^\trans\bigr]^\trans \in \compl^{KRN_c \times LN}$ contains the coding structure and the sensing phase shifts, which are known at the HRIS, and $\vet{\nu}^\mathrm{RC} = \mathrm{vec}\bigl\{\bigl[\ten{V}^\mathrm{RC}\bigr]_{(3)}^\trans\bigr\} \in \compl^{KTN_c \times 1}$ is the corresponding noise term. A least-squares (LS) estimate of the UT-HRIS channel can be found by solving the problem
\begin{equation}
    \hat{\vet{g}} = \argmin_{\vet{g}} \Bigl\|\vet{y}^\mathrm{RC} - \bigl(\mat{I}_{K} \kron \mat{X}^\trans \kron \mat{I}_{N_c}\bigr)\mat{F}_\mathrm{g}\vet{g}\Bigr\|_\mathrm{F}^2,
\end{equation}
whose analytical solution is given by
\begin{equation}\label{hriswgest}
	\hat{\mat{G}} = \mathrm{unvec}_{N \times L}\bigl\{\bigl[\bigl(\mat{I}_{K} \kron \mat{X}^\trans \kron \mat{I}_{N_c}\bigr)\mat{F}_\mathrm{g}\bigr]^\pinv\vet{y}^\mathrm{RC}\bigr\},
\end{equation}
Exploiting $\bigl[\ten{Y}^\mathrm{RC}\bigr]_{(2)}\! \doteq \! \bigl[(\mat{Y}^\mathrm{RC}_1)^\trans,\!\cdots\!,(\mat{Y}^\mathrm{RC}_K)^\trans\bigr] \!\in \!\compl^{T \times KN_c}$, corresponding to the 2-mode unfolding of $\ten{Y}^\mathrm{RC}$, we have
\begin{equation}\label{hriswyrcet}
	\bigl[\ten{Y}^\mathrm{RC}\bigr]_{(2)}^\trans = \mat{F}_\mathrm{x}\mat{X} + \bigl[\ten{V}^\mathrm{RC}\bigr]_{(2)}^\trans\in \compl^{KN_c \times T},
\end{equation}
where $\mat{F}_\mathrm{x} \doteq \bigl[(\ma{\Phi}_1\mat{G}\mat{W}_1)^\trans,\cdots,(\ma{\Phi}_K\mat{G}\mat{W}_k)^\trans\bigr]^\trans \in \compl^{KN_c \times R}$.
The symbol matrix can be found by solving
\begin{equation}
    \hat{\mat{X}} = \argmin_{\mat{X}} \left\|\unf{Y^\mathrm{RC}}{2}^\trans - \mat{F}_\mathrm{x}\mat{X}\right\|^2_{\textrm{F}},
\end{equation}
the solution of which is given by
\begin{equation}\label{hriswxest}
	\hat{\mat{X}} = \mat{F}_\mathrm{x}^\pinv\unf{Y^\mathrm{RC}}{2}^\trans.
\end{equation}

Note that \eqref{hriswgest} and \eqref{hriswxest} are jointly used to iteratively estimate the UT-HRIS channel and symbols \emph{via} a bilinear alternating LS (BALS) algorithm, herein referred to as HRIS-BALS receiver. The algorithm consists of estimating $\mat{G}$ and $\mat{X}$ iteratively, starting from a random initialization until convergence is achieved. As discussed in previous works \cite{comon2009tensor}, \cite{dearaujo2021channel}, such a BALS procedure converges after a few iterations and provides unique estimates of the channel and symbol matrices up to trivial scaling ambiguities, as will be discussed later. The HRIS-BALS receiver is summarized in Algorithm \ref{alg:hrisbalsreceiverw}. 
\begin{algorithm}[t]
	\caption{HRIS-BALS receiver (TSTC)}
	\label{alg:hrisbalsreceiverw}
    \renewcommand{\baselinestretch}{0.9}
	\begin{algorithmic}
    	\STATE \hspace{-4ex} 1. Set $i=0$ and initialize $\hat{\mat{X}}_{(i=0)}$ randomly;\\
		\STATE \hspace{-4ex} 2. $i = i + 1$;\\
		\STATE \hspace{-4ex} 3. Get $\hat{\mat{G}}_{(i)} \!=\! \mathrm{unvec}_{N \times L}\!\Bigl\{\!\Bigl[\bigl(\mat{I}_{K} \!\kron\! \mat{X}^\trans_{(i-1)} \!\kron\! \mat{I}_{N_c}\bigr)\mat{F}_\mathrm{g}\Bigr]^\pinv\!\vet{y}^\mathrm{RC}\!\Bigr\}$;\\
		\STATE \hspace{-4ex} 4. Get $\hat{\mat{X}}_{(i)} = \mat{F}_{\mathrm{x}(i)}^\pinv\unf{Y^\mathrm{RC}}{2}^\trans$;\\
		\STATE \hspace{-4ex} 5. Repeat steps 2-5 until convergence;\\
		\STATE \hspace{-4ex} 6. Remove scaling ambiguities.
	\end{algorithmic}
\end{algorithm}

\subsection{HRIS-KronF Receiver (TSTC)}\label{hriskronf}
\begin{algorithm}[t]
	\caption{HRIS-KronF receiver (TSTC)}
	\label{alg:hriskronfreceiver}
    \renewcommand{\baselinestretch}{0.9}
	\begin{algorithmic}
        \STATE \hspace{-4ex} 1. Using \eqref{hriswxgest}, find a LS estimate of $\hat{\mat{Q}}$;\\
		\STATE \hspace{-4ex} 2. Construct $\bar{\mat{Q}}$ by rearranging $\hat{\mat{Q}}$;\\
        \STATE ~\textbf{for} $l=1,\cdots,L$\\
		\STATE \quad ~~\textbf{for} $n=1,\cdots,N$\\
		\STATE \qquad\quad $\mathbb{S}_n = \{(n-1)T+1,\ldots,nT\}$\\
		\STATE \qquad\quad $\mathbb{S}_l = \{(l-1)R+1,\ldots,lR\}$\\
        \STATE \qquad\quad $\hat{\mat{Q}}^{n,l} \longleftarrow \hat{\mat{Q}}_{[\mathbb{S}_n,\mathbb{S}_l]}$\\
		\STATE \qquad\quad $\vet{q}^{n,l} \doteq \mathrm{vec}\bigl\{\hat{\mat{Q}}^{n,l}\bigr\} \in \compl^{RT \times 1}$\\
		\STATE \quad ~~\textbf{end}\\
		\STATE ~\textbf{end}\\
        \STATE ~$\bar{\mat{Q}} = \left[\vet{q}^{1,1},\ldots,\vet{q}^{N,1},\ldots,\vet{q}^{N,L}\right]$;\\
		\STATE \hspace{-4ex} 3. Compute $[\vet{u}_1,\sigma_1,\vet{v}_1] \longleftarrow$ truncated-SVD$(\bar{\mat{Q}})$;\\
	    \STATE \hspace{-4ex} 4. Reconstruct $\hat{\mat{G}}$ and $\hat{\mat{X}}$:\\
        \STATE \hspace{-4ex} \quad $\hat{\mat{G}} \!\leftarrow\! \mathrm{unvec}_{N \times L}\{\!\sqrt{\sigma_1}\vet{v}_1^\ast\}$,~$\hat{\mat{X}} \!\leftarrow\! (\mathrm{unvec}_{T \times R}\{\!\sqrt{\sigma_1}\vet{u}_1\})^\trans$;\\
	  \STATE \hspace{-4ex} 5. Remove scaling ambiguities.\\   
	\end{algorithmic}
\end{algorithm}

Define $\bar{\vet{y}}^\mathrm{RC}_k \doteq \mathrm{vec}\{(\mat{Y}^\mathrm{RC}_k)^\trans\} \in \compl^{N_cT \times 1}$, given by
\begin{equation}
    \bar{\vet{y}}^\mathrm{RC}_k = (\ma{\Phi}_k \kron \mat{I}_T)\mat{Q}\vet{w}_k + \bar{\vet{\nu}}^\mathrm{RC}_k,
\end{equation}
where $\vet{w}_k \!\doteq \!\opvec{\mat{W}_k^\trans} \!\in \!\compl^{LR \times 1}$, $\bar{\vet{\nu}}^\mathrm{RC}_k \doteq \mathrm{vec}\{(\mat{V}^\mathrm{RC}_k)^\trans\}$, and $\mat{Q}\!\doteq\! \mat{G} \kron \mat{X}^\trans \!\in \!\compl^{NT \times LR}$. Applying $\mathrm{vec}\{\cdot\}$ again, we obtain
\begin{equation}
    \bar{\vet{y}}^\mathrm{RC}_k = (\vet{w}_k^\trans \kron \ma{\Phi}_k \kron \mat{I}_T)\vet{q} + \bar{\vet{\nu}}^\mathrm{RC}_k,
\end{equation}
where $\vet{q} = \opvec{\mat{Q}} \in \compl^{LRNT \times 1}$. Defining $\bar{\vet{y}}^\mathrm{RC} \doteq \bigl[(\bar{\vet{y}}^\mathrm{RC}_1)^\trans,\cdots,(\bar{\vet{y}}^\mathrm{RC}_K)^\trans\bigr]^\trans = \mathrm{vec}\bigl\{\bigl[\ten{Y}^\mathrm{RC}\bigr]_{(2)}\bigr\}$ by collecting the sensed signals during the $K$ sub-frames, we get
\begin{equation}
     \bar{\vet{y}}^\mathrm{RC} = (\mat{F}_\mathrm{xg} \kron \mat{I}_T)\vet{q} \in \compl^{KN_cT \times 1} + \bar{\vet{\nu}}^\mathrm{RC},
\end{equation}
where $\bar{\vet{\nu}}^\mathrm{RC} \in \compl^{N_cT \times 1}$ is the related noise component, and $\mat{F}_\mathrm{xg} \in \compl^{KN_c \times LRN}$ is 
\begin{equation}\label{fxg}
    \mat{F}_\mathrm{xg} \doteq \bigl[\vet{w}_1 \kron \ma{\Phi}_1^\trans,\cdots,\vet{w}_K \kron \ma{\Phi}_K^\trans\bigr]^\trans.
\end{equation}
Let us consider the following problem
\begin{equation}
	\hat{\vet{q}} = \argmin_{\vet{q}} \left\|\bar{\vet{y}}^\mathrm{RC} - (\mat{F}_\mathrm{xg} \kron \mat{I}_T)\vet{q}\right\|^2,
\end{equation}
from which we can determine an LS estimate for the matrix $\mat{Q}$ through its corresponding solution
\begin{equation}\label{hriswxgest}
	\hat{\mat{Q}} = \mathrm{unvec}_{NT \times LR}\left\{(\mat{F}_\mathrm{xg}^\pinv \kron \mat{I}_T)\bar{\vet{y}}^\mathrm{RC}\right\}.
\end{equation}
Upon obtaining $\hat{\mat{Q}}$, the next step involves finding an estimate of $\mat{X}$ and $\mat{G}$ from it. Since the matrix $\hat{\mat{Q}}$ follows a Kronecker construction, it can be interpreted as a block matrix, whose each sub-matrix is $\hat{\mat{Q}}^{n,l} = g_{n,l}\mat{X}^\trans \in \compl^{T \times R}$. Therefore, we address the problem 
\begin{equation}\label{eq:kronFcost}
    \min_{\scriptscriptstyle \mat{X},\mat{G}} \left\|\hat{\mat{Q}} - \mat{G} \kron \mat{X}^\trans\right\|_\mathrm{F}^2,
\end{equation}
whose solution is found by the so-called Kronecker Factorization (KronF) algorithm \cite{van1993approximation}. The solution to this problem is found by recasting the problem \eqref{eq:kronFcost} as a rank-1 matrix approximation problem 
\begin{equation}
    \min_{\scriptscriptstyle \mat{X},\mat{G}} \bigl\|{\bar{\mat{Q}} - \vet{x}\vet{g}^\trans\bigr\|}_\mathrm{F}^2,
\end{equation}
where $\bar{\mat{Q}} \in \compl^{RT \times LN}$ is a matrix rearrangement of the blocks contained in $\hat{\mat{Q}}$, as follows 
\begin{equation*}
    \bar{\mat{Q}} = \bigl[g_{1,1}\vet{x},\ldots,g_{N,1}\vet{x},\ldots,g_{1,L}\vet{x},\ldots,g_{N,L}\vet{x}\bigr]\\
    = \vet{x}\vet{g}^\trans,
\end{equation*}
where $\vet{x} \doteq \opvec{\mat{X}^\trans} \in \compl^{RT \times 1}$, and $g_{n,l}\vet{x} = \mathrm{vec}\bigl\{\hat{\mat{Q}}^{n,l}\bigr\} \in \compl^{RT \times 1}$. From this problem, the estimates of $\mat{X}$ and $\mat{G}$ are given by the dominant left and right singular vectors of $\bar{\mat{Q}}$, respectively. This procedure leads to the HRIS-KronF receiver, whose key steps are summarized in Algorithm \ref{alg:hriskronfreceiver}, wherein we define the index sets $\mathbb{S}_n$ and $\mathbb{S}_l$ to indicate, respectively, the row and column ranges of ($n,l$)-th sub-matrix of $\hat{\mat{Q}}$.

\begin{remark}
    The HRIS does not need to estimate the full information contained in the symbol matrix. It may be of interest to only decode a subset of columns of $\mat{X}$, leaving the remaining subset to be decoded by the BS. As elucidated in Section \ref{sec:discussion}, $\mat{X}$ can be partitioned into user data and control data submatrices during the transmission time structure. We consider that HRIS and BS fully estimate the symbol matrix for the convenience of exposition. Note also that even in a scenario where the HRIS does not need to perform data decoding, the proposed semi-blind receivers provide data-aided CE capabilities at the HRIS. In this scenario, data symbols intended for the BS are exploited at the HRIS to estimate the associated channel matrix, unlike existing methods, which accomplish this by using only pilot symbols.
\end{remark}

\textit{Optimization problem for the BS}: As for the HRIS, joint symbol and CE can be achieved at the BS by exploiting the tensor structure of the received signal $\ten{Y}^\mathrm{BS}$ as well as the estimated UT-HRIS channel matrix obtained at the HRIS and conveyed \emph{via} the CL. Recall that the BS knows the coding tensor and the reflection phase shifts. We consider the following LS tensor fitting problem
\begin{equation}\label{optbsxh}
    \hspace{-0.3cm}\min \limits_{\mat{H},\mat{X}} \left\|\ten{Y}^\mathrm{BS} \!\! - \!\left(\!\ten{I}_{3,N} \! \times_1 \! \mat{H} \! \times_2 \! \hat{\mat{G}}^\trans \! \times_3 \! \ma{\Psi}\!\right) \! \mwcontract{3}{2}{1} \! \left(\ten{W} \! \times_2 \! \mat{X}^\trans\right)\right\|^2_{\textrm{F}}\!\!\!.
\end{equation}
In what follows, we exploit the different reshapings of the received tensor $\ten{Y^\mathrm{BS}}$ to derive the corresponding iterative and closed-form semi-blind receivers at the BS. 

\subsection{BS-BALS Receiver (TSTC)}\label{bswbalsxh}
Taking into account the tensor $\ten{Y}^\mathrm{BS}$, defined in \eqref{tenybs}, we concatenate its frontal slices $(\ten{T}_{\ma{\Omega}}\mwcontract{3}{2}{1}\ten{S})_{\cdot\cdot k} + \ten{V}^\mathrm{BS}_{\cdot\cdot k} = \mat{Y}^\mathrm{BS}_k$ for $k=1,2,\cdots,K$ to obtain its 1-mode and 2-mode unfoldings, given by $\bigl[\ten{Y}^\mathrm{BS}\bigr]_{(1)}\! =\!\bigl[\mat{Y}^{\mathrm{BS}}_1,\cdots,\mat{Y}^{\mathrm{BS}}_K\bigr] \in \compl^{M \times KT}$ and $\bigl[\ten{Y}^\mathrm{BS}\bigr]_{(2)} = \bigl[(\mat{Y}^\mathrm{BS}_1)^\trans,\cdots,(\mat{Y}^\mathrm{BS}_K)^\trans\bigr] \in \compl^{T \times KM}$, respectively. These unfoldings can be expressed in compact form as
\begin{align}
    &&\bigl[\ten{Y}^\mathrm{BS}\bigr]_{(1)} = \mat{H}\mat{E}_\mathrm{h}(\mat{I}_K \kron \mat{X}) + \bigl[\ten{V}^\mathrm{BS}\bigr]_{(1)},\label{unf1ybs}\\
    &&\bigl[\ten{Y}^\mathrm{BS}\bigr]_{(2)}^\trans = (\mat{I}_K \kron \mat{H})\mat{E}_\mathrm{x}\mat{X} + \bigl[\ten{V}^\mathrm{BS}\bigr]_{(2)}^\trans,\label{unf2ybs}
\end{align}
where $\mat{E}_\mathrm{h} \in \compl^{N \times KR}$ and $\mat{E}_\mathrm{x} \in \compl^{KN \times R}$ are defined as
\begin{align*}
    &&\mat{E}_\mathrm{h} \doteq \big[\diagof{\vet{\psi}_1}\hat{\mat{G}}\mat{W}_1,\cdots,\diagof{\vet{\psi}_K}\hat{\mat{G}}\mat{W}_K\big],\\    &&\mat{E}_\mathrm{x}\!\doteq\!\bigl[\mat{W}_1^\trans\hat{\mat{G}}^\trans\diagof{\vet{\psi}_1}\!,\!\cdots\!,\mat{W}_K^\trans\hat{\mat{G}}^\trans\diagof{\vet{\psi}_K}\bigr]^\trans.
\end{align*}
respectively. From \eqref{unf1ybs} and  \eqref{unf2ybs}, the estimation of the HRIS-BS channel $\mat{H}$ and the symbol matrix $\mat{X}$ can be obtained by solving the following LS problems
\begin{equation}
	\hat{\mat{H}} = \argmin_{\mat{H}}\fronormbig{\unf{Y^{\mathrm{BS}}}{1} - \mat{H}\mat{E}_\mathrm{h}(\mat{I}_K \kron \mat{X})}^2,
\end{equation}
\begin{equation}
	\hat{\mat{X}} = \argmin_{\mat{X}}\fronormbig{\unf{Y^{\mathrm{BS}}}{2}^\trans - (\mat{I}_K \kron \mat{H})\mat{E}_\mathrm{x}\mat{X}}^2,
\end{equation}
the solutions of which are respectively given by
\begin{equation}\label{bswhest}
	\hat{\mat{H}} = \unf{Y^{\mathrm{BS}}}{1}\bigl[\mat{E}_\mathrm{h}(\mat{I}_K \kron \mat{X})\bigr]^\pinv,
\end{equation}
\begin{equation}\label{bswxest}
	\hat{\mat{X}} = \bigl[(\mat{I}_K \kron \mat{H})\mat{E}_\mathrm{x}\bigr]^\pinv\unf{Y^{\mathrm{BS}}}{2}^\trans.
\end{equation}
Similarly to the HRIS side, the estimate of the HRIS-BS channel and transmitted symbols can be obtained by solving (\ref{bswhest}) and (\ref{bswxest}) iteratively using alternating least-squares. This algorithm is referred to as the BS-BALS receiver and is summarized in Algorithm \ref{alg:bsbalsreceiverw}.

\begin{algorithm}[t]
	\caption{BS-BALS receiver (TSTC)}
	\label{alg:bsbalsreceiverw}
    \renewcommand{\baselinestretch}{0.9}
	\begin{algorithmic}
        \STATE \hspace{-4ex} 1. Get $\hat{\mat{G}}$ from the feedback control link;
		\STATE \hspace{-4ex} 2. Set $i=0$ and initialize $\hat{\mat{X}}_{(i=0)}$ randomly;\\
		\STATE \hspace{-4ex} 3. $i = i + 1$;\\
		\STATE \hspace{-4ex} 4. Get $\hat{\mat{H}}_{(i)} = \unf{Y^{\mathrm{BS}}}{1}\bigl[\mat{E}_{\mathrm{h}(i-1)}(\mat{I}_K \kron \hat{\mat{X}}_{(i-1)})\bigr]^\pinv$;\\
		\STATE \hspace{-4ex} 5. Get $\hat{\mat{X}}_{(i)} = \bigl[(\mat{I}_K \kron \hat{\mat{H}}_{(i)})\mat{E}_\mathrm{x}\bigr]^\pinv\unf{Y^{\mathrm{BS}}}{2}^\trans$;\\
		\STATE \hspace{-4ex} 6. Repeat steps 3-6 until convergence;\\
		\STATE \hspace{-4ex} 7. Remove scaling ambiguities.
	\end{algorithmic}
\end{algorithm}

\subsection{BS-KronF Receiver (TSTC)}
We now derive the expressions for the closed-form estimation of $\mat{H}$ and $\mat{X}$ are the BS. The procedure is analogous to that discussed on the HRIS side. First, by applying $\mathrm{vec}\{\cdot\}$ to the $k$-th frontal slice of $\ten{Y}^\mathrm{BS}$, we define $\vet{y}^{\mathrm{BS}}_k\! \doteq \!\mathrm{vec}\{(\ten{T}_{\ma{\Omega}}\mwcontract{3}{2}{1}\ten{S})_{\cdot\cdot k} + \ten{V}^\mathrm{BS}_{\cdot\cdot k}\} \in \compl^{TM \times 1}$, or
\begin{equation}
    \vet{y}^{\mathrm{BS}}_k = (\mat{X}^\trans \kron \mat{H})(\mat{W}_k^\trans \kron \diagof{\vet{\psi}_k})\hat{\vet{g}} + \vet{\nu}^\mathrm{BS},
\end{equation}
where $\vet{\nu}^\mathrm{BS} \doteq \opvec{\mat{V}^\mathrm{BS}_k} \in \compl^{TM \times 1}$. Stacking column-wise the received signal vectors $\vet{y}^\mathrm{BS}_k$, for $k=1,2,\cdots,K$, we obtain the 3-mode unfolding of $\ten{Y}^\mathrm{BS}$, given by $\bigl[\ten{Y}^\mathrm{BS}\bigr]_{(3)} = \bigl[\vet{y}^{\mathrm{BS}}_1,\cdots,\vet{y}^{\mathrm{BS}}_K\bigr]^\trans \in \compl^{K \times TM}$. The transposed version of this unfolding can be expressed as
\begin{equation}\label{ybs3t}
    \bigl[\ten{Y}^\mathrm{BS}\bigr]_{(3)}^\trans = \left(\mat{X}^\trans \kron \mat{H}\right)\mat{E}_\mathrm{xh}(\mat{I}_K \kron \hat{\vet{g}}) + \bigl[\ten{V}^\mathrm{BS}\bigr]_{(3)}^\trans,
\end{equation}
where $\mat{E}_\mathrm{xh} \in \compl^{RN \times KLN}$ is defined as
\begin{equation}\label{omegaxhbscomb}
	\mat{E}_\mathrm{xh} \doteq \bigl[\mat{W}_1^\trans \kron \diagof{\vet{\psi}_1},\cdots,\mat{W}_K^\trans \kron \diagof{\vet{\psi}_K}\bigr].
\end{equation}
Defining $\mat{Z} \doteq \mat{X}^\trans \kron \mat{H} \in \compl^{TM \times RN}$, we first find its compound estimate by solving the following LS problem
\begin{equation}
	\hat{\mat{Z}} = \argmin_{\mat{Z}} \left\|\unf{Y^{\mathrm{BS}}}{3}^\trans - \mat{Z}\mat{E}_\mathrm{xh}(\mat{I}_K \kron \hat{\vet{g}})\right\|^2_{\textrm{F}},
\end{equation}
the solution of which is given by
\begin{equation}\label{bswxhest}
	\hat{\mat{Z}} = \unf{Y^{\mathrm{BS}}}{3}^\trans\bigl[\mat{E}_\mathrm{xh}(\mat{I}_K \kron \hat{\vet{g}})\bigr]^\pinv.
\end{equation}
From the estimate $\hat{\mat{Z}}$, we can jointly find the individual estimates of the $\mat{X}$ and $\mat{H}$ by solving the problem \begin{equation}\label{kronFXh}
    \min_{\scriptscriptstyle \mat{X},\mat{H}} \left\|\hat{\mat{Z}} - \mat{X}^\trans \kron \mat{H}\right\|_\mathrm{F}^2,
\end{equation}
which is solved via the KronF algorithm \cite{van1993approximation}. In our context, this is accomplished by solving the following rank-1 matrix approximation problem
\begin{equation}
    \min_{\scriptscriptstyle \mat{X},\mat{H}} \bigl\|\bar{\mat{Z}} - \vet{h}\vet{x}^\trans\bigr\|_{\mathrm{F}}^2,
\end{equation}
where $\vet{h} \doteq \opvec{\mat{H}} \in \compl^{NM \times 1}$. The rank-1 matrix $\bar{\mat{Z}} \in \compl^{RT \times LN}$ is obtained by rearranging the matrix blocks of $\hat{\mat{Z}}$ similarly to the method used for the HRIS-KronF receiver (see Section \ref{hriskronf}). The main steps of the BS-KronF receiver are summarized in Algorithm \ref{alg:bskronfreceiver}.
\begin{algorithm}[t]
    \caption{BS-KronF receiver (TSTC)}
	\label{alg:bskronfreceiver}
    \renewcommand{\baselinestretch}{0.9}
	\begin{algorithmic}
        \STATE \hspace{-4ex} 1. Get $\hat{\mat{G}}$ from the feedback control link;
        \STATE \hspace{-4ex} 2. Using \eqref{bswxhest}, find an LS estimate of $\hat{\mat{Z}}$;\\
        \STATE \hspace{-4ex} 3. Construct $\bar{\mat{Z}}$ by rearranging $\hat{\mat{Z}}$;\\
		\STATE \hspace{-4ex} 4. Compute $[\vet{u}_1,\sigma_1,\vet{v}_1] \longleftarrow$ truncated-SVD$(\bar{\mat{Z}})$;\\
        \STATE \hspace{-4ex} 5. Reconstruct $\hat{\mat{X}}$ and $\hat{\mat{H}}$:\\
        \STATE \hspace{-4ex} \quad $\hat{\mat{X}} \!\leftarrow\! [\mathrm{unvec}_{T \times R}\{\!\sqrt{\sigma_1}\vet{v}_1^\ast\}]^\trans$, $\hat{\mat{H}} \!\leftarrow\! \mathrm{unvec}_{M \times N}\{\!\sqrt{\sigma_1}\vet{u}_1\}$;\\
		\STATE \hspace{-4ex} 6. Remove scaling ambiguities.\\   
	\end{algorithmic}
\end{algorithm}

\section{Semi-Blind Receivers Exploiting KRSTC Scheme}\label{sec:receiverskrstc}
In this section, we briefly present the semi-blind receivers involved in this coding scheme, referring, when appropriate, to equations and algorithms from the TSTC case for the sake of brevity, since the algebraic steps to derive the semi-blind receivers for the KRSTC scheme are, in most cases, similar to those for the TSTC. In those cases, we replace $\mat{W}_k$ by $\diagof{\vet{\lambda}_k}$ in $\mat{F}_\mathrm{g}$, $\mat{F}_\mathrm{x}$, $\mat{F}_\mathrm{xg}$, $\mat{E}_\mathrm{h}$ and $\mat{E}_\mathrm{x}$, except in $\mat{E}_\mathrm{xh}$. In addition, $R=L$ is assumed in $\mat{X}$.

\subsection{HRIS-BALS Receiver (KRSTC)}
The BALS receiver at the HRIS for KRSTC capitalizes on the BALS's framework in Algorithm \ref{alg:hrisbalsreceiverw}, by exploting \eqref{hriswgest} and \eqref{hriswxest} while redefining $\mat{F}_\mathrm{g}$ and $\mat{F}_\mathrm{x}$ as $\mat{F}_\mathrm{g} \! \doteq \! \bigl[\diagof{\vet{\lambda}_1} \kron \ma{\Phi}_1^\trans,\!\cdots\!,\diagof{\!\vet{\lambda}_K\!} \kron \ma{\Phi}_K^\trans\bigr]^\trans \!\! \in \! \compl^{KLN_c \times LN}$ and $\mat{F}_\mathrm{x} \!\doteq\! \bigl[\diagof{\!\vet{\lambda}_1\!}\!\mat{G}^\trans\ma{\Phi}_1^\trans,\!\cdots\!,\diagof{\!\vet{\lambda}_K\!}\!\mat{G}^\trans\ma{\Phi}_K^\trans\bigr]^\trans \!\in \!\compl^{KN_c \times L}$, respectively.

\subsection{HRIS-KRF Receiver (KRSTC)}\label{hriskrf}
Closed-form estimates of $\mat{G}$ and $\mat{X}$ at the HRIS are obtained after employing steps similar to those adopted in Section \ref{hriskronf} after minor algebraic modifications. As a result, the previously Kronecker structured combined matrix $\mat{Q}$, estimated in \eqref{hriswxgest}, becomes Khatri-Rao structured, i.e., $\mat{Q} \doteq \mat{G} \krp \mat{X}^\trans \in \compl^{NT \times L}$. Therefore, we rewrite \eqref{hriswxgest} to get
\begin{equation}\label{xgesthriskr}
	\hat{\mat{Q}} = \mathrm{unvec}_{NT \times L}\{(\mat{F}_\mathrm{xg}^\pinv \kron \mat{I}_T)\bar{\vet{y}}^\mathrm{RC}\},
\end{equation}
where $\mat{F}_\mathrm{xg}$ is redefined as $\mat{F}_\mathrm{xg}\doteq\bigl[\vet{\lambda}_1 \kron \ma{\Phi}_1^\trans,\cdots,\vet{\lambda}_K \kron \ma{\Phi}_K^\trans\bigr]^\trans \in \compl^{KN_c \times LN}$. Once $\hat{\mat{Q}}$ is found, we consider to solve the problem
\begin{equation}\label{minkrf}
    \min_{\scriptscriptstyle \mat{X},\mat{G}} \bigl\|\hat{\mat{Q}} - \mat{G} \krp \mat{X}^\trans\bigr\|_\mathrm{F}^2,
\end{equation}
for which the Khatri-Rao Factorization (KRF) algorithm \cite{Kibangou2009,dearaujo2021channel,magalhaes2025reducing} can be applied to decouple estimates of $\mat{X}$ and $\mat{G}$. According \cite{Kibangou2009}, this can be accomplished by reshaping each $l$-th column of $\hat{\mat{Q}}$, defined as $\hat{\vet{x}}_{{\mathrm{g}}_l}$, into a rank-1 matrix $\mat{Q}_l \in \compl^{T \times N}$. Defining the $l$-th column of $\mat{G}$ and $\mat{X}^\trans$ as, respectively, $\vet{g}_l \in \compl^{N \times 1}$ and $\vet{x}_l \in \compl^{T \times 1}$ and by considering the property \eqref{prop:kronabvec}, $\mat{Q}_l$ can be further approximated by $\vet{x}_l\vet{g}_l^\trans$. Therefore, we can tackle the problem \eqref{minkrf} and get closed-form estimates of $\mat{G}$ and $\mat{X}$ by solving $L$ rank-1 matrix approximation subproblems, i.e.,
\begin{equation}\label{ghestkrf}
	[\hat{\mat{G}},\hat{\mat{X}}] = \argmin_{\vet{g}_l,\vet{x}_l}\sum\limits_{l=1}^L\fronorm{\mat{Q}_l - \vet{x}_l\vet{g}_l^\trans}^2,
\end{equation}
where each subproblem operates on the reshaping of the $l$-th column of $\hat{\mat{Q}}$ into a rank-1 matrix $\mat{Q}_l \in \compl^{T \times N}$. The $l$-th columns of $\hat{\mat{X}}^\trans$ and $\hat{\mat{G}}$ are respectively found from the dominant left and right singular vectors of $\mat{Q}_l$. A relevant feature of the KRF method is that by distributing the $L$ estimation steps across parallel processors, the processing delay can be reduced. The HRIS-KRF receiver is summarized in Algorithm \ref{alg:hriskronfreceiver}.

\begin{algorithm}[t]
	\caption{HRIS-KRF receiver (KRSTC)}
	\label{alg:hriskrfreceiver}
    \renewcommand{\baselinestretch}{0.9}
	\begin{algorithmic}
        \STATE \hspace{-4ex} 1. Using \eqref{xgesthriskr}, find a LS estimate of $\hat{\mat{Q}}$;\\
		\STATE \hspace{-4ex} 2. Estimate the columns of $\mat{G}$ and $\mat{X}$:\\
		\STATE \hspace{-4ex} \quad \textbf{for} $l=1,\cdots,L$\\
        \STATE \hspace{-4ex} \quad\quad $\mat{Q}_l = \mathrm{unvec}_{T \times N}\{\vet{q}_l\}$\\
        \STATE \hspace{-4ex} \quad\quad $[\vet{u}_1,\sigma_1,\vet{v}_1] \longleftarrow$ truncated-SVD$(\mat{Q}_l)$\\
 	\STATE \hspace{-4ex} \quad\quad $\hat{\vet{g}}_l = \sqrt{\sigma_1}\vet{v}_1^\ast$,~ $\hat{\vet{x}}_l = \sqrt{\sigma_1}\vet{u}_1$\\
		\STATE \hspace{-4ex} \quad \textbf{end};\\
		\STATE \hspace{-4ex} 3. $\hat{\mat{G}} \longleftarrow \left[\hat{\vet{g}}_1,\cdots,\hat{\vet{g}}_L\right]$, ~$\hat{\mat{X}} \longleftarrow \left[\hat{\vet{x}}_1,\cdots,\hat{\vet{x}}_L\right]^\trans$;\\
		\STATE \hspace{-4ex} 4. Remove scaling ambiguities.\\
	\end{algorithmic}
\end{algorithm}

\subsection{BS-BALS Receiver (KRSTC)}
At the BS, the BALS receiver for KRSTC exploits the BALS’s framework in Algorithm \ref{alg:bsbalsreceiverw} by rewriting \eqref{bswhest} and \eqref{bswxest} while redefining $\mat{E}_\mathrm{h} \in \compl^{N \times KL}$ and $\mat{E}_\mathrm{x} \in \compl^{KN \times L}$ as
$\mat{E}_\mathrm{h}\!\doteq \!\!\bigl[\diagof{\!\vet{\psi}_1\!}\!\hat{\mat{G}}\diagof{\!\vet{\lambda}_1\!}\!,\!\cdots\!,\!\diagof{\!\vet{\psi}_K\!}\!\hat{\mat{G}}\diagof{\!\vet{\lambda}_K\!}\bigr]$ and $\mat{E}_\mathrm{x}\!\doteq \!\!\bigl[\diagof{\!\vet{\lambda}_1\!}\!\hat{\mat{G}}^\trans\!\diagof{\!\vet{\psi}_1\!}\!,\!\cdots\!,\!\diagof{\!\vet{\lambda}_K\!}\!\hat{\mat{G}}^\trans\diagof{\!\vet{\psi}_K\!}\bigr]^\trans\!$.

\subsection{BS-KronF Receiver (KRSTC)}
The closed-form receiver following KRSTC can be derived by first redefining $\mat{Z} \doteq \mat{X}^\trans \kron \mat{H} \in \compl^{TM \times LN}$ and applying the $\mathrm{vec}\{\cdot\}$ operator to \eqref{matybskkr}. Then, one can use the properties \eqref{prop:vecd} and \eqref{prop:diagab} to obtain $\bigl[\ten{Y}^\mathrm{BS}\bigr]_{(3)}^\trans = \mat{Z}\diagof{\hat{\vet{g}}}\mat{E}_\mathrm{xh} + \bigl[\ten{V}^\mathrm{BS}\bigr]_{(3)}^\trans$, where $\mat{E}_\mathrm{xh}$ assumes the form $\mat{E}_\mathrm{xh} \doteq \ma{\Lambda}^\trans \krp \ma{\Psi}^\trans \in \compl^{LN \times K}$, constructed from the definitions of $\ma{\Psi}$ and $\ma{\Lambda}$ (in Section \ref{sec:tenmodel}). An estimate of the composite matrix $\mat{Z}$ can be found by solving
\begin{equation}
	\hat{\mat{Z}} = \argmin_{\mat{Z}} \left\|\unf{Y^{\mathrm{BS}}}{3}^\trans - \mat{Z}\diagof{\hat{\vet{g}}}\mat{E}_\mathrm{xh})\right\|^2_{\textrm{F}}.
\end{equation}
This way, we get 
\begin{equation}\label{bsxhestkr}
    \hat{\mat{Z}} = \unf{Y^\mathrm{BS}}{3}^\trans\bigl(\diagof{\hat{\vet{g}}}\mat{E}_\mathrm{xh}\bigr)^\pinv \in \compl^{TM \times LN}.
\end{equation}
Then, we replace \eqref{bswxhest} by \eqref{bsxhestkr} and invoke Algorithm \ref{alg:bskronfreceiver} to estimate $\mat{X} \in \mathbb{C}^{L \times T}$ and $\mat{H} \in \mathbb{C}^{M \times N}$.

\section{Identifiability}\label{sec:identifiability}
Estimating $\mat{X}$, $\mat{G}$, and $\mat{H}$ at the HRIS and BS requires solving estimation steps that include right and/or left-matrix inverses and should ensure unique solutions. For the TSTC scheme, this takes into account satisfying the identifiability conditions of \eqref{hriswgest} and \eqref{hriswxest} for the iterative receiver BALS, and \eqref{hriswxgest} for the closed-form receiver KronF at the HRIS. Similarly, at the BS, we need to ensure unique estimates of \eqref{bswhest} and \eqref{bswxest} for the BALS receiver, as \eqref{bswxhest} for the KronF. Analogous considerations should be made for the KRSTC scheme. For simplicity, we base our identifiability assessment on the assumption that $\ten{T}_{\ma{\Phi}}$, $\ten{W}$ and $\ma{\Psi}$ are designed to ensure that the full-rank property is preserved in all $K$ blocks that comprise the matrices to be left/right inverted in the TSTC scheme. For this purpose, we assume $\ma{\Phi}_k$, $\mat{W}_k$, and $\diagof{\vet{\psi}_k}$ have full-rank in \eqref{matyrckw} and \eqref{matybskw}, for $k=1,\cdots,K$. Analogously, the same assumption is applied in the KRSTC for $\ten{T}_{\ma{\Phi}}$, $\ma{\Lambda}$ and $\ma{\Psi}$, where $\ma{\Phi}_k$, $\diagof{\vet{\lambda}_k}$ and $\diagof{\vet{\psi}_k}$ in \eqref{matyrckkr} and \eqref{matybskkr} are also assumed to be full-rank. The design optimization of these phase-shifts and coding coefficients is beyond the scope of this paper and will be left for future work. Hereafter, we discuss the conditions for identifiability of the channel and symbol matrices and their implication for receiver design. 

\subsection{Identifiability of $\hat{\mat{X}}$}
The uniqueness of $\hat{\mat{X}}$ requires that $\mat{F}_\mathrm{x}$ (at the HRIS) and $(\mat{I}_K \kron \mat{H})\mat{E}_\mathrm{x}$ (at the BS) are left-invertible in, respectively, \eqref{hriswxest} and \eqref{bswxest}, which entails that both of them must have full column rank. To achieve this at the HRIS, the necessary conditions are $KN_c \geq R$ for the TSTC and $KN_c \geq L$ for the KRSTC. On the other hand, the necessary conditions are $KM \geq R$ (TSTC) and $KM \geq L$ (KRSTC) at the BS.

\begin{proposition}\label{proposition:x1}
Doing $\kappa_\mathrm{g} = \mathrm{rank}(\mat{G})$, $\mat{F}_\mathrm{x}$ has full column-rank if $K\zeta_\mathrm{x} \geq R$, where $\zeta_\mathrm{x}$ is the rank of $\mat{F}_{\mathrm{x}_k}$, i.e., the $k$-th block of $\mat{F}_\mathrm{x}$. In addition, we have $\zeta_\mathrm{x} \leq \mathrm{min}\{N_c,\kappa_\mathrm{g},R\}$ for the TSTC. This inequality becomes $\zeta_\mathrm{x} \leq \mathrm{min}\{N_c,\kappa_\mathrm{g}\}$ for the KRSTC, and $\mat{F}_\mathrm{x}$ would have full column-rank if $K\zeta_\mathrm{x} \geq L$. \textit{Proof:} see Appendix \ref{ap:ident1}.
\end{proposition}

\begin{proposition}\label{proposition:x2}
Doing $\kappa_\mathrm{h} \!=\! \mathrm{rank}(\mat{H})$ and $\xi_\mathrm{x} \!=\! \mathrm{rank}(\mat{H}\mat{E}_{\mathrm{x}_k})$, the product $(\mat{I}_K \kron \mat{H})\mat{E}_\mathrm{x}$ has full column-rank if $K\xi_\mathrm{x} \geq R$, where $\xi_\mathrm{x} \leq \mathrm{min}\{\kappa_\mathrm{h},\kappa_\mathrm{g},R\}$ for the TSTC. For the KRSTC, $(\mat{I}_K \kron \mat{H})\mat{E}_\mathrm{x}$ would have full column-rank if $K\xi_\mathrm{x} \geq L$, with $\xi_\mathrm{x} \leq \mathrm{min}\{\kappa_\mathrm{h},\kappa_\mathrm{g}\}$. \textit{Proof:} see Appendix \ref{ap:ident1}.
\end{proposition}

At the HRIS, the necessary conditions become sufficient when $\mat{G}$ has full-rank, which corresponds to a rich scattering wireless propagation for the UT-HRIS channel, e.g., Rayleigh fading. In practical scenarios where $N \gg L$, the full rank of $\mat{G}$ indicates that it has full column-rank, meaning $\kappa_\mathrm{g} = L$. Otherwise, when a poor scattering is considered, e.g., for millimeter wave or Terahertz communications, $\mat{G}$ may be rank-deficient ($\kappa_\mathrm{g} < L$). Note that a more restrictive condition occurs when $\zeta_\mathrm{x} = 1$, and then we would have $K \geq R$ and $K \geq L$ for TSTC and KRTC, respectively. This last condition corresponds to the sufficient condition that guarantees the unique solution of $\hat{\mat{X}}$ in any scenario.

On the other hand, the necessary conditions to get $\hat{\mat{X}}$ at the BS are sufficient when $\hat{\mat{G}}$ and $\mat{H}$ have full rank, which implies a rich scattering scenario to both the UT-HRIS and HRIS-BS channels. In contrast, under a poor scattering scenario in one or both channels, $\hat{\mat{G}}$ and/or $\mat{H}$ may have rank-deficient (in this case, $\kappa_\mathrm{h} < \mathrm{min}\{M,N\}$). Similar to the HRIS case, the sufficient conditions for any scenario are $K \geq R$ and $K \geq L$ for TSTC and KRSTC, respectively, once they also cover the rank possibility $\xi_\mathrm{x} = 1$ in the blocks.

\subsection{Identifiability of $\hat{\mat{G}}$}
The uniqueness of $\hat{\mat{G}}$ requires that $\bar{\mat{F}}_\mathrm{g}$ in \eqref{hriswgest} has full column-rank to be left-invertible, where $\bar{\mat{F}}_\mathrm{g} = (\mat{I}_{K} \kron \mat{X}^\trans \kron \mat{I}_{N_c})\mat{F}_\mathrm{g}$ in \eqref{hriswgest}. For this purpose, it is necessary that $KTN_c \geq LN$ for both transmission schemes.

\begin{proposition}\label{proposition:g}
Doing $\kappa_\mathrm{x} = \mathrm{rank}(\mat{X}) \leq \mathrm{min}\{R,T\}$, $\bar{\mat{F}}_\mathrm{g}$ has full column-rank if $KN_c\kappa_\mathrm{x} \geq LN$. \textit{Proof:} see Appendix \ref{ap:ident2}.
\end{proposition}

If $\mat{X}$ has full row-rank, i.e., $\kappa_\mathrm{x} = R$ (TSTC), or $\kappa_\mathrm{x} = L$ (KRSTC), the necessary condition is sufficient. However, if $\mat{X}$ is column-rank ($\kappa_\mathrm{x} = T$), the UT must compensate for this by transmitting additional sub-frames to ensure a rank equal to $LN$ for $\bar{\mat{F}}_\mathrm{g}$. Even so, designing $\mat{F}_\mathrm{g}$ as full column-rank, as assumed in Appendix \ref{ap:ident2}, already induces such compensation.

Even though $\mat{X}$ can be assumed to have full rank, recall that $\mat{G}$ is estimated alternately with $\mat{X}$ in the iterative receiver BALS at the HRIS. Due to the random initialization of one of these matrices, the initial iterations may yield poorly conditioned solutions in very low SNR regimes, potentially resulting in a low rank for $\hat{\mat{X}}$. In the worst case, more sub-frames must be transmitted to guarantee $KN_c \geq LN$ if $\kappa_\mathrm{x} = 1$.

\subsection{Identifiability of $\hat{\mat{H}}$}
Estimating $\hat{\mat{H}}$ uniquely requires that $\mat{E}_\mathrm{h}(\mat{I}_K \kron \mat{X})$ in \eqref{bswhest} have full row-rank, i.e., right-invertible. This holds if the necessary condition $KT \geq N$ is satisfied. 

\begin{proposition}\label{proposition:h}
$\mat{E}_\mathrm{h}(\mat{I}_K \kron \mat{X})$ has full row-rank if $K\xi_\mathrm{h} \geq N$, where $\xi_\mathrm{h}$ is the rank of $\mat{E}_{\mathrm{h}_k}\mat{X}$. \textit{Proof:} see Appendix \ref{ap:ident3}.
\end{proposition}

Assuming $\hat{\mat{G}}$ has full rank, the matrix $\mat{E}_\mathrm{h}$ would have full row rank if $KR \geq N$ for TSTC and $KL \geq N$ for KRSTC. Accordingly, the necessary condition above is sufficient if $\mat{X}$ has full row-rank. However, if $\mat{X}$ is column-rank, the UT must transmit additional sub-frames to ensure $\mat{E}_\mathrm{h}(\mat{I}_K \kron \mat{X})$ to have full row-rank. Therefore, the sufficient condition is $K\xi_\mathrm{h} \geq N$. The more restrictive scenario in this concern requires $K \geq N$.

\subsection{Identifiability of $\hat{\mat{Q}}$}
Estimating the combined matrix $\mat{Q}$ is mandatory before applying the Kronecker factorization in the TSTC scheme and the Khatri-Rao factorization in the KRSTC one, both of which are closed-form solutions for estimating $\mat{X}$ and $\mat{G}$. To $\hat{\mat{Q}}$ be unique, it is sufficient that $\mat{F}_\mathrm{xg}$ in \eqref{hriswxgest} and \eqref{xgesthriskr} to be full column-rank. This requires $KN_c \geq RLN$ and $KN_c \geq LN$ for TSTC and KRSTC, respectively. Indeed, estimating $\hat{\mat{Q}}$ in the TSTC is $R$ times more restrictive than in KRSTC. In contrast, KRSTC does not provide multiplexing of multiple streams at the transmitter. In addition, to recover $\mat{X}$ and $\mat{G}$ without scaling ambiguities, the UT must transmit $L$ known symbols per sub-frame, unlike TSTC, which requires only 1 symbol, as we will see in the next section.

\subsection{Identifiability of $\hat{\mat{Z}}$}
The estimate of the composite matrix $\hat{\mat{Z}}$, required before applying the Kronecker factorization, is unique if $\mat{E}_\mathrm{xh}(\mat{I}_K \kron \hat{\vet{g}})$ in \eqref{bswxhest} is full-row rank, which implies $K \geq RN$ for TSTC and $K \geq LN$ for KRSTC. These solutions provide unique estimates of $\mat{E}_\mathrm{xh}$ even in more challenging scenarios where $\mat{G}$ is rank-deficient. Note that the design requirements for executing those closed-form methods to estimate $\mat{X}$ and $\mat{H}$ are the more restrictive ones in terms of the number of sub-frames and, hence, time overhead.

The identifiability conditions required to satisfy each receiver (at both the HRIS and BS) are summarized in Table \ref{tab:idcomp} for both TSTC and KRSTC schemes, which are presented in terms of the minimum number $K$ of sub-frames necessary to ensure the estimation of the corresponding channel and symbol matrices. For BALS receivers, the required conditions to estimate both matrices must be satisfied simultaneously. To address this, simplified conditions meeting this requirement are provided in Table \ref{tab:idcomp}. In addition, Table \ref{tab:idcomp} exhibits the computational complexity of each receiver, which will be covered in Section \ref{sec:cost}.

\begin{table}[!t]
    \setlength{\tabcolsep}{0.7pt}
    \caption{Identifiability conditions and computational complexities.}
	\label{tab:idcomp}
	\centering
	\renewcommand{\arraystretch}{0.9}
	\begin{tabular}{>{\centering\arraybackslash}m{0.9cm} >{\centering\arraybackslash}m{0.9cm} >{\centering\arraybackslash}m{0.9cm} >{\centering\arraybackslash}m{3cm} >{\centering\arraybackslash}m{2.7cm}}
        \midrule
		Receiver&Entity&Coding&Condition $K \geq \lowint{\cdot}$&Complexity $\mathcal{O}(\cdot)$\\
		\midrule 
		BALS&HRIS&TSTC&$(1/N_c)\mathrm{max}\{R,\!LN/T\}$&$KN_c(R^2\!+\!L^2N^2T)$\\
        KronF&HRIS&TSTC&$LRN/N_c$&$LRN(LRNKN_c\! +\! T)$\\
        BALS&BS&TSTC&$\mathrm{max}\{R/M,N/T\}$&$K(R^2M\!+\!N^2T)$\\
        KronF&BS&TSTC&$RN$&$RN(RNK\! +\! TM)$\\
		BALS&HRIS&KRSTC&$(1/N_c)\mathrm{max}\{\!L,\!LN/T\}$&$L^2KN_c(1\! + \!N^2T)$\\
        KRF&HRIS&KRSTC&$LN/N_c$&$LN(LNKN_c\! +\! T)$\\
        BALS&BS&KRSTC&$\mathrm{max}\{L/M,N/T\}$&$K(L^2M\!+\!N^2T)$\\
		KronF&BS&KRSTC&$LN$&$LN(LNK \!+\! TM)$\\
        H&BS&TSTC&$N/T$&$KN^2T$\\
        H&BS&KRSTC&$N/T$&$KN^2T$\\
        \midrule
	\end{tabular}
\end{table}

\section{Uniqueness}\label{sec:uniqueness}
Once the conditions outlined in Table \ref{tab:idcomp} are met, the estimated matrices $\hat{\mat{G}}$ and $\hat{\mat{X}}$ (at the HRIS) and $\hat{\mat{H}}$ and $\hat{\mat{X}}$ (at the BS) share scaling ambiguities that mutually compensate each other. As mentioned in Section \ref{sec:tenmodel}, the received signal at the HRIS following TSTC can be interpreted as a double-Tucker tensor structure with one of the factor matrices being $\mat{I}_{N_c}$. We can see in Section \ref{sec:receiverststc} that $\mat{G}$ and $\mat{X}$ interact with each other through a Kronecker product. In this way, we can study the uniqueness of the double-Tucker with one known matrix through one of its unfoldings. Let us assuming that $\mat{G}$ and $\mat{X}$ are linked to their estimates as $\mat{X} = \mat{U}_\mathrm{x}^\trans\hat{\mat{X}}$ and $\mat{G} = \hat{\mat{G}}\mat{U}_\mathrm{g}$, where $\mat{U}_\mathrm{x} \in \compl^{R \times R}$ and $\mat{U}_\mathrm{g} \in \compl^{L \times L}$ are non-singular transformation matrices. Replacing both matrices in the noiseless part of \eqref{matyrckw} yields
\begin{equation}\label{uniqdt1}
	\mat{Y}^\mathrm{RC}_k = \ma{\Phi}_k\hat{\mat{G}}\mat{U}_\mathrm{g}\mat{W}_k\mat{U}_\mathrm{x}^\trans\hat{\mat{X}} + \mat{V}^\mathrm{RC}_k.
\end{equation}
After applying $\opvec{\cdot}$ to the transpose of \eqref{uniqdt1}, we obtain
\begin{equation}\label{uniqdt2}
    \bar{\vet{y}}^\mathrm{RC}_k = (\ma{\Phi}_k \kron \mat{I}_T)(\hat{\mat{G}}\mat{U}_\mathrm{g} \kron \hat{\mat{X}}^\trans\mat{U}_\mathrm{x})\vet{w}_k + \bar{\vet{\nu}}^\mathrm{RC}_k,
\end{equation}
and using the property \eqref{prop:kronkron}, we have
\begin{equation}
    \bar{\vet{y}}^\mathrm{RC}_k = (\ma{\Phi}_k \kron \mat{I}_T)(\hat{\mat{G}} \kron \hat{\mat{X}}^\trans)(\mat{U}_\mathrm{g} \kron \mat{U}_\mathrm{x})\vet{w}_k + \bar{\vet{\nu}}^\mathrm{RC}_k.
\end{equation}
Applying $\opvec{\cdot}$ again, followed by stacking row-wise $\bar{\vet{y}}^\mathrm{RC}_k$ for $k=1,\cdots,K$, we finally get
\begin{equation}
    \bar{\vet{y}}^\mathrm{RC} = (\mat{F}_\mathrm{xg} \kron \mat{I}_T)(\mat{U}_\mathrm{g}^\trans \kron \mat{U}_\mathrm{x}^\trans \kron \mat{I}_{NT})\vet{q} + \bar{\vet{\nu}}^\mathrm{RC}.
\end{equation}
If $\mat{F}_\mathrm{xg}$ has full column-rank, we can do
\begin{equation}
    (\mat{F}_\mathrm{xg}^\pinv \kron \mat{I}_T)\bar{\vet{y}}^\mathrm{RC} = (\mat{F}_\mathrm{xg}^\pinv\mat{F}_\mathrm{xg} \kron \mat{I}_T)(\mat{U}_\mathrm{g}^\trans \kron \mat{U}_\mathrm{x}^\trans \kron \mat{I}_{NT})\vet{q},
\end{equation}
which leads to
\begin{equation}
    \mat{F}_\mathrm{xg}^\pinv\mat{F}_\mathrm{xg} \kron \mat{I}_T = \mat{U}_\mathrm{g}^\trans \kron \mat{U}_\mathrm{x}^\trans \kron \mat{I}_{NT} = \mat{I}_{LRNT}
\end{equation}
Therefore, $\mat{U}_\mathrm{x}^\trans \kron \mat{U}_\mathrm{h}^\trans = \mat{I}_{LR}$. The unique solution for this occurs when $\mat{U}_\mathrm{x}$ and $\mat{U}_\mathrm{g}$ are scaled identity matrices, i.e., $\mat{U}_\mathrm{x} = \alpha\mat{I}_R$ and $\mat{U}_\mathrm{g} = (1/\alpha)\mat{I}_L$. By this means, $\hat{\mat{X}}$ and $\hat{\mat{G}}$ are unique up to scaling factors that cancel each other.

On the other hand, $\mat{G}$ and $\mat{X}$ entangle through a Khatri-Rao product in the KRSTC scheme. This happens due to the algebraic structure that follows the Tucker-PARAFAC structure. Let us rewrite \eqref{uniqdt2} as
\begin{equation}
    \bar{\vet{y}}^\mathrm{RC}_k = (\ma{\Phi}_k \kron \mat{I}_T)(\hat{\mat{G}}\mat{U}_\mathrm{g} \krp \hat{\mat{X}}^\trans\mat{U}_\mathrm{x})\vet{\lambda}_k + \bar{\vet{\nu}}^\mathrm{RC}_k,
\end{equation}
wherein the symbol matrix is recast as $\mat{X} \in \compl^{L \times T}$. Using the property \eqref{prop:vecd}, we obtain
\begin{equation}
    \bar{\vet{y}}^\mathrm{RC}_k = (\ma{\Phi}_k \kron \mat{I}_T)(\hat{\mat{G}} \kron \hat{\mat{X}}^\trans)(\mat{U}_\mathrm{g} \krp \mat{U}_\mathrm{x})\vet{\lambda}_k + \bar{\vet{\nu}}^\mathrm{RC}_k.
\end{equation}
We now apply $\opvec{\cdot}$ and we stack row-wise $\bar{\vet{y}}^\mathrm{RC}_k$ for $k=1,\cdots,K$, we have
\begin{equation*}
    \bar{\vet{y}}^\mathrm{RC} = (\mat{F}_\mathrm{xg} \kron \mat{I}_T)\bigl[(\mat{U}_\mathrm{g} \krp \mat{U}_\mathrm{x})^\trans \kron \mat{I}_{NT} \bigr]\mathrm{vec}\bigl\{\hat{\mat{G}} \kron \hat{\mat{X}}^\trans\bigr\} + \bar{\vet{\nu}}^\mathrm{RC}.
\end{equation*}
Stated $\mat{A} \in \compl^{I \times P}$ and $\mat{B} \in \compl^{J \times P}$, the Khatri-Rao product $\mat{A} \krp \mat{B} \in \compl^{IJ \times P}$ can be computed from the Kronecker one $\mat{A} \kron \mat{B} \in \compl^{IJ \times P^2}$ using a reduction matrix \cite{teseflorian} such that $\mat{A} \krp \mat{B} = (\mat{A} \kron \mat{B})\ma{\Xi}$. By applying $\opvec{\cdot}$ to this and using the property \eqref{prop:vec}, we obtain $\mathrm{vec}\bigl\{\mat{A} \krp \mat{B}\bigr\} = (\ma{\Xi}^\trans \kron \mat{I}_{IJ})\mathrm{vec}\bigl\{\mat{G} \kron \mat{X}\bigr\}$. This way,
\begin{equation}
    \mathrm{vec}\bigl\{\hat{\mat{G}} \krp \hat{\mat{X}}^\trans\bigr\} = \bigl[(\mat{I}_L \krp \mat{I}_L)^\trans \kron \mat{I}_{NT} \bigr]\mathrm{vec}\bigl\{\hat{\mat{G}} \kron \hat{\mat{X}}^\trans\bigr\}.
\end{equation}
Note that, if $\mat{U}_\mathrm{g}$ and $\mat{U}_\mathrm{x}$ are diagonal matrices, then, $\mat{U}_\mathrm{g} \krp \mat{U}_\mathrm{x} = \mat{I}_L \krp \mat{U}_\mathrm{g}\mat{U}_\mathrm{x} = \mat{U}_\mathrm{g}\mat{U}_\mathrm{x} \krp \mat{I}_L$. Therefore, to recover the Khatri-Rao product from the Kronecker one, it is mandatory that $\mat{U}_\mathrm{g}\mat{U}_\mathrm{x} = \mat{I}_L$. This means that the estimates $\hat{\mat{G}}$ and $\hat{\mat{X}}$ are unique up to diagonal scaling matrices that mutually cancel.

Stated $\mat{U}_\mathrm{x}^\trans \in \compl^{R \times R}$ and $\mat{U}_\mathrm{h} \in \compl^{N \times N}$ are non-singular transformation matrices, $\mat{X}$ and $\mat{H}$ are related to their estimates as $\mat{X} = \mat{U}_\mathrm{x}^\trans\hat{\mat{X}}$ and $\mat{H} = \hat{\mat{H}}\mat{U}_\mathrm{h}$, respectively, so that we can rewrite \eqref{ybs3t} as
\begin{equation}
    \bigl[\ten{Y}^\mathrm{BS}\bigr]_{(3)}^\trans = (\hat{\mat{X}}^\trans\mat{U}_\mathrm{x} \kron \hat{\mat{H}}\mat{U}_\mathrm{h})\bar{\mat{E}}_\mathrm{xh} + \bigl[\ten{V}^\mathrm{BS}\bigr]_{(3)}^\trans,
\end{equation}
where $\bar{\mat{E}}_\mathrm{xh} = \mat{E}_\mathrm{xh}(\mat{I}_K \kron \hat{\vet{g}})$. Using the property \eqref{prop:kronkron}, we have
\begin{equation}
    \bigl[\ten{Y}^\mathrm{BS}\bigr]_{(3)}^\trans = (\hat{\mat{X}}^\trans \kron \hat{\mat{H}})(\mat{U}_\mathrm{x} \kron \mat{U}_\mathrm{h})\bar{\mat{E}}_\mathrm{xh} + \bigl[\ten{V}^\mathrm{BS}\bigr]_{(3)}^\trans.
\end{equation}
If $\bar{\mat{E}}_\mathrm{xh}$ is full row-rank,
\begin{equation}
    \bigl[\ten{Y}^\mathrm{BS}\bigr]_{(3)}^\trans\bar{\mat{E}}_\mathrm{xh}^\pinv = (\hat{\mat{X}}^\trans \kron \hat{\mat{H}})(\mat{U}_\mathrm{x} \kron \mat{U}_\mathrm{h})\bar{\mat{E}}_\mathrm{xh}\bar{\mat{E}}_\mathrm{xh}^\pinv.
\end{equation}
Then,
\begin{equation}
    \mat{U}_\mathrm{x} \kron \mat{U}_\mathrm{h} = \bar{\mat{E}}_\mathrm{xh}\bar{\mat{E}}_\mathrm{xh}^\pinv = \mat{I}_{RN}.
\end{equation}
To be possible $\mat{U}_\mathrm{x} \kron \mat{U}_\mathrm{h} = \mat{I}_{RN}$, the unique solutions are $\mat{U}_\mathrm{x} = \alpha\mat{I}_R$ and $\mat{U}_\mathrm{h} = (1/\alpha)\mat{I}_N$. Therefore, $\hat{\mat{X}}$ and $\hat{\mat{H}}$ are unique up to scaling factors that cancel each other.

Note that the interaction between the HRIS-BS channel and symbol matrices $\mat{H}$ and $\mat{X}$ is dictated utilizing a Kronecker product for both TSTC and KRSTC. Therefore, this ambiguity pattern is valid for both transmission schemes at the BS. 

\subsection{Scaling Ambiguity Removal}
On the HRIS side, for TSTC, scaling ambiguities can be mitigated by simply sending a single pilot embedded into the transmitted data. A simple choice is to set $\mat{X}_{1,1} = 1$. This knowledge allows us to determine $\alpha$ to eliminate the scaling ambiguity through normalization. On the other hand, for KRSTC, computing $\ma{\Delta}_\mathrm{x}$ implies the knowledge of one column of $\mat{X} \in \compl^{L \times T}$ to eliminate the scaling ambiguities. In this case, the UT can send a pilot embedded in the first symbol period of each data stream. A straightforward option is to consider $\mat{X}_{\cdot,1} = [1,\cdots,1]^\trans$. For both coding schemes, the scaling ambiguities affecting the estimated channel and symbol matrices at the BS are given by $\hat{\mat{X}} = \beta\mat{X}$ and $\hat{\mat{H}} = (1/\beta)\mat{H}$, which can also be eliminated using the same procedure discussed for the HRIS side. 

\section{Computational Complexity}\label{sec:cost}
As far as computational complexity is concerned, let us first recall the complexity of the matrix inverse. We consider a complexity of $\mathcal{O}(I^2J)$ to calculate the pseudo-inverse of a wide matrix $\mat{A} \in \compl^{I \times J}$, where $\mathrm{rank}\{\mat{A}\}=I$. For the iterative BALS algorithms, the complexity of each iteration is dominated by the two matrix inverses in \eqref{hriswgest} and \eqref{hriswxest} (for the HRIS-BALS receiver) and in \eqref{bswhest} and \eqref{bswxest} (for the BS-BALS receiver). The overall complexity is given by multiplying the complexity of a single iteration by the number of iterations to convergence. Moreover, note that the complexity of computing the truncated-SVD$(\mat{A})$ is assumed to be $\mathcal{O}(IJ\mathrm{rank}\{\mat{A}\})$. In the particular case of the KronF algorithms, the complexity is given by that of the LS estimation step in the first stage, given by \eqref{hriswxgest} for the HRIS-KronF receiver and by \eqref{bswxhest} for the BS-KronF receiver, followed by the complexity associated with computing a rank-1 matrix approximation step associated with the Kronecker factorization problems in \eqref{eq:kronFcost} and \eqref{kronFXh}, respectively. Finally, the KRF algorithm (considered at the HRIS under the KRSTC scheme) involves solving \eqref{xgesthriskr} followed by $L$ parallel rank-1 matrix approximation routines. Table \ref{tab:idcomp} lists the complexity of all receivers discussed in this work, with the complexity of all BALS receivers provided per iteration.

\begin{table*}[!t]
    \setlength{\tabcolsep}{1pt}
    \caption{Receiver pairs.}
	\label{tab:pairs}
	\centering
	\begin{tabular}{>{\centering\arraybackslash}m{1.5cm} >{\centering\arraybackslash}m{2cm} >{\centering\arraybackslash}m{3cm} >{\centering\arraybackslash}m{2cm} >{\centering\arraybackslash}m{2cm} >{\centering\arraybackslash}m{3cm} >{\centering\arraybackslash}m{1.8cm}}
        \midrule
		CL scenario&Feedback set&Bits fed back&Receiver pair&Coding scheme&Equations&Algorithms\\
		\midrule 
		1&$\mathscr{C}=\{\hat{\mat{G}}\}$&$LN\eta$&BALS-BALS&TSTC&\eqref{hriswgest}, \eqref{hriswxest}, \eqref{bswhest}, \eqref{bswxest}&\ref{alg:hrisbalsreceiverw} \& \ref{alg:bsbalsreceiverw}\\
        1&$\mathscr{C}=\{\hat{\mat{G}}\}$&$LN\eta$&BALS-KronF&TSTC&\eqref{hriswgest}, \eqref{hriswxest}, \eqref{bswxhest}&\ref{alg:hrisbalsreceiverw} \& \ref{alg:bskronfreceiver}\\
        1&$\mathscr{C}=\{\hat{\mat{G}}\}$&$LN\eta$&KronF-BALS&TSTC&\eqref{hriswxgest}, \eqref{bswhest}, \eqref{bswxest}&\ref{alg:hriskronfreceiver} \& \ref{alg:bsbalsreceiverw}\\
        1&$\mathscr{C}=\{\hat{\mat{G}}\}$&$LN\eta$&KronF-kronF&TSTC&\eqref{hriswxgest}, \eqref{bswxhest}&\ref{alg:hriskronfreceiver} \& \ref{alg:bskronfreceiver}\\
        1&$\mathscr{C}=\{\hat{\mat{G}}\}$&$LN\eta$&BALS-BALS&KRSTC&\eqref{hriswgest}$^\ast$, \eqref{hriswxest}$^\ast$, \eqref{bswhest}$^\ast$, \eqref{bswxest}$^\ast$&\ref{alg:hrisbalsreceiverw}$^\ast$ \& \ref{alg:bsbalsreceiverw}$^\ast$\\
        1&$\mathscr{C}=\{\hat{\mat{G}}\}$&$LN\eta$&BALS-KronF&KRSTC&\eqref{hriswgest}$^\ast$, \eqref{hriswxest}$^\ast$, \eqref{bsxhestkr}&\ref{alg:hrisbalsreceiverw}$^\ast$ \& \ref{alg:bskronfreceiver}$^\ast$\\
        1&$\mathscr{C}=\{\hat{\mat{G}}\}$&$LN\eta$&KRF-BALS&KRSTC&\eqref{xgesthriskr}, \eqref{bswhest}$^\ast$, \eqref{bswxest}$^\ast$&\ref{alg:hriskrfreceiver} \& \ref{alg:bsbalsreceiverw}$^\ast$\\
        1&$\mathscr{C}=\{\hat{\mat{G}}\}$&$LN\eta$&KRF-KronF&KRSTC&\eqref{xgesthriskr}, \eqref{bsxhestkr}&\ref{alg:hriskrfreceiver} \& \ref{alg:bskronfreceiver}$^\ast$\\
        2&$\mathscr{C}=\{\hat{\mat{G}},\hat{\mat{X}}\}$&$(RT\!-\!1)\mathrm{log}_2\varrho + LN\eta$&BALS-H&TSTC&\eqref{hriswgest}, \eqref{hriswxest}, \eqref{bswhest}&\ref{alg:hrisbalsreceiverw} \& eq. \eqref{bswhest}\\
        2&$\mathscr{C}=\{\hat{\mat{G}},\hat{\mat{X}}\}$&$(RT\!-\!1)\mathrm{log}_2\varrho + LN\eta$&KronF-H&TSTC&\eqref{hriswxgest}, \eqref{bswhest}&\ref{alg:hriskronfreceiver} \& eq. \eqref{bswhest}\\
        2&$\mathscr{C}=\{\hat{\mat{G}},\hat{\mat{X}}\}$&$L(T\!-\!1)\mathrm{log}_2\varrho + LN\eta$&BALS-H&KRSTC&\eqref{hriswgest}$^\ast$, \eqref{hriswxest}$^\ast$, \eqref{bswhest}$^\ast$&\ref{alg:hrisbalsreceiverw}$^\ast$\& eq. \eqref{bswhest}$^\ast$\\
        2&$\mathscr{C}=\{\hat{\mat{G}},\hat{\mat{X}}\}$&$L(T\!-\!1)\mathrm{log}_2\varrho + LN\eta$&KRF-H&KRSTC&\eqref{xgesthriskr}, \eqref{bswhest}$^\ast$&\ref{alg:hriskrfreceiver} \& eq. \eqref{bswhest}$^\ast$\\
        \midrule
	\end{tabular}
    \flushleft $^\ast$ Equations and algorithms from TSC that can be reused for KRSTC with necessary adaptations.
\end{table*}

\section{HRIS-BS Receiver Pairs}\label{sec:pairs}
\begin{figure}[!t]
	\centering
	\includegraphics[width=0.46\textwidth]{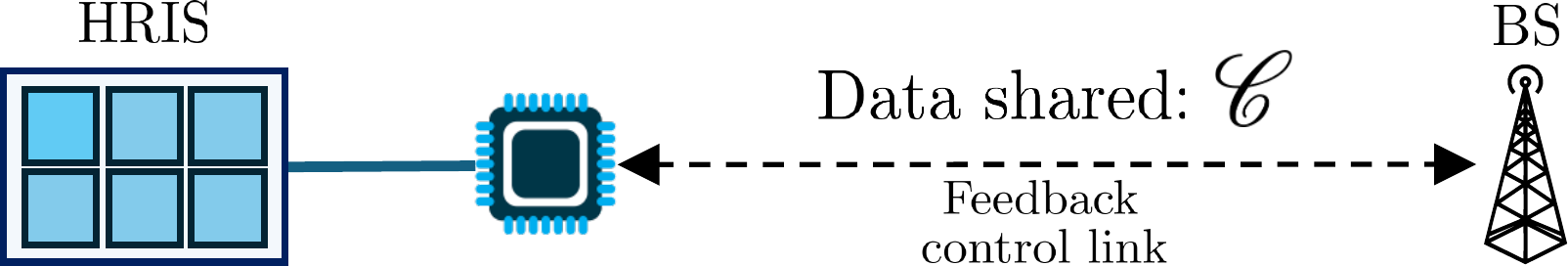}\\
	\caption{Information shared from HRIS to BS \emph{via} control link.}
	\label{fig:cl}
\end{figure}

At the HRIS, two matrices are estimated by following our semi-blind strategy: the symbol matrix, which carries useful data from the user terminal, and the UT-HRIS channel matrix. Depending on the system design and control overhead constraints, either or both of these matrices can be conveyed to the BS \emph{via} the control link. Fig. \ref{fig:cl} illustrates this process. This shared information defines the operational mode of the receiver at the BS to determine whether full, partial, or no prior knowledge is assumed during the processing of the received signals. The specific choice influences the strategy employed at the base station for joint channel and symbol estimation, ultimately affecting the receiver's performance, design constraints, and complexity. Capitalizing on the methods adopted in both HRIS and BS, we can define semi-blind receiver pairs by simply associating the receivers used in both, whose formation depends on the information shared between the HRIS and the BS through the control channel. Henceforth, we refer to these receiver pairs by adopting labels in the form ``HRIS-BS'' pair. Defining the set of data conveyed from the HRIS to the BS through the CL as $\mathscr{C}$, we can envision different operation modes at the BS. 

The first possible scenario would be $\mathscr{C} = \{\hat{\mat{G}},\hat{\mat{X}}\}$, i.e., both the estimated symbol and the UT-HRIS channel are reported to the BS. In this case, the BS would only need to calculate $\hat{\mat{H}}$ by utilizing \eqref{bswhest}. We refer to this simplified method as simply replacing ``BS'' by ``H'' in the `HRIS-BS'' label. Accordingly, as the joint symbol and CE can be accomplished iteratively or in closed form, this scenario leads to the following possibilities of receiver pairs: BALS-H and KronF-H for TSTC, and BALS-H and KRF-H for KRSTC.

\noindent \textbf{Discussion:} Recall that the semi-blind receivers provide continuous-valued estimations of the symbols and channel coefficients, so that before conveying the entries of the estimated matrices through the feedback control channel, these must be quantized. Considering a fixed resolution to represent the actual estimated data symbols, depending on the adopted modulation, the feedback duration related to sending $\hat{\mat{X}}$ \emph{via} CL is proportional to $(RT-1)\mathrm{log}_2\varrho$ for TSTC, and $L(T-1)\mathrm{log}_2\varrho$ for KRSTC, where $\varrho$ is the cardinality of the constellation. These quantities already account for the number of known symbols to remove scaling ambiguities. In addition, a level of resolution must be set to represent the estimated UT-HRIS channel coefficients, namely, the number of bits to quantize them. This way, the feedback duration related to sending $\hat{\mat{G}}$ is proportional to $LN\eta$, with $\eta$ being the resolution of each $\hat{\mat{G}}$ entry in bits. Indeed, this first scenario provides a lower computational cost at the BS by operating under a receiver processing simplified compared to KronF and BALS, and may offer better estimation performances. Despite the potential benefits, this approach would increase the control link load. In the context of the proposed data-aided semi-blind approach, this scenario serves as a performance reference that will be considered in our numerical evaluations.

A second scenario corresponds to $\mathscr{C} = \{\hat{\mat{G}}\}$. This was initially exploited in the work \cite{zhang2023channel} in the pilot-aided approach for CE. In line with this control link usage, there are four possible combinations of semi-blind receiver pairs for each coding scheme (TSTC or KRSTC). In the TSTC scheme, the four receiver pairs are BALS-BALS, BALS-KronF, KronF-BALS, and KronF-KronF. In the KRSTC scheme, we have BALS-BALS, BALS-KronF, KRF-BALS, and KRF-KronF. These terminologies (and those of the first scenario) will be used in Section \ref{sec:results}, where the numerical results of the different receiver pairs will be evaluated. By sending to the BS just one matrix, this scenario reduces the feedback overhead, decreasing the HRIS-to-BS latency, and thereby conserving network resources, which can be an advantage in high-mobility or fast-fading environments. This strategy also preserves network resources by limiting the use of the control channel to signaling exchange, thereby avoiding its occupation with data transmission.

For reference, we provide in Table \ref{tab:pairs} all the possible combinations of ``HRIS-BS'' estimation methods by considering the scenario of the CL usage, related equations, and algorithms. When considering all the receiver options available at both the HRIS and the BS, there is flexibility in selecting the receiver pair to be used for the joint estimation of the channel and symbols. For instance, one may opt for combinations that offer low computational complexity at the cost of being more restrictive regarding system design. Conversely, more costly schemes can provide greater design flexibility. Moreover, a balance between complexity and design constraints can be achieved by selecting mixed receiver pairs with iterative and closed-form solutions. For example, using KronF at the HRIS and BALS at the BS would provide low computational cost for the HRIS, while it would be more costly for the BS. In contrast, the design requirements would be more restrictive for the HRIS than the BS.

\section{Simulation Results}\label{sec:results}
We adopt a distance-dependent path loss (PL) model, given by $\mathrm{PL} = \mathrm{PL}_0(d/d_0)^{-\alpha}$, in which $\mathrm{PL}_0 = -20$ dB is the path loss at the reference distance $d_0 = 1$m, $d$ is the individual link distance, and $\alpha$ denotes the path loss exponent. We consider $d_u = 40$ m, $d_h = 10$ m, and we set $\alpha_g =2.5$ and $\alpha_h=2$ as, respectively, the UT-HRIS and HRIS-BS link distances and path loss exponents. We assume the Rayleigh fading channel model, in which the UT-HRIS and HRIS-BS channels are taken from a zero-mean independent and identically distributed (i.i.d.) complex-valued Gaussian distribution with variances $\gamma$ and $\beta$, respectively, corresponding to the path losses of these links. To keep the analyses simple, both the HRIS and the BS have the same noise power level $\sigma_v^2 = -90$ dBm. Given that the energy coupling level is dictated through meta-atom design, as highlighted in \cite{alexandropoulos2024hybrid}, we allocate the same coupling level to all meta-atoms, and we assume the $\rho$ parameter is non-reconfigurable to simplify the assessment. We design the reflecting and sensing phase shifts as well as the coding (for both TSTC and KRSTC) according to Appendix \ref{apdesign}. The symbol matrix $\mat{X}$ is based on a 64-QAM constellation. We evaluate joint symbol and CE accuracies employing the symbol error rate (SER) and the normalized mean square error (NMSE), respectively. Each result is an average over at least $10^4$ independent Monte Carlo runs. Each run considers different realizations of the symbols, channels, and noise. To ensure a fair comparison between the proposed TSTC and KRSTC schemes, we set $R = L$ and dismiss the entire first column of $\hat{\mat{X}} \in \compl^{R \times T}$ to calculate the SER (not only $\hat{\mat{X}}_{1,1}$). Unless otherwise stated, we assume the parameter set $\{M,N,N_c,L,R,T,K\} = \{8,32,2,2,2,4,64\}$.
\begin{figure}[!t]
	\centering
	\includegraphics[width=0.4\textwidth]{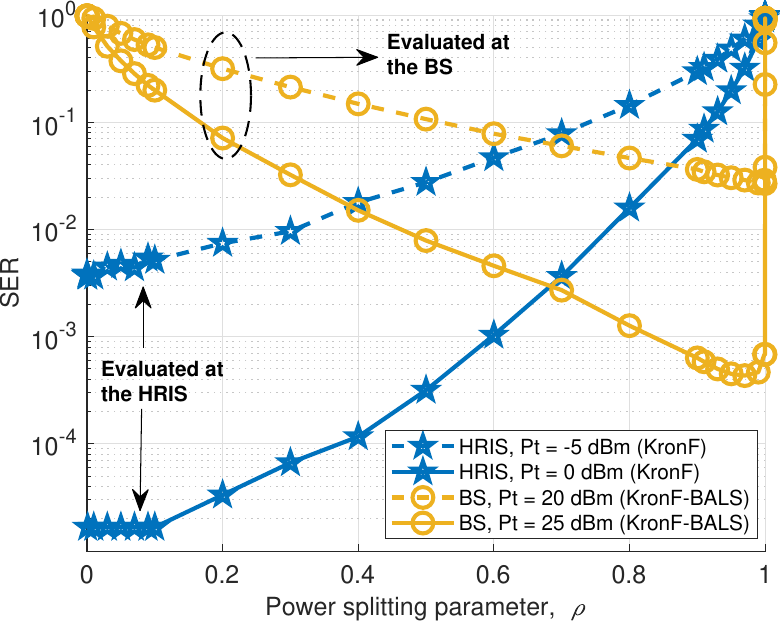}
 \caption{Behavior of SER vs. $\rho$, evaluated at both the HRIS and BS.}
	\label{fig:serrho}
\end{figure}
\begin{figure}[!t]
	\centering
	\includegraphics[width=0.4\textwidth]{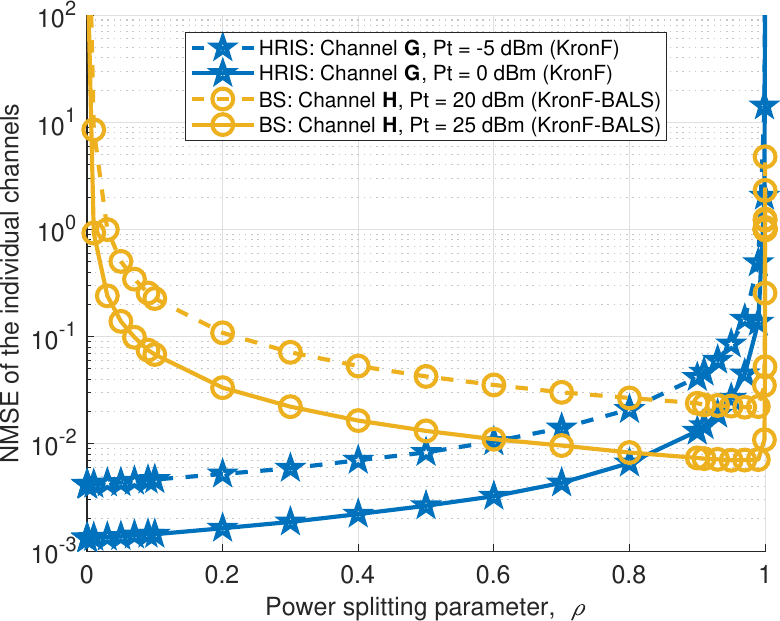}
 \caption{NMSE of individual channels vs. $\rho$.}
	\label{fig:nmserho}
\end{figure}

\subsection{Impact of the Power Splitting Parameter on the Symbol Estimation Performance}
Firstly, we examine the trade-off between the data-aided semi-blind CE accuracy and the power splitting parameter $\rho$, shown in Figs. \ref{fig:serrho} and \ref{fig:nmserho}. We provide NMSE and SER results by fixing transmit power values in dBm, denoted by $P_t$, and changing $\rho$. In particular, we retain the focus of this study on symbol estimation performance as the CE evaluation was previously reported in \cite{zhang2023channel}, which introduced a pilot-aided approach. To inspect symbol estimation at both HRIS and BS, we selected the scenario in which only the estimated channel matrix $\hat{\mat{G}}$ is received at the BS from the CL (\textit{CL scenario 1}), which implies using semi-blind receivers for joint symbol and CE at both ends (BS and HRIS). We chose the KronF and BALS receivers following the TSTC scheme for the HRIS and BS, respectively.

From the results, we can observe that when $\rho$ approaches 0, the HRIS becomes a quasi ``only detecting (not reflecting) RIS'', causing low SER values for the symbol estimation performed at the HRIS, while they approach 1 at the BS. As $\rho$ increases, the symbol estimation performance is degraded at the HRIS due to the decreased sensing/detection capability. In contrast, the estimation accuracy at the BS side is enhanced as the reflected signal arrives at the BS with greater strength. Likewise, higher values of $\rho$ imply an increase in the NMSE of $\mat{G}$ while decreasing the NMSE of $\mat{H}$, as predicted in \cite{zhang2023channel}. Since the CE capability at the HRIS is less affected by path loss in our setup (the HRIS is closer to the UT than the BS), the estimation accuracy for the channel $\mat{G}$ remains much higher even by decreasing the sensing capability (i.e., increasing $\rho$). Nevertheless, when $\rho$ approaches 1, and the HRIS behaves closer to a ``pure reflecting RIS'', the improvement on the estimates of $\hat{\mat{H}}$ at the BS stops, since the accuracy of $\hat{\mat{G}}$ becomes compromised. In addition, from the Fig. \ref{fig:serrho}, we can note that for smaller values of $\rho$, the SER performance at the HRIS is more sensitive to transmit power variations. The same happens with the SER performance at the BS for higher values of $\rho$. On the other hand, the Fig. \ref{fig:nmserho} shows that the channel estimation performances are less sensitive to the variation of the $P_t$ compared to the SER ones.

In the next experiments, all the results consider $\rho=90$\%. Since the BS experiences higher path loss due to the cascaded (UT-HRIS-BS) link, this choice allocates more power to the reflected signal part. Despite the lower power allocated to the sensed signal part, the low path loss associated with the UT-HRIS link still ensures reliable symbol detection and CE at the HRIS.

\subsection{Performance of Individual Channel Estimation and Symbol Decoding at the HRIS}
\begin{figure}[!t]
	\centering
	\includegraphics[width=0.4\textwidth]{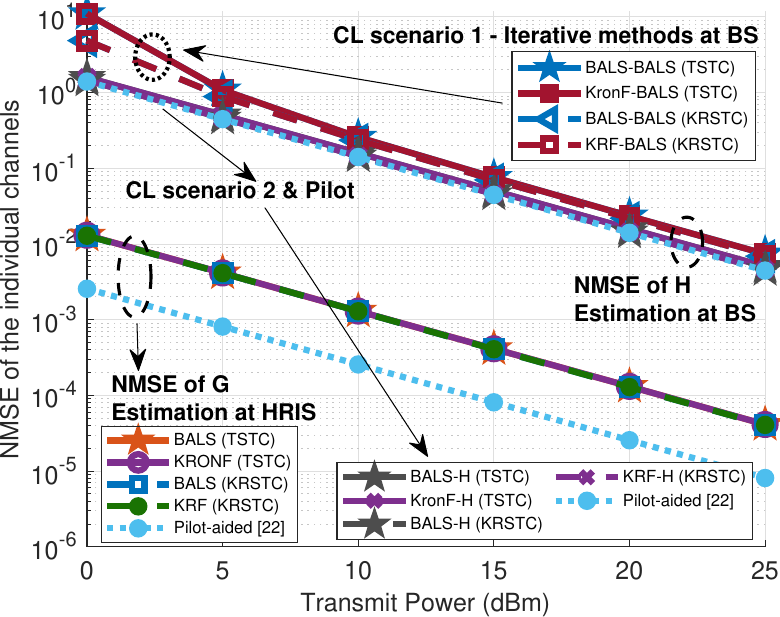}
 \caption{NMSE of the individual channels vs. transmit power (dBm).}
	\label{fig:nmsegh}
\end{figure}

\begin{figure}[!t]
	\centering
	\includegraphics[width=0.4\textwidth]{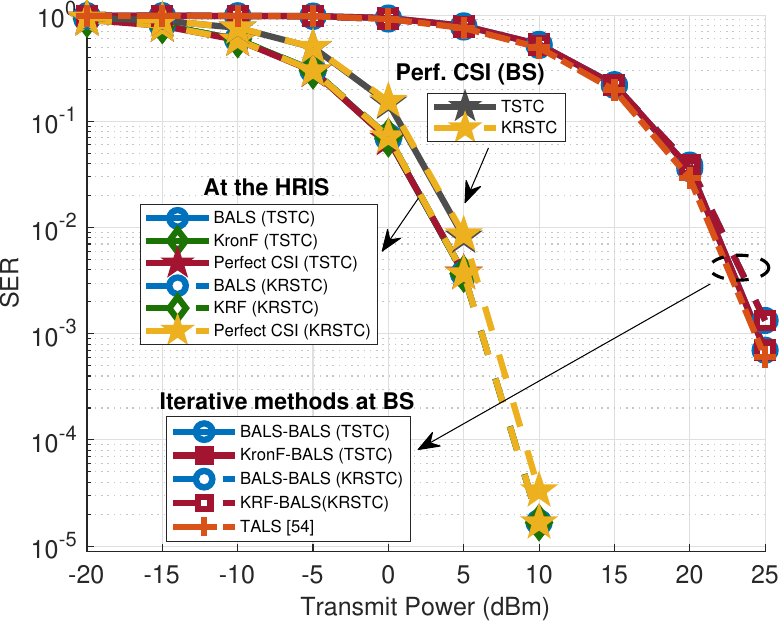}
 \caption{Symbol error rate vs. transmit power (dBm).}
	\label{fig:ser}
\end{figure}

In Fig. \ref{fig:nmsegh}, we study the NMSE performances at the HRIS and the BS as a function of the transmit power. We depict the NMSE of the individual channels $\mat{G}$ (estimated at the HRIS) and $\mat{H}$ (estimated at the BS). On the other hand, Fig. \ref{fig:ser} displays the SER results associated with the symbol detection at both HRIS and BS. In both figures, we compare the performances of the proposed semi-blind receivers as follows: i) we evaluate the results of the proposed receivers designed for the HRIS (see ``NMSE of $\mat{G}$'' curves); ii) by selecting a receiver for the HRIS and another for the BS, we account the performance of ``HRIS-BS'' receiver pairs (see ``NMSE of $\mat{H}$'' curves). This is made for both coding schemes, TSTC and KRSTC. Let us first start with the performance on the HRIS side. As a reference for comparisons, in Fig. \ref{fig:nmsegh}, we also plot the performance of the pilot-assisted case based on \cite{zhang2023channel} using the same set of parameters\footnote{\linespread{0.7}\selectfont We adapted the signal model of \cite{zhang2023channel} to the single-user case. In this case, we design $\mat{X}$ as a truncated discrete Fourier transform (DFT) matrix. Since HRIS optimization is out of the scope of our work, and to keep the fair comparison, we do not leverage the optimization procedure proposed in \cite{zhang2023channel}.}, wherein we apply a simple LS solution to estimate $\mat{G}$ at the HRIS and $\mat{H}$ at the BS. 


As shown in Figs. \ref{fig:nmsegh} and \ref{fig:ser}, all semi-blind receivers operating at the HRIS exhibit the same NMSE and SER performances (for both TSTC and KRSTC), with the HRIS offering higher accuracy in symbol and CE compared to the BS due to its positioning. Such results align with those reported in \cite{zhang2023channel} in the CE scope (see Fig. 11 therein). Moreover, the spatial diversity introduced by analog combining at the HRIS also contributes to improved performance. However, from Fig. \ref{fig:nmsegh}, comparing the pilot-aided CE \cite{zhang2023channel} with all the data-aided semi-blind receivers, we observe that the former can be regarded as a lower bound for the proposed methods in terms of CE performance. This is because estimating the symbol matrix in addition to the channel introduces an additional challenge compared to the baseline pilot-assisted method, which is restricted to only estimating the channel $\mat{G}$ with full knowledge of $\mat{X}$ (pilots, in this case). Nevertheless, as we will see later, this performance difference at the HRIS will not be significant for certain receiver combinations (``HRIS-BS'' pairs), meaning that the estimation at the BS will not be substantially affected. Furthermore, the semi-blind approach allows the UT to transmit data symbols instead of only pilot sequences within the same time-division structure. Indeed, adding a joint symbol and CE functionality at the HRIS can unlock new potential for RIS-aided wireless communication systems, which will be further discussed in Section \ref{sec:discussion}. Regarding symbol estimation, Fig. \ref{fig:ser} shows that all the receivers performed competitively for both coding schemes, corroborating our numerical results shown previously. These results represent a remarkable milestone in symbol estimation utilizing the hybrid architecture proposed by \cite{alexandropoulos2024hybrid} using only two RF-chains out of $N=32$ HRIS elements. 

\subsection{Individual Channel Estimation at the BS}
Still considering Figs. \ref{fig:nmsegh} and \ref{fig:ser}, let us now focus on the BS performance by considering pairs of ``HRIS-BS'' receivers. Recall all the possibilities of ``HRIS-BS'' receiver pairs/combinations by referring back to Table \ref{tab:pairs}. Here, we do not consider receiver pairs using closed-form schemes at the BS for \textit{CL scenario 2} (i.e., KronF for both coding schemes) since we found unsatisfactory results. The adoption of such closed-form receivers at the BS will be discussed later in the topic  , at the end of this Section. To assist us in the discussion, Figs. \ref{fig:time} and \ref{fig:it} display, respectively, the average runtime (in seconds) and the number of iterations required for convergence of the iterative algorithms as a function of $P_t$ for all receiver pairs. Additionally, Fig. \ref{fig:flops} shows the evolution of the computational complexity with respect to the number of HRIS elements.

Regarding the estimation of the channel $\mat{H}$ (HRIS-BS channel), whose results are also exhibited in Fig. \ref{fig:nmsegh}, all receiver pairs arising from the \textit{CL scenario 2} and all pairs from the \textit{CL scenario 1} using iterative BALS at the BS performed similarly and demonstrated improved estimation accuracy. Although a difference in accuracy was noted between pilot-aided and data-aided approaches when estimating $\mat{G}$ at the HRIS, the NMSE curves related to estimating $\mat{H}$ at the BS closely resemble the baseline pilot-aided method, particularly for KronF-H, KRF-H, and BALS-H (in both coding schemes). The accurate estimation previously obtained at the HRIS effectively narrowed the performance gap at the BS between pilot-aided and the early-mentioned data-aided semi-blind methods, significantly reducing the performance disparities. Note that scenarios in which $\hat{\mat{X}}$ and $\hat{\mat{G}}$ are jointly conveyed \emph{via} the CL (\textit{CL scenario 2}) result in solutions with lower computational complexities and less restrictive design requirements, reaching the best results in terms of joint symbol and CE. However, this method requires more feedback associated with the additional conveyance of $\hat{\mat{X}}$ estimated at the HRIS, especially when it has larger dimensions. This way, when the choice is to save on feedback, i.e., send only the estimated matrix $\hat{\mat{G}}$, the BS would jointly estimate channel and symbols. By inspecting the NMSE results for the individual channel $\mat{H}$ in Fig. \ref{fig:nmsegh}, applying iterative receivers at the BS (\textit{CL scenario 1}) implies just a ~2 dB gap in performance, being an exciting finding. Recall that BALS receivers have the distinguishing feature of refining the channel and symbol estimates at each iteration, based only on the estimated channel $\hat{\mat{G}}$ reported by the HRIS. In contrast, their overall complexity depends on the SNR, since the number of iterations required for convergence increases for lower transmit power levels, as indicated in Fig. \ref{fig:it}.

\begin{figure}[!t]
	\centering
	\includegraphics[width=0.4\textwidth]{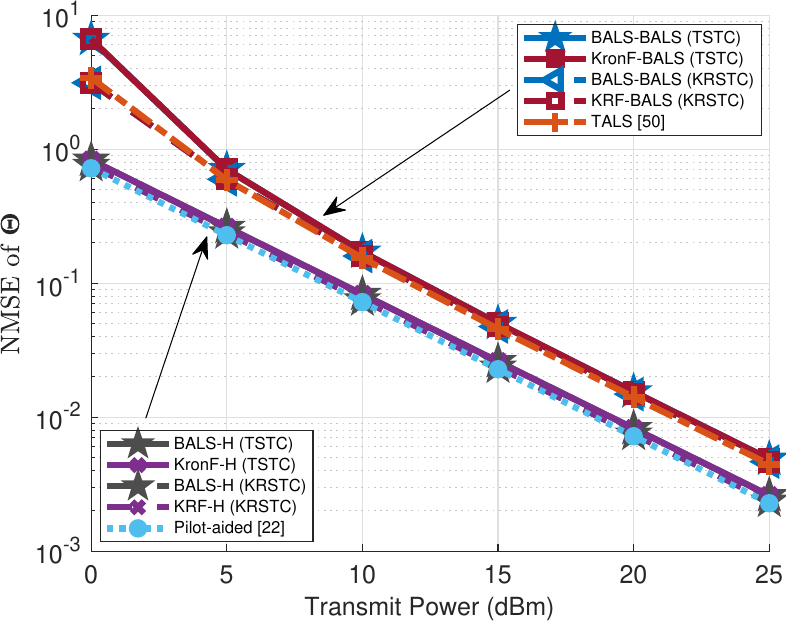}
 \caption{NMSE of the combined channel vs. transmit power (dBm).}
	\label{fig:nmsetheta}
\end{figure}

\subsection{Evaluation of the Combined Channel Estimation}
In addition, we plot the NMSE of the combined channel in Fig. \ref{fig:nmsetheta}, evaluated at the BS by adopting the Khatri-Rao structured matrix $\ma{\Theta} = \mat{G}^\trans \krp \mat{H} \in \compl^{LM \times N}$. We observe that the performance of the combined CE follows similar patterns to those obtained from the estimations of $\mat{H}$. This emphasizes the similarity in performance between the two groups of methods regarding the estimation at the BS: one first group that employs BALS-based iterative semi-blind receivers (from \textit{CL scenario 1}, to estimate $\mat{H}$ and $\mat{X}$), and a second group that estimates only $\mat{H}$ (the data-aided semi-blind approach from \textit{CL scenario 2}, and the pilot-aided one). In the context of the estimation of $\ma{\Theta}$, the Fig. \ref{fig:nmsetheta} reveals a more distinguishable and clear separation between such groups, previously observed only for estimation of $\mat{H}$ in Fig. \ref{fig:nmsegh}. Note that the performance gap between pilot and data-aided strategies was practically eliminated when we considered the combined channel estimation $\ma{\Theta}$. In addition to the proposed methods previously discussed, we include in Fig. \ref{fig:nmsetheta} the iterative trilinear ALS (TALS) semi-blind receiver proposed in \cite{dearaujo2023semiblind}, representing the baseline method for data-aided semi-blind CE in the context of PRIS-assisted communications. Recall that in the HRIS approach using the KRSTC scheme, the received signal at the BS differs from that of the PRIS one adopted in \cite{dearaujo2023semiblind} only by introducing the factor $\rho$, causing the HRIS to reflect an impinging wave's fraction instead of its totality ($\rho = 1$). Although a small performance gap was observed between groups 1 and 2, the methods in group 2 demonstrated similar performance to that of the baseline TALS, highlighting their effectiveness in solving the problem of joint channel and symbol estimation semi-blindly in a data-aided CE viewpoint.

\subsection{Symbol Estimation Performance at the BS}
To assess symbol estimation performance at the BS, Fig. \ref{fig:ser} depicts the SER results of the receiver pairs discussed earlier, and those are compared to the symbol estimation provided by the TALS receiver \cite{dearaujo2023semiblind}. The results show that ``HRIS-BS'' receiver pairs executing BALS (for both TSTC and KRSTC) at the BS perform similarly to the PRIS case using TALS. These results support our findings in Fig. \ref{fig:nmsetheta} for the combined CE related to the receivers of group 2 and the baseline TALS in the PRIS approach. It is essential to highlight that the HRIS absorbs 10\% of the incident signal's energy. This is significant, as the joint symbol and channel estimation remains nearly unaffected compared to the PRIS case when employing iterative BALS receivers. Moreover, it is worth noting that the proposed receivers in the hybrid approach offer a scaling ambiguity-free separate CE while decentralizing the CE task, which was previously performed only at the BS in the passive approach. However, one should consider the trade-off between hardware complexity and power consumption when opting for the HRIS architecture.

\begin{figure*}[!t]
\minipage[t]{0.31\linewidth}
    \includegraphics[width=\textwidth]{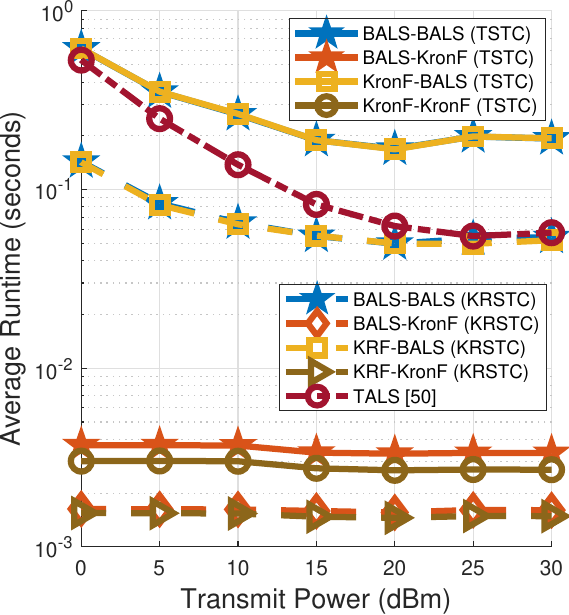}
    \caption{Average runtime of the receiver pairs (\textit{CL scenario 1}) vs. transmit power.}
    \label{fig:time}
\endminipage\hspace{2ex}
\minipage[t]{0.31\linewidth}
    \includegraphics[width=\textwidth]{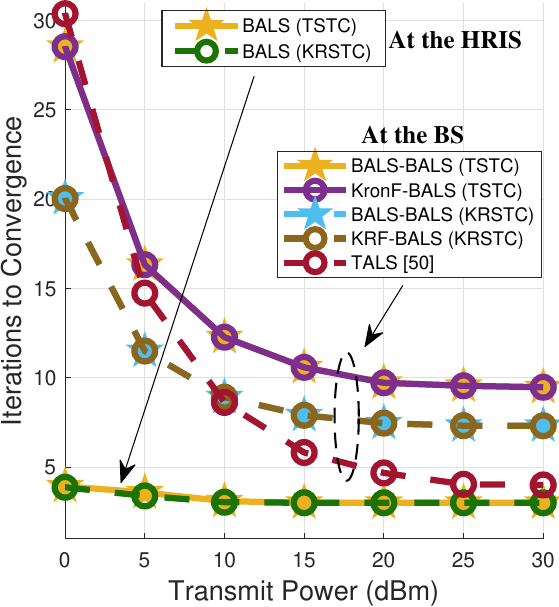}
    \caption{Iterations to convergence for iterative methods vs. transmit power.}
    \label{fig:it}
\endminipage\hspace{2ex}
\minipage[t]{0.34\linewidth}
\raggedright
    \includegraphics[width=0.92\textwidth]{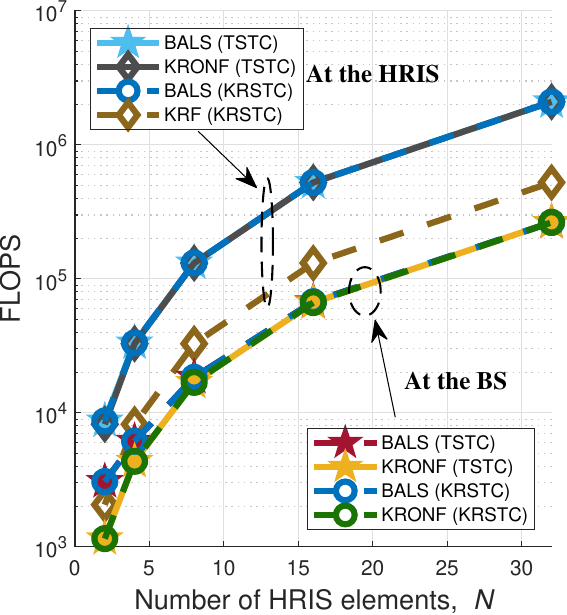}
    \caption{Number of FLOPS vs. number of HRIS elements.}
    \label{fig:flops}
\endminipage
\end{figure*}

\subsection{Employing Closed-Form Semi-Blind Receivers at the BS}\label{kronfresults}
\begin{figure}[!t]
	\centering
	\includegraphics[width=0.4\textwidth]{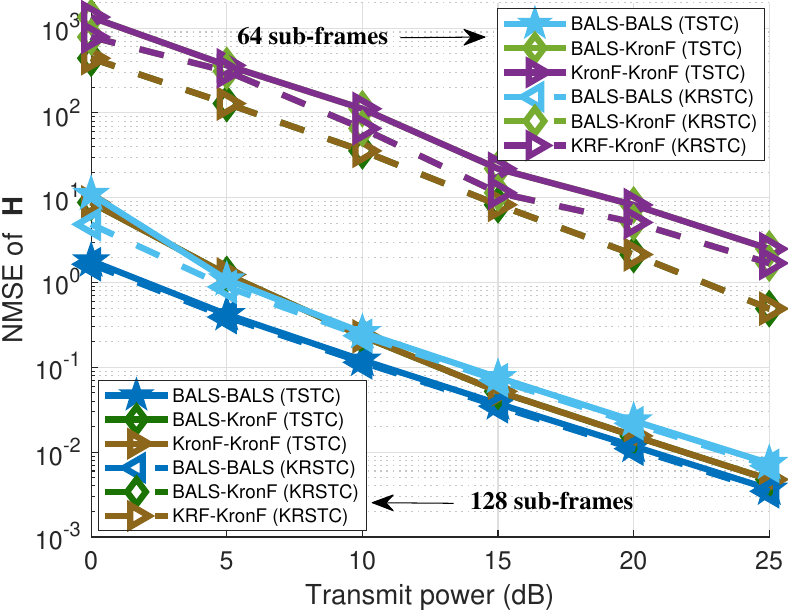}
 \caption{Evaluation of CE performance using closed-form receivers at the BS for $K=64$ and $K=128$.}
	\label{fig:nmseh128comp64}
\end{figure}
\begin{figure}[!t]
	\centering
	\includegraphics[width=0.4\textwidth]{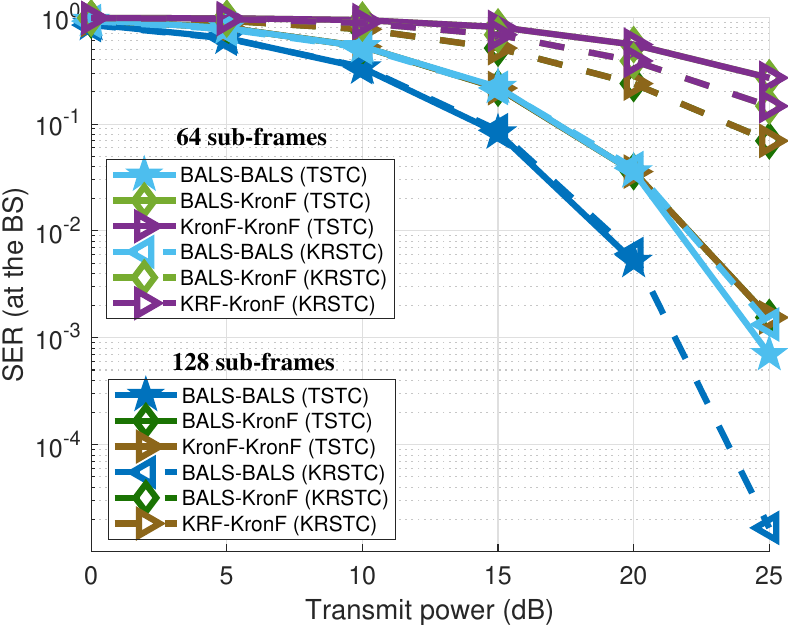}
 \caption{Evaluation of symbol estimation performance using closed-form receivers at the BS for $K=64$ and $K=128$.}
	\label{fig:serbs128comp64}
\end{figure}

To provide simulation results under the umbrella of a fair comparison, we have assigned the minimum number of sub-frames that simultaneously meet the identifiability condition for all matrix inverses present in the considered estimators. This choice led to the results of the joint symbol, and CE presented in Figs. \ref{fig:nmsegh} and \ref{fig:ser}. In this topic, we show that the use of semi-blind closed-form KronF receivers at the BS can be leveraged at the cost of paying additional time overhead.

From Table \ref{tab:idcomp}, note that $K=64$ sub-frames are more than enough to meet the identifiability conditions of the proposed iterative BALS receivers. In contrast, the KronF ones operate tightly at their minimum identifiability boundary ($K=RN$ for TSTC and $K=LN$ for KRSTC). To complement our discussion, we present additional simulation results by setting $K=128$, i.e., beyond the minimum $K$ required for all the closed-form receivers at the BS (recall Table \ref{tab:idcomp}). We show in Figs. \ref{fig:nmseh128comp64} and \ref{fig:serbs128comp64} the NMSE of the HRIS-BS channel (channel $\mat{H}$) and SER at the BS, respectively, for all the receiver pairs that apply closed-form receivers KronF at the BS, i.e, BALS-KronF and KRF-KronF for KRSTC, and BALS-KronF and KronF-KronF for TSTC. We provide comparisons between the formers and the pairs BALS-BALS for both TSTC and KRSTC (which apply an iterative receiver at the BS).

We can see that such an increase of $K$ was not sufficient to bring improved performance at the BS when the Khatri-Rao coding scheme and KronF are exploited (BALS-KronF and KRF-KronF for KRSTC). Therefore, we do not recommend using such a closed-form receiver at the BS when adopting the KRSTC transmission scheme. In contrast, KronF, using the TSTC scheme, achieves a significant performance improvement at the BS since the BALS-KronF and KronF-KronF pairs performed similarly to the iterative ones and the baseline methods. Note that the benefits provided by the KronF receiver under the TSTC come with the transmission of more sub-frames. Such findings were initially reported in \cite{magalhaes2025closed}. It is worth mentioning that under the same number of UT antennas, the TSTC scheme can allocate more streams than the number $L$ of transmit antennas, which is an interesting setup not possible with the KRSTC scheme.

Concerning the choice of iterative or closed-form receivers at the BS for joint symbol and CE in the \textit{CL scenario 1}, let us recall that BALS-based receivers can operate under more flexible system setups compared to the closed-form ones. Conversely, KronF performs only a single matrix inverse followed by a rank-1 matrix approximation step, which is much less complex than BALS (for both coding schemes). As mentioned earlier, in the low transmit power regime, the ALS procedure may require several iterations to converge, thereby increasing the overall computational cost. Hence, the runtime of receiver pairs associated with KronF at the BS in Fig. \ref{fig:time} is relatively low compared to those of BALS-based ones. However, it is worth mentioning that although the computational complexity of KronF is lower than that of BALS, it implies more restrictive system setups, as demonstrated in Table \ref{tab:idcomp}.

This illustrates the interesting tradeoffs offered by the proposed semi-blind HRIS-BS receiver pairs involving performance, complexity, and operating conditions.

\section{Discussion}\label{sec:discussion}
In the following, we discuss a few examples of application scenarios and use cases that can potentially benefit from a joint channel and symbol estimation at the HRIS. Relying on direct estimation, \textit{uplink sensing} was considered in \cite{rahman2019framework} in a perceptive mobile network \cite{zhang2020perceptive} employing joint communication and sensing, which involves the detection of UTs and environmental characteristics between them and remote radio units (RRUs). Therein, symbols are embedded into a sensing matrix, and compressed sensing is applied to estimate delay, Doppler, and angular parameters. The semi-blind approach is suitable for the mentioned joint communication and radar sensing scenario since it dispenses pilot-based training and jointly recovers symbols and channel estimates in a one-way time protocol employing simple receiver algorithms. Furthermore, multiple HRISs can be distributed to serve as decentralized uplink sensing points while alleviating the processing load at the BS. It is worth pointing out that recent works have considered estimating channel parameters at a hybrid RIS. As examples, please refer to \cite{alexandropoulos2022localization, hu2022semipassive, shao2022target, luo2023reconfigurable}.

In \cite{yigit2022overtheair}, a multi-antenna UT conveys extra bits to the RIS controller \emph{via} a CL while sending data symbols to the BS \emph{via} the UT-RIS channel during the uplink transmission. The RIS controller uses such extra bits to apply an \emph{over-the-air beamforming} technique to improve the transmission. This relies on the prerequisite that there is a CL between the UTs and the RIS. However, maintaining a CL with each active UT can result in a scalability problem since the RIS controller may need to support simultaneous connections with many UTs. This is another suitable application for the proposed semi-blind HRIS processing since control signals can be embedded directly into the data and sent over the UT-RIS link, thereby reducing or eliminating the dependency on multiple CLs between the HRIS and the UTs. More specifically, the transmitted signals may contain a payload containing both data and control symbols, i.e., the symbol matrix $\mat{X}$ can be partitioned into $\mat{X} = [\mat{X}_c, \mat{X}_d]$, where $\mat{X}_c$ contains control symbols to fulfill the mentioned purpose. In contrast, $\mat{X}_d$ contains data symbols. This way, our semi-blind approach eliminates the need for CLs between UTs and the RIS since $\mat{X}$ (or part of it) can be found at the HRIS. Otherwise stated, leveraging the information contained in $\mat{X}_c$ allows the HRIS to decode control signals in a stand-alone fashion directly. 

Another application that can potentially benefit from a joint symbol and channel estimation at the HRIS is vehicular communications. For instance, consider a scenario with multiple roadside RISs placed to serve high-mobility vehicles, as shown in \cite{huang2023roadside}. In \cite{huang2023roadside}, assuming that the RIS-BS channel is static, the time-varying UT-RIS channel can be estimated/predicted at the RIS in a decentralized manner, i.e., without the assistance of the BS, minimizing the usage of the CL and avoiding feedback delays and outdated beamforming optimization as a consequence of high UT's mobility. To this, the RIS controller transmits the pilots by the UT (vehicle) to the BS during the uplink transmission protocol. Bringing this problem to our proposed semi-blind approach, we foresee new use cases allowing UTs to directly share valuable information with their serving HRIS (and the adjacent ones) by embedding control data such as position, speed, and handover commands into the symbol matrix, which can be decoded at each HRIS and exchanged between in a decentralized way without requiring feedback with their serving BSs. For example, such control data can include speed/position \cite{ahmed2023vehicular}, following a similar perspective to that used in \emph{active road safety} \cite{karagiannis2011vehicular} applications in vehicular networking.

Finally, we can also envisage a useful scenario where the BS sends control data to (re)configure the HRIS autonomously. In that case, the symbol matrix to be estimated/decoded at the HRIS contains control commands for HRIS configuration purposes or any other relevant network signaling information. This scenario corresponds to an over-the-air HRIS reconfiguration or standalone operation without using the control link.

\section{Conclusion}\label{sec:conclusion}
This paper proposed semi-blind joint channel and symbol estimation solutions for a hybrid simultaneous reflecting and sensing RIS. Adopting a tensor modeling approach, we revealed the tensor structures of the transmitted signals and the received signals at the HRIS and BS as combinations of PARAFAC and Tucker models, from which novel semi-blind receiver pairs for combined HRIS-BS processing are derived. The proposed tensor-based receivers provide data-aided estimations of the involved channels at both the HRIS and the BS without an \emph{a priori} pilot transmission stage, reducing the symbol decoding delay and improving the data rate. We derived both iterative and closed-form algorithms for joint channel and symbol estimation. We also studied identifiability conditions for guaranteed channel and symbol recovery for each semi-blind receiver pair, revealing the competitive performances of the proposed solutions in comparison with reference methods. Extensive simulation results showcased the performance trends and tradeoffs for the different HRIS-BS receiver pairs. Despite their higher computational complexity, receiver pairs using iterative BALS at the BS offer better estimation accuracy compared to the closed-form (KronF) ones, regardless of the receiver chosen at the HRIS. On the other hand, we have demonstrated that KronF can achieve performance comparable to that of iterative methods and further reduce computational complexity at the expense of additional overhead. Our discussion also illuminates the opportunities and use cases arising from empowering HRIS with symbol detection capability. Our numerical results also clarified the impacts of power splitting and tensor coding on channel estimation accuracy and symbol error rates for HRIS-assisted communications. These insights are pivotal for optimizing the system performance in future HRIS deployments. Perspectives include extending the proposed semi-blind receivers to multi-user scenarios and studying alternative tensor-based estimation algorithms.

\renewcommand{\appendixname}{Appendices}
\appendix

\subsection{Proof of Propositions \ref{proposition:x1} and \ref{proposition:x2}}\label{ap:ident1}
For the HRIS, presuming the design of $\ten{T}_{\ma{\Phi}}$ and $\ten{W}$ ensures $\mathrm{dim}[\mathtt{R}(\mat{F}_{\mathrm{x}_1}) \cap \cdots \cap \mathtt{R}(\mat{F}_{\mathrm{x}_K})]=0$, where $\mathtt{R}(\mat{F}_{\mathrm{x}_k})$ denotes the row-space of $\mat{F}_{\mathrm{x}_k}$, $\mat{F}_\mathrm{x}$ has full column-rank if the sum of the ranks of all blocks \cite{favier2020algebraic},\!\cite{matsaglia1974equalities} $\mat{F}_{\mathrm{x}_1},\!\cdots\!,\mat{F}_{\mathrm{x}_K}$ is no less than $R$ when considering the TSTC scheme, i.e,
\begin{align}
    \mathrm{rank}\bigl(\mat{F}_\mathrm{x}\bigr) =& \mathrm{min}\{\textstyle \sum_{k=1}^K{\mathrm{rank}(\mat{F}_{\mathrm{x}_k})},R\}\nonumber\\
    =& \mathrm{min}\{K\zeta_\mathrm{x},R\},\label{rankfx}
\end{align}
which in turn requires $K\zeta_\mathrm{x} \geq R$. Assuming $N \gg N_c$, we can determine the upper-bound of $\zeta_\mathrm{x}$ by exploiting the well-known property $\mathrm{rank}(\mat{ABC}) \leq \mathrm{min}\{\mathrm{rank}(\mat{A}),\mathrm{rank}(\mat{B}),\mathrm{rank}(\mat{C})\}$. In this way, $\zeta_\mathrm{x} \leq \mathrm{min}\{\mathrm{rank}(\ma{\Phi}_k), \mathrm{rank}(\mat{G}), \mathrm{rank}(\mat{W}_k)\}$, or $\zeta_\mathrm{x} \leq \mathrm{min}\{N_c,\kappa_\mathrm{g},\mathrm{min}\{L,R\}\}$. Since $\kappa_\mathrm{g} \leq L$, the upper-bound inequality can be simplified by simply replacing $\mathrm{min}\{L,R\}$ by the number of data streams that the UT encodes through TSTC, i.e.,
\begin{equation}\label{lemmax1}
    \zeta_\mathrm{x} \leq \mathrm{min}\{N_c,\kappa_\mathrm{g},R\}.
\end{equation}
When KRSTC is employed, recall that $\mat{W}_k$ is replaced by $\diagof{\vet{\lambda}_k}$. This way, the inequality of \eqref{lemmax1} becomes $\zeta_\mathrm{x} \leq \mathrm{min}\{N_c,\kappa_\mathrm{g}\}$, and following the same reasoning, \eqref{rankfx} turns into $\mathrm{rank}\bigl(\mat{F}_\mathrm{x}\bigr) = \mathrm{min}\{K\zeta_\mathrm{x},L\}$, requiring $K\zeta_\mathrm{x} \geq L$.

For the BS, assuming the design of $\ma{\Psi}$ and $\ten{W}$ guarantees $\mathrm{dim}[\mathtt{R}(\mat{H}\mat{E}_{\mathrm{x}_1}) \cap \cdots \cap \mathtt{R}(\mat{H}\mat{E}_{\mathrm{x}_K})]=0$, to $(\mat{I}_K \kron \mat{H})\mat{E}_\mathrm{x}$ have full column-rank, the sum of the ranks of all blocks must be at least $R$ in the TSTC or, equivalently,
\begin{align}\label{rankex}
    \mathrm{rank}\bigl((\mat{I}_K \kron \mat{H})\mat{E}_\mathrm{x}\bigr) =& \mathrm{min}\{\textstyle \sum_{k=1}^K{\mathrm{rank}(\mat{H}\mat{E}_{\mathrm{x}_k})},R\}\nonumber\\
    =& \mathrm{min}\{K\xi_\mathrm{x},R\}.
\end{align}
Therefore, $K\xi_\mathrm{x} \geq R$, and the upper-bound of $\xi_\mathrm{x}$ is obtained as $\xi_\mathrm{x} \leq \mathrm{min}\{\mathrm{rank}(\mat{H}), \mathrm{rank}(\diagof{\vet{\psi}_k}),\mathrm{rank}(\mat{G}),$ $\mathrm{rank}(\mat{W}_k)\}$, or simply
\begin{equation}\label{lemmax2}
    \xi_\mathrm{x} \leq \mathrm{min}\{\kappa_\mathrm{h},\kappa_\mathrm{g},R\}.
\end{equation}
Likewise, $\mathrm{rank}\bigl((\mat{I}_K \kron \mat{H})\mat{E}_\mathrm{x}\bigr) = \mathrm{min}\{K\xi_\mathrm{x},L\}$ for KRSTC, implying on $K\xi_\mathrm{x} \geq L$ and $\xi_\mathrm{x} \leq \mathrm{min}\{\kappa_\mathrm{h},\kappa_\mathrm{g}\}$.

\subsection{Proof of Proposition \ref{proposition:g}}\label{ap:ident2}
Using the property $\mathrm{rank}(\mat{AB}) \leq \mathrm{min}\{\mathrm{rank}(\mat{A}),\mathrm{rank}(\mat{B})\}$, we have
\begin{equation}
    \mathrm{rank}\bigl(\bar{\mat{F}}_\mathrm{g}\bigr) \leq \mathrm{min}\{\mathrm{rank}(\mat{I}_{K} \kron \mat{X}^\trans \kron \mat{I}_{N_c}),\mathrm{rank}(\mat{F}_\mathrm{g})\}.
\end{equation}
Since $\mat{F}_\mathrm{g}$ depends only on $\ten{T}_{\ma{\Phi}}$ and $\ten{W}$ (or $\mat{\Lambda}$), we consider it is designed to have full column rank, which implies $KRN_c \geq LN$ for TSTC, and $KN_c \geq N$ for KRSTC. Doing $\mathrm{rank}(\mat{F}_\mathrm{g}) = LN$ and applying the property $\mathrm{rank}(\mat{A} \kron \mat{B}) = \mathrm{rank}(\mat{A})\mathrm{rank}(\mat{B})$, we obtain
\begin{equation}\label{lemmag1}
    \mathrm{rank}\bigl(\bar{\mat{F}}_\mathrm{g}\bigr) \leq \mathrm{min}\{KN_c\kappa_\mathrm{x},LN\}.
\end{equation}

\subsection{Proof of Proposition \ref{proposition:h}}\label{ap:ident3}
For both TSTC and KRSTC transmission schemes, we can express the rank of the $k$-th block of the column-wise stacking $\mat{E}_\mathrm{h}(\mat{I}_K \kron \mat{X})$ as
\begin{equation}\label{lemmah2}
    \mathrm{rank}(\mat{E}_{\mathrm{h}_k}\mat{X})  = \xi_\mathrm{h} \leq \mathrm{min}\{\kappa_\mathrm{g},\kappa_\mathrm{x}\}.
\end{equation}
Following \cite{favier2020algebraic, matsaglia1974equalities}, to ensure that $\mat{E}_\mathrm{h}(\mat{I}_K \kron \mat{X})$ has full row-rank, the sum of the ranks of all blocks $\mat{E}_{\mathrm{h}_k}\mat{X}$ must be no less than $N$. On the assumption that the design of $\ten{W}$ and $\ma{\Psi}$ enforces $\mathrm{dim}[\mathtt{C}(\mat{E}_{\mathrm{h}_1}\mat{X}) \cap \cdots \cap \mathtt{C}(\mat{E}_{\mathrm{h}_K}\mat{X})]=0$, where $\mathtt{C}(\mat{E}_{\mathrm{h}_k}\mat{X})$ denotes the column-space of $\mat{E}_{\mathrm{h}_k}\mat{X}$, we have
\begin{align}\label{lemmah3}
    \mathrm{rank}\bigl(\mat{E}_\mathrm{h}(\mat{I}_K \kron \mat{X})\bigr) =& \mathrm{min}\Bigl\{N, \textstyle \sum_{k=1}^K{\mathrm{rank}(\mat{E}_{\mathrm{h}_k}\mat{X})}\Bigr\}\nonumber\\
    =& \mathrm{min}\{N, K\xi_\mathrm{h}\}.
\end{align}
Therefore, $K\xi_\mathrm{h} \geq N$.

\subsection{Design of Coding/Phase-Shifts}\label{apdesign}
We jointly design the sensing and reflecting phase shifts by adapting a procedure proposed in \cite{choi2023ajoint} while designing the tensor coding separately. We adopt an index vector $\tau_i^J \triangleq [(i-1)J+1,(i-1)J+2,\cdots,iJ] \in {\integer^\ast_{+}}^{J \times 1}$ for $i=1,\cdots,I$ to denote the $i$-th block of an $IJ$-dimensional column vector, in which each block has length of $J$. Consider a $KN_c$-dimensional DFT matrix $\mat{D} = [\vet{d}_1,\cdots,\vet{d}_{KN_c}]$. By sampling $\mat{D}$, the 3-mode fibers of $\ten{T}_{\ma{\Phi}}$ and the columns of $\ma{\Psi}$ are filled by, respectively, $\ma{\Phi}_{n_cn\cdot} = \vet{d}_n(\tau_{n_c}^K) \in \compl^{K \times 1}$ and $\ma{\Psi}_{\cdot n} = \vet{d}_{(n-1)N_c+1}(\tau_1^K) \in \compl^{K \times 1}$, for $n_c=1,\cdots,N_c$ and $n=1,\cdots,N$, where the constraint $KN_c \geq N$ is assumed. This yields respectively the following equivalent constructions for $\ten{T}_{\ma{\Phi}}$ and $\ma{\Psi}$, defined in Section \ref{sec:tenmodel}: 
\begin{equation*}
    \bigl[\ten{T}_{\ma{\Phi}}\bigr]_{(3)} \!=\!\! 
    \left[\hspace{-1.5ex}\begin{array}{*{7}{c@{\hspace{4pt}}}}
        d_{1,1}&\!\cdots\!&d_{(N_c\!-\!1)K\!+\!1,1}&\cdots&d_{1,N}&\!\cdots\!&d_{(N_c\!-\!1)K\!+\!1,1}\\
        \vdots&\!\cdots\!&\vdots&\!\cdots\!&\vdots&\!\cdots\!&\vdots\\
        d_{K,1}&\!\cdots\!&d_{N_cK,1}&\cdots&d_{K,N}&\!\cdots\!&d_{N_cK,N}
    \end{array}\hspace{-1.5ex}\right]\!\!,
\end{equation*}
\begin{equation*}
\ma{\Psi} = \begin{bmatrix}
       d_{1,1}&\!\cdots\!&d_{1,(N-1)N_c+1}\\
       \vdots&\!\cdots\!&\vdots\\
       d_{K,1}&\!\cdots\!&d_{K,(N-1)N_c+1}
   \end{bmatrix}.
\end{equation*}
To design the coding tensor $\ten{W}$ (TSTC), we first construct the matrix $\ma{\Upsilon} \in \real^{K \times RL}$ by truncating a $K$-dimensional Hadamard matrix to its first $RL$ columns, such that $\ma{\Upsilon} = \unf{W}{3}$. Then, we get $\ten{W}$ by tensorizing $\ma{\Upsilon}$ or, simply, by doing $\mat{W}_k = \mathrm{unvec}_{L \times R}\{\ma{\Upsilon}_{k\cdot}^\trans\}$, $k=1,\cdots,K$. For the KRSTC scheme, the coding matrix $\ma{\Lambda}$ is designed as a truncated Hadamard matrix, where $K \geq L$. This design can prevent the generalized inverses mentioned in Section \ref{sec:cost} by replacing them with matrix multiplications through simplified expressions. For \ KRSTC, this is achieved when $K \geq LN$, and for TSTC, when $K \geq RLN$. Herein, however, we prioritize choosing the minimum number of sub-frames required to ensure joint symbol and CE uniqueness for all semi-blind receivers.

\bibliographystyle{IEEEtran}
\bibliography{references}

\begin{thebibliography}{10}
\providecommand{\url}[1]{#1}
\csname url@samestyle\endcsname
\providecommand{\newblock}{\relax}
\providecommand{\bibinfo}[2]{#2}
\providecommand{\BIBentrySTDinterwordspacing}{\spaceskip=0pt\relax}
\providecommand{\BIBentryALTinterwordstretchfactor}{4}
\providecommand{\BIBentryALTinterwordspacing}{\spaceskip=\fontdimen2\font plus
\BIBentryALTinterwordstretchfactor\fontdimen3\font minus \fontdimen4\font\relax}
\providecommand{\BIBforeignlanguage}[2]{{%
\expandafter\ifx\csname l@#1\endcsname\relax
\typeout{** WARNING: IEEEtran.bst: No hyphenation pattern has been}%
\typeout{** loaded for the language `#1'. Using the pattern for}%
\typeout{** the default language instead.}%
\else
\language=\csname l@#1\endcsname
\fi
#2}}
\providecommand{\BIBdecl}{\relax}
\BIBdecl

\bibitem{tishchenko2025theemergence}
A.~Tishchenko \emph{et~al.}, ``The emergence of multi-functional and hybrid reconfigurable intelligent surfaces for integrated sensing and communications - a survey,'' \emph{IEEE Commun. Surveys Tuts.}, pp. 1--1, 2025.

\bibitem{katwe2024overview}
M.~V. Katwe \emph{et~al.}, ``An overview of intelligent meta-surfaces for 6{G} and beyond: opportunities, trends, and challenges,'' \emph{IEEE Commun. Stand. Mag.}, vol.~8, no.~4, pp. 62--69, 2024.

\bibitem{alexandropoulos2024hybrid}
G.~C. Alexandropoulos, N.~Shlezinger, I.~Alamzadeh, M.~F. Imani, H.~Zhang, and Y.~C. Eldar, ``Hybrid reconfigurable intelligent metasurfaces: Enabling simultaneous tunable reflections and sensing for 6{G} wireless communications,'' \emph{IEEE Veh. Technol Mag.}, vol.~19, no.~1, pp. 75--84, 2024.

\bibitem{basar2024reconfigurable}
E.~Basar \emph{et~al.}, ``Reconfigurable intelligent surfaces for 6{G}: Emerging hardware architectures, applications, and open challenges,'' \emph{IEEE Veh. Technol Mag.}, vol.~19, no.~3, pp. 27--47, 2024.

\bibitem{chen2023reconfigurable}
Z.~Chen \emph{et~al.}, ``Reconfigurable-intelligent-surface-assisted {B5G/6G} wireless communications: Challenges, solution, and future opportunities,'' \emph{IEEE Commun. Mag.}, vol.~61, no.~1, pp. 16--22, 2023.

\bibitem{pan2021reconfigurable}
C.~Pan \emph{et~al.}, ``Reconfigurable intelligent surfaces for 6{G} systems: Principles, applications, and research directions,'' \emph{IEEE Commun. Mag.}, vol.~59, no.~6, pp. 14--20, 2021.

\bibitem{you2025next}
C.~You \emph{et~al.}, ``Next generation advanced transceiver technologies for 6{G} and beyond,'' \emph{IEEE J. Sel. Areas Commun.}, 2025.

\bibitem{huang2020holographic}
C.~Huang \emph{et~al.}, ``Holographic {MIMO} surfaces for {6G} wireless networks: Opportunities, challenges, and trends,'' \emph{IEEE Wireless Commun.}, vol.~27, no.~5, pp. 118--125, 2020.

\bibitem{wu2020towards}
Q.~Wu and R.~Zhang, ``Towards smart and reconfigurable environment: Intelligent reflecting surface aided wireless network,'' \emph{IEEE Commun. Mag.}, vol.~58, no.~1, pp. 106--112, 2020.

\bibitem{wu2020beamforming}
------, ``Beamforming optimization for wireless network aided by intelligent reflecting surface with discrete phase shifts,'' \emph{IEEE Trans. Commun.}, vol.~68, no.~3, pp. 1838--1851, 2020.

\bibitem{dong2020secure}
L.~Dong and H.-M. Wang, ``Secure {MIMO} transmission via intelligent reflecting surface,'' \emph{IEEE Wireless Commun. Lett.}, vol.~9, no.~6, pp. 787--790, 2020.

\bibitem{alfattani2021aerial}
S.~Alfattani \emph{et~al.}, ``Aerial\! platforms\! with\! reconfigurable\! smart\! surfaces\! for\! {5G}\! and\! beyond,'' \emph{IEEE Commun. Mag.}, vol.~59, no.~1, pp. 96--102, 2021.

\bibitem{chepuri2023integrated}
S.~P. Chepuri, N.~Shlezinger, F.~Liu, G.~C. Alexandropoulos, S.~Buzzi, and Y.~C. Eldar, ``Integrated sensing and communications with reconfigurable intelligent surfaces: From signal modeling to processing,'' \emph{IEEE Signal Process. Mag.}, vol.~40, no.~6, pp. 41--62, 2023.

\bibitem{han2023csi}
Y.~Han, W.~Tang, X.~Li, M.~Matthaiou, and S.~Jin, ``{CSI} acquisition in {RIS}-assisted mobile communication systems,'' \emph{Natl. Sci. Rev.}, vol.~10, no.~8, p. nwad127, 2023.

\bibitem{renzo2020smart}
M.~Di~Renzo \emph{et~al.}, ``Smart\! radio\! environments\! empowered\! by\! reconfigurable\! intelligent\! surfaces:\! how\! it\! works,\! state\! of\! research,\! and \!the\! road\! ahead,'' \emph{\hspace{-0.5ex}IEEE\! J.\! Sel.\! Areas\! Commun.}, vol. \hspace{-0.5ex}38, no. \hspace{-0.5ex}11, pp. \hspace{--0.5ex}2450--2525, \hspace{-0.5ex}2020.

\bibitem{swindlehurst2022channel}
A.~L. Swindlehurst, G.~Zhou, R.~Liu, C.~Pan, and M.~Li, ``Channel estimation with reconfigurable intelligent surfaces — a general framework,'' \emph{Proc. IEEE}, vol. 110, no.~9, pp. 1312--1338, 2022.

\bibitem{schroeder2022two}
R.~Schroeder, J.~He, G.~Brante, and M.~Juntti, ``Two-stage channel estimation for hybrid {RIS} assisted {MIMO} systems,'' \emph{IEEE Trans. Commun.}, vol.~70, no.~7, pp. 4793--4806, 2022.

\bibitem{jin2021channel}
Y.~Jin, J.~Zhang, X.~Zhang, H.~Xiao, B.~Ai, and D.~W.~K. Ng, ``Channel estimation for semi-passive reconfigurable intelligent surfaces with enhanced deep residual networks,'' \emph{IEEE Trans. Veh. Technol.}, vol.~70, no.~10, pp. 11\,083--11\,088, 2021.

\bibitem{zhu2023sensing}
J.~Zhu, K.~Liu, Z.~Wan, L.~Dai, T.~J. Cui, and H.~V. Poor, ``Sensing {RIS}s: Enabling dimension-independent {CSI} acquisition for beamforming,'' \emph{IEEE Trans. Inf. Theory}, vol.~69, no.~6, pp. 3795--3813, 2023.

\bibitem{yang2023separate}
S.~Yang, W.~Lyu, D.~Wang, and Z.~Zhang, ``Separate channel estimation with hybrid {RIS}-aided multi-user communications,'' \emph{IEEE Trans. Veh. Technol.}, vol.~72, no.~1, pp. 1318--1324, 2023.

\bibitem{luo2023reconfigurable}
C.~Luo, J.~Hu, L.~Xiang, and K.~Yang, ``Reconfigurable intelligent sensing surface aided wireless powered communication networks: A sensing-then-reflecting approach,'' \emph{IEEE Trans. Commun.}, 2023.

\bibitem{zhang2023channel}
H.~Zhang \emph{et~al.}, ``Channel estimation with hybrid reconfigurable intelligent metasurfaces,'' \emph{\hspace{-0.5ex}IEEE Trans. Commun.}, vol.~71, no.~4, pp. \hspace{--0.5ex}2441--2456, \hspace{-0.5ex}2023.

\bibitem{hu2022semipassive}
X.~Hu, R.~Zhang, and C.~Zhong, ``Semi-passive elements assisted channel estimation for intelligent reflecting surface-aided communications,'' \emph{IEEE Trans. Wireless Commun.}, vol.~21, no.~2, pp. 1132--1142, 2022.

\bibitem{lin2021tensor}
Y.~Lin, S.~Jin, M.~Matthaiou, and X.~You, ``Tensor-based algebraic channel estimation for hybrid {IRS}-assisted {MIMO}-{OFDM},'' \emph{IEEE Trans. Wireless Commun.}, vol.~20, no.~6, pp. 3770--3784, 2021.

\bibitem{liu2020deep}
S.~Liu, Z.~Gao, J.~Zhang, M.~Di~Renzo, and M.-S. Alouini, ``Deep denoising neural network assisted compressive channel estimation for mm{W}ave intelligent reflecting surfaces,'' \emph{IEEE Trans. Veh. Technol.}, vol.~69, no.~8, pp. 9223--9228, 2020.

\bibitem{wei2021channel}
L.~Wei, C.~Huang, G.~C. Alexandropoulos, C.~Yuen, Z.~Zhang, and M.~Debbah, ``Channel estimation for {RIS}-empowered multi-user {MISO} wireless communications,'' \emph{IEEE Trans. Commun.}, vol.~69, no.~6, pp. 4144--4157, 2021.

\bibitem{dearaujo2021channel}
G.~T. de~Ara{\'u}jo, A.~L.~F. de~Almeida, and R.~Boyer, ``Channel estimation for intelligent reflecting surface assisted {MIMO} systems: A tensor modeling approach,'' \emph{IEEE J. Sel. Topics Signal Process.}, vol.~15, no.~3, pp. 789--802, 2021.

\bibitem{choi2023ajoint}
J.~Choi and J.~H. Cho, ``A joint optimization of pilot and phase shifts in uplink channel estimation for hybrid {RIS}-aided multi-user communication systems,'' \emph{IEEE Trans. Veh. Technol.}, 2023.

\bibitem{zegrar2020ageneral}
S.~E. Zegrar, L.~Afeef, and H.~Arslan, ``A general framework for {RIS}-aided mm{W}ave communication networks: Channel estimation and mobile user tracking,'' \emph{arXiv preprint arXiv:2009.01180}, 2020.

\bibitem{nadeem2020asymptotic}
Q.-U.-A. Nadeem, A.~Kammoun, A.~Chaaban, M.~Debbah, and M.-S. Alouini, ``Asymptotic max-min {SINR} analysis of reconfigurable intelligent surface assisted {MISO} systems,'' \emph{IEEE Trans. Wireless Commun.}, vol.~19, no.~12, pp. 7748--7764, 2020.

\bibitem{li2022joint}
R.~Li, B.~Guo, M.~Tao, Y.-F. Liu, and W.~Yu, ``Joint design of hybrid beamforming and reflection coefficients in {RIS}-aided mm{W}ave {MIMO} systems,'' \emph{IEEE Trans. Commun.}, vol.~70, no.~4, pp. 2404--2416, 2022.

\bibitem{ye2020joint}
J.~Ye, S.~Guo, and M.-S. Alouini, ``Joint reflecting and precoding designs for {SER} minimization in reconfigurable intelligent surfaces assisted {MIMO} systems,'' \emph{IEEE Trans. Wireless Commun.}, vol.~19, no.~8, pp. 5561--5574, 2020.

\bibitem{sun2022energy}
Y.~Sun \emph{et~al.}, ``Energy-efficient hybrid beamforming for multilayer ris-assisted secure integrated terrestrial-aerial networks,'' \emph{IEEE Trans. Commun.}, vol.~70, no.~6, pp. 4189--4210, 2022.

\bibitem{boiadjieva2023joint}
B.~Boiadjieva and M.~Vu, ``Joint multi-user channel estimation for hybrid reconfigurable intelligent surfaces,'' in \emph{IEEE Int. Conf. Commun.}\hskip 1em plus 0.5em minus 0.4em\relax IEEE, 2023, pp. 877--882.

\bibitem{yang2023active}
S.~Yang, W.~Lyu, Y.~Xiu, Z.~Zhang, and C.~Yuen, ``Active {3D} double-{RIS}-aided multi-user communications: Two-timescale-based separate channel estimation via bayesian learning,'' \emph{IEEE Trans. Commun.}, 2023.

\bibitem{hu2021twotimescale}
C.~Hu, L.~Dai, S.~Han, and X.~Wang, ``Two-timescale channel estimation for reconfigurable intelligent surface aided wireless communications,'' \emph{IEEE Trans. Commun.}, vol.~69, no.~11, pp. 7736--7747, 2021.

\bibitem{chen2021low}
X.~Chen, J.~Shi, Z.~Yang, and L.~Wu, ``Low-complexity channel estimation for intelligent reflecting surface-enhanced massive {MIMO},'' \emph{IEEE Wireless Commun. Lett.}, vol.~10, no.~5, pp. 996--1000, 2021.

\bibitem{taha2021enabling}
A.~Taha, M.~Alrabeiah, and A.~Alkhateeb, ``Enabling large intelligent surfaces with compressive sensing and deep learning,'' \emph{IEEE Access}, vol.~9, pp. 44\,304--44\,321, 2021.

\bibitem{alexandropoulos2020ahardware}
G.~C. Alexandropoulos and E.~Vlachos, ``A hardware architecture for reconfigurable intelligent surfaces with minimal active elements for explicit channel estimation,'' in \emph{Proc. IEEE Int. Conf. Acoust. Speech Signal Process.}\hskip 1em plus 0.5em minus 0.4em\relax IEEE, 2020, pp. 9175--9179.

\bibitem{shlezinger2021dynamic}
N.~Shlezinger, G.~C. Alexandropoulos, M.~F. Imani, Y.~C. Eldar, and D.~R. Smith, ``Dynamic metasurface antennas for {6G} extreme massive {MIMO} communications,'' \emph{IEEE Wireless Commun.}, vol.~28, no.~2, pp. 106--113, 2021.

\bibitem{alamzadeh2021reconfigurable}
I.~Alamzadeh, G.~C. Alexandropoulos, N.~Shlezinger, and M.~F. Imani, ``A reconfigurable intelligent surface with integrated sensing capability,'' \emph{Sci. Rep.}, vol.~11, no.~1, p. 20737, 2021.

\bibitem{Sokal_2023}
B.~Sokal, P.~R.~B. Gomes, A.~L.~F. de~Almeida, B.~Makki, and G.~Fodor, ``Reducing the control overhead of intelligent reconfigurable surfaces via a tensor-based low-rank factorization approach,'' \emph{IEEE Trans. Wireless Commun.}, vol.~22, no.~10, pp. 6578--6593, 2023.

\bibitem{Almeida_Elsevier_2007}
A.~L.~F. {de Almeida}, G.~Favier, and J.~C.~M. Mota, ``{PARAFAC}-based unified tensor modeling for wireless communication systems with application to blind multiuser equalization,'' \emph{Signal Process.}, vol.~87, no.~2, pp. 337--351, 2007.

\bibitem{teseandre}
A.~L.~F. de~Almeida, ``Tensor modeling and signal processing for wireless communication systems,'' Ph.D. dissertation, Université de Nice-Sophia Antipolis, 11 2007.

\bibitem{sidiropoulos2000blind}
N.~D. Sidiropoulos, G.~B. Giannakis, and R.~Bro, ``Blind {PARAFAC} receivers for {DS}-{CDMA} systems,'' \emph{IEEE Trans. Signal Process.}, vol.~48, no.~3, pp. 810--823, 2000.

\bibitem{favier2012tensor}
G.~Favier, M.~N. da~Costa, A.~L.~F. de~Almeida, and J.~M.~T. Romano, ``Tensor space--time ({TST}) coding for {MIMO} wireless communication systems,'' \emph{Signal Process.}, vol.~92, no.~4, pp. 1079--1092, 2012.

\bibitem{favier2014tensor}
G.~Favier and A.~L.~F. de~Almeida, ``Tensor space-time-frequency coding with semi-blind receivers for {MIMO} wireless communication systems,'' \emph{IEEE Trans. Signal Process.}, vol.~62, no.~22, pp. 5987--6002, 2014.

\bibitem{Chen2021}
H.~Chen, F.~Ahmad, S.~Vorobyov, and F.~Porikli, ``Tensor decompositions in wireless communications and {MIMO} radar,'' \emph{IEEE J. Sel. Topics Signal Process.}, vol.~15, no.~3, pp. 438--453, 2021.

\bibitem{Miron2020}
S.~Miron \emph{et~al.}, ``Tensor methods for multisensor signal processing,'' \emph{IET Signal Process.}, vol.~14, no.~10, pp. 693--709, 2020.

\bibitem{Sidiropoulos2017}
N.~D. Sidiropoulos, L.~De~Lathauwer, X.~Fu, K.~Huang, E.~E. Papalexakis, and C.~Faloutsos, ``Tensor decomposition for signal processing and machine learning,'' \emph{IEEE Trans. Signal Process.}, vol.~65, no.~13, pp. 3551--3582, 2017.

\bibitem{Almeida2016}
A.~L.~F. de~Almeida, G.~Favier, J.~P. da~Costa, and J.~C.~M. Mota, ``{Overview of tensor decompositions with applications to communications},'' in \emph{{Signals and Images: Adv. Results in Speech, Estimation, Compression, Recognition, Filtering, and Processing}}.\hskip 1em plus 0.5em minus 0.4em\relax {CRC-Press}, 1 2016, no. Chapter 12, pp. 325--356.

\bibitem{ardah2021trice}
K.~Ardah, S.~Gherekhloo, A.~L.~F. de~Almeida, and M.~Haardt, ``{TRICE}: A channel estimation framework for {RIS}-aided millimeter-wave {MIMO} systems,'' \emph{IEEE Signal Process. Lett.}, vol.~28, pp. 513--517, 2021.

\bibitem{dearaujo2022semiblind}
G.~T. de~Ara{\'u}jo, P.~R.~B. Gomes, A.~L.~F. de~Almeida, G.~Fodor, and B.~Makki, ``Semi-blind joint channel and symbol estimation in {IRS}-assisted multiuser {MIMO} networks,'' \emph{IEEE Wireless Commun. Lett.}, vol.~11, no.~7, pp. 1553--1557, 2022.

\bibitem{dearaujo2023semiblind}
G.~T. de~Ara{\'u}jo, A.~L.~F. de~Almeida, R.~Boyer, and G.~Fodor, ``Semi-blind joint channel and symbol estimation for {IRS}-assisted {MIMO} systems,'' \emph{IEEE Trans. Signal Process.}, vol.~71, pp. 1184--1199, 2023.

\bibitem{gomes2023channel}
P.~R.~B. Gomes, G.~T. de~Ara{\'u}jo, B.~Sokal, A.~L.~F. de~Almeida, B.~Makki, and G.~Fodor, ``Channel\! estimation\! in\! {RIS}-assisted\! {MIMO}\! systems \!operating \!under\! imperfections,'' \emph{IEEE Trans. Veh. Technol.}, 2023.

\bibitem{comon2009tensor}
P.~Comon, X.~Luciani, and A.~L.~F. de~Almeida, ``{Tensor decompositions, alternating least squares and other tales},'' \emph{J. Chemom.}, vol.~23, no. 7-8, pp. 393--405, 2009.

\bibitem{sepideh2024anefficient}
S.~Gherekhloo, K.~Ardah, A.~L.~F. de~Almeida, and M.~Haardt, ``An efficient channel training protocol for channel estimation in double {RIS}-aided {MIMO} systems,'' in \emph{2024 32nd European Signal Processing Conference (EUSIPCO)}, 2024, pp. 2067--2071.

\bibitem{magalhaes2025closed}
A.~L. Magalhães, P.~R.~B. Gomes, A.~L.~F. de~Almeida, and L.~Deneire, ``Closed-form receivers for {MIMO} communications assisted by hybrid sensing and reflecting {RIS},'' in \emph{Proc. IEEE International Conference on Communications (ICC)}, 2025, pp. 6628--6633, to appear.

\bibitem{harshman1996uniqueness}
R.~A. Harshman and M.~E. Lundy, ``Uniqueness proof for a family of models sharing features of {T}ucker's three-mode factor analysis and {PARAFAC/CANDECOMP},'' \emph{Psychometrika}, vol.~61, pp. 133--154, 1996.

\bibitem{dealmeida2009space}
A.~L.~F. de~Almeida, G.~Favier, and J.~C.~M. Mota, ``Space--time spreading--multiplexing for {MIMO} wireless communication systems using the {PARATUCK}-2 tensor model,'' \emph{Signal Process.}, vol.~89, no.~11, pp. 2103--2116, 2009.

\bibitem{magalhaes2025reducing}
A.~L. Magalhães, A.~L.~F. de~Almeida, and G.~T. de~Araújo, ``Reducing complexity of data-aided channel estimation in {RIS}-assisted communications,'' \emph{IEEE Wireless Commun. Lett.}, pp. 1--1, 2025.

\bibitem{nguyen2024decision}
L.~V. Nguyen and A.~L. Swindlehurst, ``Decision-directed hybrid {RIS} channel estimation with minimal pilot overhead,'' \emph{IEEE Trans. Commun.}, vol.~72, no.~10, pp. 6505--6519, 2024.

\bibitem{gavras2025simultaneous}
I.~Gavras and G.~C. Alexandropoulos, ``Simultaneous communications and sensing with hybrid reconfigurable intelligent surfaces,'' in \emph{19th European Conference on Antennas and Propagation (EuCAP)}.\hskip 1em plus 0.5em minus 0.4em\relax IEEE, 2025, pp. 1--5.

\bibitem{huang2022coordinated}
Y.~Huang, Y.~Fang, X.~Li, and J.~Xu, ``Coordinated power control for network integrated sensing and communication,'' \emph{IEEE Trans. Veh. Technol.}, vol.~71, no.~12, pp. 13\,361--13\,365, 2022.

\bibitem{taghvaee2024fully}
H.~Taghvaee, M.~Khodadadi, G.~Gradoni, and M.~Khalily, ``Fully autonomous reconfigurable metasurfaces with integrated sensing and communication,'' in \emph{18th European Conference on Antennas and Propagation (EuCAP)}.\hskip 1em plus 0.5em minus 0.4em\relax IEEE, 2024, pp. 1--5.

\bibitem{strinati2024distributed}
E.~C. Strinati \emph{et~al.}, ``Distributed intelligent integrated sensing and communications: The {6G}-{DISAC} approach,'' in \emph{Joint European Conference on Networks and Communications \& {6G} Summit (EuCNC/6G Summit)}.\hskip 1em plus 0.5em minus 0.4em\relax IEEE, 2024, pp. 392--397.

\bibitem{ximenes2014parafac}
L.~R. Ximenes, G.~Favier, A.~L.~F. de~Almeida, and Y.~C.~B. Silva, ``{PARAFAC}-{PARATUCK} semi-blind receivers for two-hop cooperative {MIMO} relay systems,'' \emph{IEEE Trans. Signal Process.}, vol.~62, no.~14, pp. 3604--3615, 2014.

\bibitem{Sidropoulos_2002_KRST}
N.~Sidiropoulos and R.~Budampati, ``Khatri-{R}ao space-time codes,'' \emph{IEEE Trans. Signal Process.}, vol.~50, no.~10, pp. 2396--2407, 2002.

\bibitem{kolda2009tensor}
T.~G. Kolda and B.~W. Bader, ``Tensor decompositions and applications,'' \emph{SIAM Review}, vol.~51, no.~3, pp. 455--500, 2009.

\bibitem{FavierAlmeida2014}
G.~Favier and A.~L.~F. de~Almeida, ``Overview of constrained {PARAFAC} models,'' \emph{EURASIP Journal of Advances in Signal Processing}, vol. 2014, no. 142, pp. 1--25, Sep. 2014.

\bibitem{van1993approximation}
C.~F. Van~Loan and N.~Pitsianis, ``Approximation with {K}ronecker products,'' in \emph{Linear algebra for large scale and real-time applications}.\hskip 1em plus 0.5em minus 0.4em\relax Springer, 1993, pp. 293--314.

\bibitem{Kibangou2009}
A.~Y. {Kibangou} and G.~{Favier}, ``Non-iterative solution for {PARAFAC} with a toeplitz matrix factor,'' in \emph{Proc. EUSIPCO}, 8 2009, pp. 691--695.

\bibitem{teseflorian}
F.~Roemer, ``Advanced algebraic concepts for efficient multi-channel signal processing,'' Ph.D. dissertation, Ilmenau University of Technology, 2013.

\bibitem{rahman2019framework}
M.~L. Rahman, J.~A. Zhang, X.~Huang, Y.~J. Guo, and R.~W. Heath, ``Framework for a perceptive mobile network using joint communication and radar sensing,'' \emph{IEEE Trans. Aerosp. Electron. Syst.}, vol.~56, no.~3, pp. 1926--1941, 2019.

\bibitem{zhang2020perceptive}
A.~Zhang, M.~L. Rahman, X.~Huang, Y.~J. Guo, S.~Chen, and R.~W. Heath, ``Perceptive mobile networks: Cellular networks with radio vision via joint communication and radar sensing,'' \emph{IEEE Veh. Technol. Mag.}, vol.~16, no.~2, pp. 20--30, 2020.

\bibitem{alexandropoulos2022localization}
G.~C. Alexandropoulos, I.~Vinieratou, and H.~Wymeersch, ``Localization via multiple reconfigurable intelligent surfaces equipped with single receive {RF} chains,'' \emph{IEEE Wireless Commun. Lett.}, vol.~11, no.~5, pp. 1072--1076, 2022.

\bibitem{shao2022target}
X.~Shao, C.~You, W.~Ma, X.~Chen, and R.~Zhang, ``Target sensing with intelligent reflecting surface: Architecture and performance,'' \emph{IEEE J. Sel. Areas Commun.}, vol.~40, no.~7, pp. 2070--2084, 2022.

\bibitem{yigit2022overtheair}
Z.~Yigit, E.~Basar, and I.~Altunbas, ``Over-the-air beamforming with reconfigurable intelligent surfaces,'' \emph{Front. Comms. Net.}, vol.~3, p. 1016270, 2022.

\bibitem{huang2023roadside}
Z.~Huang, B.~Zheng, and R.~Zhang, ``Roadside {IRS}-aided vehicular communication: Efficient channel estimation and low-complexity beamforming design,'' \emph{IEEE Trans. Wireless Commun.}, 2023.

\bibitem{ahmed2023vehicular}
M.~Ahmed \emph{et~al.}, ``Vehicular communication network enabled {CAV} data offloading: A review,'' \emph{IEEE Trans. Intell. Transp. Syst.}, 2023.

\bibitem{karagiannis2011vehicular}
G.~Karagiannis \emph{et~al.}, ``Vehicular networking: A survey and tutorial on requirements, architectures, challenges, standards and solutions,'' \emph{IEEE Commun. Surveys Tuts.}, vol.~13, no.~4, pp. 584--616, 2011.

\bibitem{favier2020algebraic}
G.~Favier, \emph{From Algebraic Structures to Tensors}.\hskip 1em plus 0.5em minus 0.4em\relax John Wiley \& Sons, 2020, vol.~1.

\bibitem{matsaglia1974equalities}
G.~Matsaglia and G.~P~H~Styan, ``Equalities and inequalities for ranks of matrices,'' \emph{Linear and multilinear Algebra}, vol.~2, no.~3, pp. 269--292, 1974.

\end{thebibliography}

\end{document}